\documentclass[a4paper,12pt]{article}
\usepackage{jheppub}
\usepackage[dvipsnames]{xcolor}
\usepackage{pgfplots}
\usepackage[T1]{fontenc}
\usepackage[]{slashed}
\usepackage[]{bm}     
\usepackage{physics}
\usepackage{dsfont}
\usetikzlibrary{shapes.geometric}

\usepackage{hyperref}
\hypersetup{colorlinks=true,linkcolor=magenta,anchorcolor=green,citecolor=cyan,filecolor=black,menucolor=black,urlcolor=brown}
\makeatletter

\usepackage{lipsum}

\newlength{\eqreserve}
\newsavebox{\alignedbox}
\newenvironment{naligned}
  {\begin{lrbox}{\alignedbox}$\displaystyle\begin{aligned}}
  {\end{aligned}$\end{lrbox}%
   \begin{equation}\makebox[\dimexpr\linewidth-\eqreserve][c]{\usebox{\alignedbox}}\end{equation}}

\usepackage{graphicx}
\usepackage{epstopdf}
\usepackage{bbm,amsmath,graphicx,amssymb,amsfonts,amsthm}

\def\tr{\textrm{tr}}

\usepackage{soul}
\definecolor{bubbles}{rgb}{0.91, 1.0, 1.0}
\definecolor{aquamarine}{rgb}{0.5, 1.0, 0.83}
\definecolor{bubblegum}{rgb}{0.99, 0.76, 0.8}
\definecolor{blackbell}{rgb}{0.64, 0.64, 0.82}
\definecolor{dollarbill}{rgb}{0.72, 0.93, 0.6}

\def\sq[#1,#2]{\left[#1\,#2\right]}
\def\an[#1,#2]{\left\langle#1\,#2\right\rangle}

\def\an[#1,#2]{\left\langle#1\,#2\right\rangle}
\def\aq[#1,#2,#3]{\left\langle#1|#2|#3\right]}
\def\qa[#1,#2,#3]{\left[#1|#2|#3\right\rangle}
\def\sq[#1,#2]{\left[#1\,#2\right]}
\def\spa#1.#2{\left\langle#1\,#2\right\rangle}
\def\spab[#1,#2,#3]{\left\langle#1|#2|#3\right]}
\def\spba[#1,#2,#3]{\left[#1|#2|#3\right\rangle}
\def\spb#1.#2{\left[#1\,#2\right]}

\def\Ttrma(#1,#2,#3,#4){{\rm tr}_{-}[\slash \!\!\!\;\!\! #1\slash  \!\!\!\;\!\! #2 \slash  \!\!\!\;\!\!#3\slash  \!\!\!\;\!\!#4]}
\def\Ttrmb(#1,#2,#3,#4,#5,#6){{\rm tr}_{-}[\slash \!\!\!\;\!\! #1\slash  \!\!\!\;\!\! #2 \slash  \!\!\!\;\!\!#3\slash  \!\!\!\;\!\!#4\slash  \!\!\!\;\!\!#5\slash  \!\!\!\;\!\!#6]}
\def\Ttrmc(#1,#2,#3,#4,#5,#6,#7,#8){{\rm tr}_{-}[\slash \!\!\!\;\!\! #1\slash  \!\!\!\;\!\! #2 \slash  \!\!\!\;\!\!#3\slash  \!\!\!\;\!\!#4\slash  
\!\!\!\;\!\!#5\slash  \!\!\!\;\!\!#6\slash  \!\!\!\;\!\!#7\slash  \!\!\!\;\!\!#8]}
\def\Dp(#1,#2){(#1\cdot #2)}

\def\triangleboxleft{\scalebox{.9}{$\triangleleft$}\kern-.1em\Box}
\def\triangleboxright{\Box\kern-.1em\scalebox{.9}{$\triangleright$}}
\def\dBox{\Box\kern-.1em\Box}
\def\dNPBoxs{\scalebox{.9}{$\bowtie$}\kern-.1em\Box}
\def\dNPBoxu{\Box\kern-.1em\scalebox{.9}{$\bowtie$}}

\def\beq{\begin{equation}}
\def\eeq{\end{equation}}
\def\bes{\begin{split}}
\def\ees{\end{split}}
\def\beqa{\begin{eqnarray}}
\def\eeqa{\end{eqnarray}}

\def\eeqa{\end{eqnarray}}
\def\ek[#1,#2]{(\varepsilon_{#1}\cdot k_{#2})}
\def\e[#1,#2]{(\varepsilon_{#1}\cdot \varepsilon_{#2})}
\def\s(#1,#2){{(\ell_#1\cdot\ell_#2)}}

\def\e{\epsilon}

\newcommand{\ord}[1]{{\scriptscriptstyle (#1)}}

\pgfdeclarelayer{bg}    
\pgfsetlayers{bg,main}  

\usepackage{float}

\usetikzlibrary{shapes.misc}
\usetikzlibrary{arrows}
\usetikzlibrary{decorations.markings}
\usetikzlibrary{positioning,fit}
\usetikzlibrary{patterns}

\tikzset{cross/.style={cross out, draw=black, minimum size=2*(#1-\pgflinewidth), inner sep=0pt, outer sep=0pt},
	cross/.default={2pt}}

\title{Recursion Relations for Classical Gravity}

\author[a,b]{\!\! Poul H. Damgaard,}
\author[c]{\!\! Kwangeon Kim,}
\author[d]{\!\! and Kanghoon Lee}

\affiliation[a]{Niels Bohr International Academy, Niels Bohr Institute, University of Copenhagen, Blegdamsvej 17, DK-2100 Copenhagen, Denmark}
\affiliation[b]{Theoretical Physics Department, CERN, 1211 Geneva 23, Switzerland}
\affiliation[c]{Department of Physics, Yonsei University, Seoul 03722, Korea}
\affiliation[d]{Quantum Universe Center, Korea Institute for Advanced Study, 85 Hoegi-ro, Dondaemun-gu, Seoul 02455, Korea}

\preprint{CERN-TH-2026-163}

\abstract{We derive a set of recursion relations for the perturbative evaluation of the scattering of two 
black holes in Einstein gravity. To illustrate, we solve the equations up to third post-Minkowskian order
in the gravitational coupling constant G, recovering the correct result, including the back-reaction 
from gravitational radiation.}

\begin{document} 
\maketitle
\flushbottom
\section{Introduction}\label{sec:intro}

In recent papers~\cite{Damgaard:2024fqj,Damgaard:2026kqg,Damgaard:2026ocb} it has been demonstrated how a systematic iterative 
formalism can be established for the solution of Einstein's field equations in perturbation theory. The recursive
nature of the solution makes it particularly suited for high-order calculations since the solution to each
new order in the coupling constant $G$ is built up from combinations of lower-order results that have already
been computed. An illuminating example is the evaluation of the Schwarzschild metric to all orders in
perturbation theory: In a suitable choice of variables~\cite{Damgaard:2024fqj} the metric takes the form of a 
purely geometric series plus a few finite-order terms. The series sums to the known non-perturbative solution 
for all radii $r$ in the region $GM/r < 1$, where $M$ is the mass of the black hole. This raises the hope that 
in larger generality the post-Minkowskian perturbation theory of general relativity may be convergent in 
specific dynamical situations, such as the important two-body problem of Einstein gravity. The purpose of the 
present paper is to explore the impact of using algebraic recursion relations similar
to those introduced in ref.~\cite{Damgaard:2024fqj} in the dynamical set-up of two-body scalar scattering 
in general relativity. 

Our framework shall thus be that of the post-Minkowskian expansion which can be described as gravitational perturbation theory in just one quantity, Newton's constant $G$, around a background Minkowskian space-time. In early days, this was pursued by perturbatively solving the Einstein equations of motion (see, $e.g.$, 
ref.~\cite{Bel:1981be}, and in particular also ref.~\cite{Westpfahl:1985tsl} which provided the first calculation of the scattering angle to second post-Minkowskian order). More recently, the amplitude-based approach has introduced powerful new methods for such calculations~\cite{Damour:2016gwp,Damour:2017zjx,Bjerrum-Bohr:2018xdl,Cheung:2018wkq,Bern:2019nnu,Bern:2019crd,Damour:2019lcq,DiVecchia:2020ymx,Damour:2020tta,DiVecchia:2021ndb,DiVecchia:2021bdo,Bjerrum-Bohr:2021vuf,Bjerrum-Bohr:2021din,DiVecchia:2022owy,DiVecchia:2022piu,Damgaard:2021rnk}. These
amplitude calculations provide different angles on the problem, and have been explored also in the Hamiltonian language for the
conservative pieces~\cite{Cheung:2018wkq,Cristofoli:2019neg,Cristofoli:2020uzm,Kalin:2019rwq,Bjerrum-Bohr:2019kec,Damgaard:2022jem}. An
all-encompassing framework, including dissipative effects due to gravitational radiation, is that of the KMOC-formalism~\cite{Kosower:2018adc,Maybee:2019jus,Cristofoli:2021vyo,Damgaard:2023vnx,Damgaard:2023ttc} which computes the change in physical observables based on the quantum field theoretic {\em in-in} formalism.

Because the quantum field theoretic amplitude methods applied to gravity are capable of describing full quantum mechanical scattering
amplitudes several schemes have been devised to subtract all pieces that do not contribute to the classical limit. One approach is to Fourier transform into impact parameter space and exploit the associated exponentiation of the scattering amplitudes in
the eikonal limit~\cite{KoemansCollado:2019ggb,Parra-Martinez:2020dzs,DiVecchia:2020ymx,DiVecchia:2021ndb,DiVecchia:2021bdo,Bjerrum-Bohr:2021vuf,Bjerrum-Bohr:2021din,DiVecchia:2022owy,DiVecchia:2022piu,Cristofoli:2020uzm,Bellazzini:2022wzv}. Quantum mechanical terms are then kept in the prefactor of the exponential while super-classical pieces by construction cannot appear in the argument of the exponential; the scattering angle thus becomes classical, as it should be. The KMOC-formalism achieves the same cancellation of super-classical pieces but includes quantum mechanical terms that can subsequently be systematically discarded as in a normal semiclassical limit of quantum mechanics. Ways to introduce recursive ideas in the amplitude formalism have also been discussed ~\cite{Cho:2021nim,Lee:2022aiu,Cho:2022faq,Adamo:2023cfp,Lee:2023zuu,Cho:2023kux,Tao:2023yxy}. 

A different approach pursues the semiclassical limit of quantum mechanical scattering by introducing an exponential representation of the $S$-matrix~\cite{Damgaard:2021ipf}, the proper way to describe the semiclassical limit of the $S$-matrix itself. Because of its exponential form all super-classical terms are absent by construction. In combination with the KMOC-formalism this yields very compact expressions for classical observables such as the momentum kick or scattering angle~\cite{Damgaard:2023ttc}. An alternative track builds on the large-mass expansion of quantum field theory amplitudes~\cite{Damgaard:2019lfh,Aoude:2020onz} to produce the needed subtractions of super-classical pieces~\cite{Brandhuber:2021eyq}.

The method we shall outline in this paper is instead a return to the perturbative solution of the classical equations of motion as described in quite some detail in the classic paper of ref.~\cite{Westpfahl:1985tsl}. As it relies on classical Green function methods, it is very closely related the post-Minkowskian wordline approach~\cite{Kalin:2020mvi,Kalin:2020fhe,Kalin:2020lmz,Mogull:2020sak,Jakobsen:2021smu,Mougiakakos:2021ckm,Riva:2021vnj,Riva:2022fru,Liu:2021zxr,Dlapa:2021npj,Jakobsen:2021lvp,Jakobsen:2021zvh,Dlapa:2021vgp,Jakobsen:2022fcj,Jakobsen:2022psy,Kalin:2022hph,Dlapa:2022lmu,Jakobsen:2022zsx,Dlapa:2023hsl,Ajith:2024fna}. There are nevertheless also differences: while the worldline formalisms of the above references are based on the conventional gravitational perturbation theory around a Minkowski background $\eta_{\mu\nu}$ by metric perturbations $g_{\mu\nu} = \eta_{\mu\nu} + h_{\mu\nu}$, we find it convenient to use what are often called Landau-Lifshitz (or `gothic') variables. An apparently small technical detail, this choice was nevertheless one of the crucial ingredients that allowed for the all-order solution to the Schwarzschild black hole metric of ref.~\cite{Damgaard:2024fqj}. In addition, we choose to work with a doubled set of metric 
variables, one being the inverse of the other, only relating them at the end. This was one of the other simplifying features of the method described in ref.~\cite{Damgaard:2024fqj}. Heuristically, we can understand this simplification from the fact that much of the complexity of the perturbatively expanded Einstein-Hilbert action arises from
the inversion of the metric: the resulting geometric series leads to an infinity of vertices in the gravitational sector. By working with a doubled set of variables we can avoid this
expansion of the action, and implement the required reciprocity relation between the metric and inverse metric variables when needed.

By returning to the classical equations of motion we also do away with all use of effective field theory. The starting point is the matter distribution and its associated energy-momentum tensor. For the two-body scattering problem we thus start the recusions for initial time at minus infinity, and let the two sources of massive bodies evolve from flat space-time towards each other. Since it is phrased perturbatively, we are not scattering two black holes off each other. Rather, we are scattering two massive sources that, if considered in vacuum, each build up the Schwarzschild metric as higher and higher orders in $G$ are included. Due to the motion and the presence of the other body, we do not first build up two Schwarzschild black holes, and then scatter those two objects. Only if we were to include all orders in $G$ would we eventually, at time plus infinity, recover two Schwarzschild black holes that had been scattered off each other. During the scattering itself (and since we are not able to compute to all orders in $G$) there are not any black holes in the problem, and there are therefore also no horizons. The scales of the Schwarzschild radii $2GM_i$ for the two massive sources are not visible at any finite order in $G$. 

It is frequently argued that effective field theory is required to tackle the scattering of two black holes due to a clash between the description of a black hole with its horizon and a point-like matter source with, apparently, no horizon. But the Schwarzschild black hole metric is precisely, in chosen coordinates, an exact vacuum solution to the Einstein field equations of a point-like matter source. The emergence of an exact Schwarzschild black hole metric in perturbation theory arises due to the convergence of the infinite-order perturbative expansion which, as mentioned above, in suitable coordinates is just a geometric series. It appears that the scattering of two massive objects in general relativity is not fundamentally different. Order by order in the $G$-expansion the space-time metric becomes increasingly complicated due to the presence of the two matter sources and their associated gravitational fields. Clearly, the convergence of the perturbative expansion of the two-body problem, so striking in the case of just a single body, is not guaranteed. Still, there are heuristic reasons for hoping that the expansion may have a region of convergence in this case, even after including gravitational radiation. By building up the trajectories of the two massive sources perturbatively, we are essentially approximating the pseudo-hyperbolic orbits by better and better trajectories. While the actual classical trajectories obviously are coordinate-dependent notions, there is hope that such approximations could converge to the full trajectories in the given coordinates as higher and higher orders of $G$ are included. An explicit evaluation of the trajectories in harmonic coordinates to order $G^2$ accuracy can be found in ref.~\cite{Bini:2024hme}.

When described with the doubled set of graviton variables mentioned above, the equations of motion still lead to a rapid proliferation of terms when solved in perturbation theory. This stems from 
the nature of the two-body problem itself since it involves not only the graviton currents but also the two massive matter currents, iteratively solved from the equations of motion. The metric in the chosen gauge thus needs to
be computed with matter sources at two distinct positions, and to counter the resulting complexity we exploit the possibility of splitting up Fourier transforms in terms of three kinds:
one for each of the two position variables separately (denoted in our notation by $[\ell_1,0]$ and $[0,\ell_2]$ for particles 1 and and 2, respectively), and one which is the 
combination of both (denoted by $[\ell_2,\ell_2]$). The associated current iterations mimic this structure so that we are led to a compact formulation of iterative relations 
which again splits into three kinds, now in terms of the convolution integrals that form the backbone of the perturbative recursion relations. The resulting momentum integrals
seem at first sight to bear no evident relation to neither the loop integrals of amplitude calculations for classical gravity nor of the integrals arising from gravitational worldline formulations. This is not surprising because our definition of metric perturbations differ and it is of course only the final integrated answers, independent of the choice of variables,
that should agree. Our approach therefore offers a third
way of performing the computations in classical gravity, and we hope that it will be of benefit to have such a new scheme.

The recursion relations for post-Minkowskian gravity lead to relatively simple iterative integrals when restricted to vacuum 
solutions for the metric. All relevant integrals become combinations of massless Green functions with
tensor numerators. This means that each new order in the recursion becomes a simple function of previous lower-order solutions, all integrals being of generalized bubble-form. This persists when we add the geodesic equations
for the matter sources but the nature of the iterated integrals in momentum space becomes more complex. While
still iterative, the combinations of lower order terms yield new integrands that are not self-reproducing in a similar way as for the vacuum solutions for the metric. Some of the integrals can become separated and are thus doable sequentially, but we are left with several new integrals that need to be performed at increasing orders in $G$. The simplifications due to the recursive
nature of the perturbative expansion are therefore not immediately apparent for all terms when applied to the two-body problem. We nevertheless
believe that the recursion relations have an advantage in that they rewrite the integrands in terms of simpler building blocks whose integrals are already known.

In this first paper we test the formalism up to third post-Minkowskian order. This is the first order at which it is crucial to include the effect of gravitational back-reaction
\cite{DiVecchia:2020ymx,Damour:2020tta,Bjerrum-Bohr:2021vuf,Bjerrum-Bohr:2021din}. The conclusion is as follows: we find full agreement with the two-body scattering angle result of third post-Minkowskian order from the references just listed. What ensures the correct result, including the dissipative pieces due to gravitational radiation, is the use of retarded Green functions. This is in agreement with general expectations and the results of worldline calculations based on the {\em in-in} formalism \cite{Jakobsen:2022psy,Kalin:2022hph}.  It is quite
remarkable that ensuring causality by the use of retarded Green functions is sufficient to capture all dissipative effects from the equations of motion. Details of how this comes about will be given below.


\section{The perturbative expansion}

We divide gravitational perturbation theory up in two parts. The first concerns the perturbative expansion
of the metric variables, the second concerns the perturbative evolution of point-like matter sources. Both are
described in terms of currents and they couple to each other through the equations of motion. In
ref.~\cite{Damgaard:2024fqj} we could simplify the calculation because the matter contribution to the
energy-momentum tensor was taken to be a static point-like source. The extension in this paper pertains to
the scattering situation of
two point-like sources initially at space-time infinity and considered in the center of mass frame where
the two massive objects move towards each other with equal and opposite momenta, separated from the origin
by impact parameters $b_\alpha$, $\alpha=1,2$. At first, the two objects are subjected only to Minkowskian space-time
but they lead to perturbations of that background metric, and those perturbative fluctuations give
rise to both self-interactions and interactions with the matter sources themselves. 
This set-up is not different from
what is pursued in the amplitude and worldline approaches but we here work directly with the equations of
motion. In the end, we must compute physical and coordinate-independent quantities such as the scattering 
angle $\chi$, the momentum kick $\Delta P_\alpha$ of the two matter particles, the angular momentum loss due to
radiation, etc. These observables can all be related to the matter sources at time minus infinity and time plus
infinity where the background space-time is flat. But a crucial intermediate
quantity during the scattering process is the metric fluctuation itself, in the chosen gauge. 
This we turn to first.

\subsection{Metric perturbations}

As mentioned in the Introduction, an important simplification comes from working with the gothic (Landau-Lifshitz) variables where, given a metric $g_{\mu\nu}$, we
introduce the tensor density 
\begin{equation}
  \mathfrak{g}^{\mu\nu} \equiv \sqrt{-g} g^{\mu\nu} \,.
\label{}\end{equation}
We consider Einstein-Hilbert action $S_{\rm EH}$ coupled to an external source $j^{\mu\nu}$,
\begin{equation}
  S[\mathfrak{g},j]
  =
  S_{\rm EH}[\mathfrak{g}] + \frac{1}{16\pi G} \int \mathrm{d}^{D}x \frac{1}{\sqrt{-\mathfrak{g}}}\mathfrak{g}_{\mu\nu} \mathfrak{j}^{\mu\nu}\,.
\label{}\end{equation}
Here $S_{\rm EH}[\mathfrak{g}]$ is given in terms of $\mathfrak{g}$-variables as follows:
\begin{equation}
\begin{aligned} 
  S_{\mathrm{EH}}[\mathfrak{g}]
  = \frac{1}{16\pi G}\int \mathrm{d}^{D} x \bigg[ & 
  	  \frac{1}{4} \mathfrak{g}^{\mu \nu} \partial_{\mu} \mathfrak{g}^{\rho \sigma} \partial_{\nu} \mathfrak{g}_{\rho \sigma}
  	- \frac{1}{2} \mathfrak{g}^{\mu \nu} \partial_{\mu} \mathfrak{g}^{\rho \sigma} \partial_{\rho} \mathfrak{g}_{\nu \sigma}+(D-2) \mathfrak{g}_{\mu \nu} d^{\mu} d^{\nu} 
  \\&
  	+\partial_{\mu}\left(2 d^{\mu} -\partial_{\nu} \mathfrak{g}^{\mu \nu}\right)\bigg]\,,
\end{aligned}
\label{}\end{equation}
where $G$ is the Newton constant, and 
\beq
  d^{\mu} \equiv - \frac{1}{2(D-2)} \mathfrak{g}^{\mu\kappa} \mathfrak{g}_{\rho \sigma}\partial_{\kappa} \mathfrak{g}^{\rho \sigma} 
\eeq 
To the action $S_{\mathrm{EH}}[\mathfrak{g}]$ will eventually be added a suitable gauge-fixing term, see below.

The two colliding black holes are sourced by point-like energy-momentum tensors. Since they are initially at very large distance from each other, each black hole feels initially 
only the asymptotically flat metric produced by the other black hole. The initial trajectories will to leading order therefore correspond to free motion in flat Minkowski space, and the energy-momentum tensor density $\mathfrak{j}^{\mu\nu}(x)$ is that of two non-interacting point-like sources,
\begin{equation}
  \mathfrak{j}^{\mu\nu}(x) 
  = 8\pi G \sum_{\alpha=1}^{2} m_{\alpha} \int \mathrm{d} \tau \dot{X}_{\alpha}^{\mu}(\tau) \dot{X}_{\alpha}^{\nu}(\tau) \delta^{D}\big[x^{\mu} - X^{\mu}_{\alpha}(\tau) \big]\,,
\label{delta_function_source}\end{equation}
where $X^{\mu}_{1} (\tau)$ and $X^{\mu}_{2} (\tau)$ are the trajectories of the two point masses and $\tau$ is proper time. We have denoted proper time derivatives by a dot, 
so that, $e.g.$,
\beq
\dot{X}_{\alpha}^{\mu}(\tau) \equiv \frac{d X_{\alpha}^{\mu}(\tau)}{d \tau}\,.
\eeq
where $v_{\alpha}^{\mu}(\tau)$ are four-velocities. This is all the same set-up, and same choice of variables, 
as was adapted by Westpfahl in ref.~\cite{Westpfahl:1985tsl}. Explicit solutions for the matter trajectories up to second post-Minkowskian order have recently been derived by
Bini and Damour~\cite{Bini:2024hme} based on the formalism of ref.~\cite{Bel:1981be}.

For further simplification, we replace the gothic metric density with lower indices by an auxiliary field $\tilde{\mathfrak{g}}_{\mu\nu}$ which we treat on equal footing by imposing the constraint
\begin{equation}
  \mathfrak{g}^{\mu\rho}\tilde{\mathfrak{g}}_{\rho\nu} = \delta^{\mu}{}_{\nu}\,.
\label{}\end{equation}
We will therefore operate with two metric perturbations which we choose to parametrize as deviations from the flat Minkowski space metric $\eta_{\mu\nu}$ in the following way:
\begin{equation}
  \mathfrak{g}^{\mu\nu} = \eta^{\mu\nu} - \mathfrak{h}^{\mu\nu} \,,
  \qquad
  \tilde{\mathfrak{g}}_{\mu\nu} = \eta_{\mu\nu} + \tilde{\mathfrak{h}}_{\mu\nu} \,.
\label{metric_perturbation}\end{equation}
Clearly, then, the variables $\mathfrak{h}^{\mu\nu}$ and $\tilde{\mathfrak{h}}_{\mu\nu}$ are related by the following constraint:
\begin{equation}
  \tilde{\mathfrak{h}}_{\mu\nu} = \mathfrak{h}_{\mu\nu} + \mathfrak{h}_{\mu}{}^{\rho} \tilde{\mathfrak{h}}_{\rho\nu} ~,
\label{}\end{equation}
and indices are here raised and lowered by the Minkowski metric.

We now expand the metric $\mathfrak{g}$ and the two trajectories $X_{\alpha}^{\mu} ~,~ \alpha =  1,2$ in Newton's constant $G$,
\begin{equation}
\begin{aligned}
  \mathfrak{h}^{\mu\nu} &= \sum_{n=1}^{\infty} \mathfrak{h}^{\mu\nu}\big|_{n} G^{n}\,,
  \qquad
  \tilde{\mathfrak{h}}_{\mu\nu} = \sum_{n=1}^{\infty} \tilde{\mathfrak{h}}_{\mu\nu}\big|_{n} G^{n}\,,
  \\
  X_{\alpha}^{\mu} &= \sum_{n=0}^{\infty} X^{\mu}_{\alpha}\big|_{n} G^{n} 
  = 
  b_{\alpha}^{\mu} + v_{\alpha}^{\mu} \tau + \sum_{n=1}^{\infty} X^{\mu}_{\alpha}\big|_{n} G^{n}\,,
\end{aligned}\label{}
\end{equation}
where we parametrize the zeroth order of the trajectories in terms of the initial positions (impact parameters) $b_{\alpha}^{\mu} $ and initial velocities of each mass so that  $X^{\mu}_{\alpha}\big|^{0} = b_{\alpha}^{\mu} + v_{\alpha}^{\mu} \tau$. Since the background is flat, the $v_{\alpha}^{\mu}$ are constants. Although written in a general manner here, we shall always consider kinematics in the center of mass frame only. For what follows, we find it useful to introduce a new product ``$*$'' between the coefficient of the PM expansion, defined in the following way:
\begin{equation}
\begin{aligned}
  A * B\big|^{m} 
  &=
  \sum_{p=1}^{m-1} A\big|^{m-p} \, B\big|^{p} \,,
  \\
  A * B * C\big|^{m} 
  &=
  \sum_{n=1}^{m-2} \sum_{p=1}^{m-n-1} A\big|^{n} \, B\big|^{p} \, C\big|^{m-n-p}\,,\qquad m\geq3\,,
\end{aligned}\label{convolution_product}
\end{equation}
corresponds to projecting the expanded products down on fixed order $m$  in $G$.

\subsection{The Einstein equations of motion}

Let us now consider variations of the action with respect to $\mathfrak{g}_{\mu\nu}$: 
\begin{equation}
\begin{aligned}
  \delta_{\mathfrak{g}_{}} S 
   &=    
  \frac{1}{16\pi G}  \int \mathrm{d}^{D} x\, \delta \mathfrak{g}_{\mu \nu}\left[ 
     \frac{\delta}{\delta \mathfrak{g}_{\mu\nu}}\Big(L_{\rm EH} \Big) 
   + \frac{1}{\sqrt{-\mathfrak{g}}} \bigg(\mathfrak{j}^{\mu\nu} -\frac{1}{2} \mathfrak{g}^{\mu\nu}\mathfrak{j}^{\rho\sigma} \mathfrak{g}_{\rho\sigma}\bigg)
   \right]\,,
  \\
  &= \frac{1}{16\pi G} \int \mathrm{d}^{D} x\, \delta \mathfrak{g}_{\mu \nu}\left[ 
   - \mathcal{G}^{\mu\nu} 
   + \frac{1}{\sqrt{-\mathfrak{g}}} \bigg(\mathfrak{j}^{\mu\nu} -\frac{1}{2} \mathfrak{g}^{\mu\nu}\mathfrak{j}^{\rho\sigma} \mathfrak{g}_{\rho\sigma}\bigg) 
   \right]\,,
\end{aligned}\label{}
\end{equation}
where $\mathcal{G}_{\mu\nu}$ is the Einstein tensor density\footnote{Note that the determinant in $\sqrt{-\mathfrak{g}}$ is defined by the lower-index object, $\mathfrak{g}= \det \mathfrak{g}_{\mu\nu}$, and its variation is given by
\begin{equation}
  \delta \frac{1}{\sqrt{-\mathfrak{g}}} 
  = -\frac{1}{2} (-\mathfrak{g})^{-\frac{3}{2}} \big(- \delta \mathfrak{g}\big) 
  = -\frac{1}{2} (-\mathfrak{g})^{-\frac{3}{2}} \big(- \mathfrak{g} \delta\mathfrak{g}_{\mu\nu} \mathfrak{g}^{\mu\nu}\big)
  = -\frac{1}{2} \frac{1}{\sqrt{-\mathfrak{g}}} \delta\mathfrak{g}_{\mu\nu} \mathfrak{g}^{\mu\nu}\,.
\label{}\end{equation}
} defined by $\mathcal{G}^{\mu\nu} = -16\pi G\frac{\delta S_{\rm EH}}{\delta \mathfrak{g}_{\mu\nu}}$ and we replace $\mathfrak{g}_{\mu\nu} \to \tilde{\mathfrak{g}}_{\mu\nu}$, as explained earlier:
\begin{equation}
\begin{aligned} 
  \mathcal{G}^{\mu\nu}= &\
    \frac{1}{2} \partial_{\kappa} \big(\mathfrak{g}^{\kappa\lambda} \partial_{\lambda} \mathfrak{g}^{\mu\nu}\big) 
  - \frac{1}{2} \partial_{\lambda} \mathfrak{g}^{\kappa\mu} \partial_{\kappa} \mathfrak{g}^{\lambda\nu}
  - \mathfrak{g}^{\kappa(\mu} \partial_{\kappa} \partial_{\lambda} \mathfrak{g}^{\nu) \lambda}
  + (D-2) d^{\mu} d^{\nu}
  + \mathfrak{g}^{\mu\nu} \partial_{\kappa} d^{\kappa}
  \\&
  +\frac{1}{2} \mathfrak{g}^{\kappa\alpha} \mathfrak{g}^{\beta(\nu}\partial_{\alpha}\mathfrak{g}^{\mu)\lambda} \partial_{\kappa}\tilde{\mathfrak{g}}_{\lambda\beta}
  - \mathfrak{g}^{(\mu|\alpha|} \mathfrak{g}^{\nu)\beta}\partial_{\alpha}\mathfrak{g}^{\kappa\lambda} \partial_{\kappa}\tilde{\mathfrak{g}}_{\lambda\beta}
  +\frac{1}{4} \mathfrak{g}^{\mu\alpha} \mathfrak{g}^{\nu\beta}\partial_{\alpha}\mathfrak{g}^{\kappa\lambda} \partial_{\beta}\tilde{\mathfrak{g}}_{\kappa\lambda}
  \,,
\end{aligned}
\label{EinsteinTensorDensity}\end{equation}
The Einstein equation with external source $\mathfrak{j}^{\mu\nu}$ is then given by
\begin{equation}
  \mathcal{G}^{\mu\nu} 
  = 
  \frac{1}{\sqrt{-\tilde{\mathfrak{g}}}} 
  \bigg( 
    	\mathfrak{j}^{\mu\nu}
  	  - \frac{1}{2} \mathfrak{g}^{\mu\nu} \mathfrak{j}^{\rho\sigma} \tilde{\mathfrak{g}}_{\rho\sigma}
  \bigg) \,.
\label{}\end{equation}

By eliminating the trace part of the above Einstein equation, the external source on the right-hand side reduces to a single term, and metric perturbations of the Einstein equation are greatly simplified:
\begin{equation}
  \tilde{\mathcal{G}}^{\mu\nu}  = \frac{1}{\sqrt{-\tilde{\mathfrak{g}}}} \mathfrak{j}^{\mu\nu} \,,
\label{}\end{equation}
where $\tilde{\mathcal{G}}^{\mu\nu} = \mathcal{G}^{\mu\nu} - \frac{1}{D-2} \mathfrak{g}^{\mu\nu} \mathcal{G}^{\rho\sigma} \tilde{\mathfrak{g}}_{\rho\sigma}$. It is now convenient to group terms having the same structure in the following manner:
\begin{equation}
\begin{aligned}
  \tilde{\mathcal{G}}^{\mu\nu}
  =
  - \frac{1}{2} \eta^{\kappa\lambda}\partial_{\kappa}\partial_{\lambda} \mathfrak{h}^{\mu\nu}
  + \frac{1}{2}  W^{\mu \nu}+\frac{1}{2} Z^{(\mu}{ }_{\kappa} \mathfrak{g}^{\nu) \kappa}+\frac{D-2}{2} d^{\mu \nu}\,,
\end{aligned}\label{}
\end{equation}
where
\begin{equation}
\begin{aligned} 
  W^{\mu \nu}
  & \equiv
  -\mathfrak{h}^{\kappa \lambda} \partial_{\kappa} \partial_{\lambda} \mathfrak{g}^{\mu \nu}-\partial_{\lambda} \mathfrak{g}^{\kappa \mu} \partial_{\kappa} \mathfrak{g}^{\lambda \nu}\,,
  \\
  Z^{\mu}{}_{\nu}
  & \equiv
  \frac{1}{2}\left(2 Z^{\kappa \mu}{}_{\kappa \nu}-4 Z^{\mu \kappa}{}_{\kappa \nu}+Z^{\mu \kappa}{}_{\nu \kappa}\right)
  +\frac{1}{4} \delta^{\mu}{}_{\nu}\left(2 Z^{\kappa \lambda}{}_{\lambda \kappa} -Z^{\kappa \lambda}{}_{\kappa \lambda}\right)\,,
  \\
  d^{\mu \nu}
  & \equiv
  2 d^{\mu} d^{\nu}-\frac{2}{D-2} \mathfrak{g}^{\mu \nu} \tilde{\mathfrak{g}}_{\kappa \lambda} d^{\kappa} d^{\lambda}\,.
\end{aligned}
\end{equation}
Here $Z^{\mu \nu}{ }_{\rho \sigma} \equiv \mathfrak{g}^{\mu \kappa} \partial_{\kappa} \mathfrak{g}^{\nu \lambda} \partial_{\rho} \tilde{\mathfrak{g}}_{\lambda \sigma}$, and we have used the harmonic gauge condition $\partial_{\mu}\mathfrak{g}^{\mu\nu} = 0$ to simplify expressions. 

Next, we consider the post-Minkowskioan expansion of the external source $\mathfrak{j}^{\mu\nu}$. Inserting a Fourier integral representation of the $\delta$-function in  eq.\eqref{delta_function_source} we write this source in the following way:
\begin{equation}
\begin{aligned}
  \mathfrak{j}^{\mu\nu}(x)
  &=
  8\pi G \sum_{\alpha=1}^{2} m_{\alpha} 
  \int \mathrm{d} \tau \int_{\ell} 
  e^{i \ell\cdot (x -X_{\alpha})}
  \dot{X}_{\alpha}^{\mu} \dot{X}_{\alpha}^{\nu} \,,
  \\
  &=
  8\pi G \sum_{\alpha=1}^{2} m_{\alpha}
  \int \mathrm{d} \tau \int_{\ell} e^{i\ell \cdot x -i \ell \cdot (b_{\alpha}+v_{\alpha}\tau)} 
  \zeta^{\mu\nu}_{\alpha}\,.
\end{aligned}\label{}
\end{equation}
where
\begin{equation}
  \zeta^{\mu\nu}_{\alpha} = e^{-i \ell \cdot (X_\alpha - X_{\alpha}|_{0})} \dot{X}^{\mu}_{\alpha} \dot{X}^{\nu}_{\alpha}
\label{}\end{equation}
and it is also expanded in powers of $G$,
\begin{equation}
  \zeta^{\mu\nu}_{\alpha}
  =
  \sum_{n=0}^{\infty} G^{n} \zeta^{\mu\nu}_{\alpha}\big|_{n}\,.
\label{}\end{equation}
Then the first few terms of this expansion are given by
\begin{equation}
\begin{aligned}
  \zeta^{\mu\nu}_{\alpha}\big|_{0}
  &=
  v^{\mu}_{\alpha} v^{\nu}_{\alpha}\,,
  \\
  \zeta^{\mu\nu}_{\alpha}\big|_{1}
  &=
  2 v^{(\mu}_{\alpha}\, \dot{X}^{\nu)}_{\alpha}\big|_{1}
  	- i v^{\mu}_{\alpha} v^{\nu}_{\alpha} \big(\ell\cdot X_{\alpha}\big)\big|_{1} \,,
  \\
  \zeta^{\mu\nu}_{\alpha}\big|_{2}
  &= 
  2 v^{(\mu}_{\alpha} \dot{X}^{\nu)}_{\alpha}\big|_{2} 
  + \big(\dot{X}^{\mu}_{\alpha}* \dot{X}^{\nu}_{\alpha}\big)\big|_{2}
  - 2i\big((\ell\cdot X_{\alpha}) *\dot{X}^{(\mu}_{\alpha}\big)\big|_{2} v^{\nu)}_{\alpha}
 \\&\quad
  - \bigg[
  	  i\ell\cdot X_{\alpha}\big|_{2}
  	+ \frac{1}{2}\Big((\ell\cdot X_{\alpha})* (\ell\cdot X_{\alpha})\Big)\Big|_{2}
  \bigg] v^{\mu}_{\alpha} v^{\nu}_{\alpha}\,.
\end{aligned}\label{}
\end{equation}
Note that there is no summation over repeated $\alpha$ indices; the expansion is performed individually on each of the particle trajectories.

Let us now consider the expansion of the source term in the Einstein equations. To do so, we need to compute the determinant of $\tilde{\mathfrak{g}}$. Expanding it in powers of $G$, we get
\begin{equation}
\begin{aligned}
  |-\operatorname{det} \tilde{\mathfrak{g}}|^{-\frac{1}{D-2}}
  &= 
  1 -\frac{1}{D-2} \mathfrak{h}
  +\frac{1}{2(D-2)}\left(\frac{\mathfrak{h}^{2}}{D-2}-\tr\!\left[\mathfrak{h}^{2}\right]\right) 
  \\&\quad
  -\frac{1}{D-2}\left(\frac{1}{6(D-2)^{2}} \mathfrak{h}^{3}-\frac{1}{2(D-2)} \mathfrak{h} \tr\!\left[\mathfrak{h}^{2}\right]+\frac{1}{3} \operatorname{tr}\left[\mathfrak{h}^{3}\right]\right) 
  + \cdots\,,
\end{aligned}\label{Expansion_j_density}
\end{equation}
where $\mathfrak{h} = \mathfrak{h}^{\kappa}{}_{\kappa}$ and the trace operation is defined for arbitrary rank 2 tensors in the obvious way,
\begin{equation}
\begin{aligned}
  \tr\!\big[A B\big] 
  &= A^{\mu}{}_{\nu} B^{\nu}{}_{\mu}\,,
  \\
  \tr\!\big[A^{n}\big] 
  &= A^{\mu_{n}}{}_{\mu_{1}} A^{\mu_{1}}{}_{\mu_{2}}A^{\mu_{2}}{}_{\mu_{3}} \cdots A^{\mu_{n-1}}{}_{\mu_{n}}\,.
\end{aligned}\label{}
\end{equation}
We are now faced with a choice. For actual calculations we will be employing dimensional regularization, and we could now choose to keep the classical equations in $D=4$
dimensions, or extend the equations into arbitrary $D$-dimensions. For simplicity we choose the former. The difference between the two would seem to correspond to different
schemes of subtractions but since there is no room for subtractions at the classical level such a potential difference cannot survive the limit $D \to 4$ in the end.
Specializing the expansion of eq. \eqref{Expansion_j_density} to the $D=4$ case we can thus introduce $G^n$-projected coefficients $H\big|^{n}$ by
\begin{equation}
\begin{aligned}
  |-\det \tilde{\mathfrak{g}}|^{-\frac{1}{2}} 
  &= 
  1 + G H\big|_{1} + G^{2} H\big|_{2} + G^{3}H\big|_{3} + \cdots\,,
\end{aligned}\label{}
\end{equation}
where 
\begin{equation}
\begin{aligned}
  H\big|_{1}
  &=
  - \frac{1}{2} \mathfrak{h}\big|_{1}\,,
  \\
  H\big|_{2}
  &=
  - \frac{1}{2} \mathfrak{h}\big|_{2} 
  + \frac{1}{8} \mathfrak{h}*\mathfrak{h}\big|_{2}
  - \frac{1}{4} \tr\!\big[\mathfrak{h}* \mathfrak{h}\big|_{2} \big]\,,
  \\
  H\big|_{3} &= 
  - \frac{1}{2} \mathfrak{h}\big|_{3}
  + \frac{1}{8} \mathfrak{h} * \mathfrak{h}\big|_{3}
  - \frac{1}{4} \!\tr\!\big[\mathfrak{h} * \mathfrak{h}\big|_{3}\big]
  - \frac{1}{48} \big(\mathfrak{h}\big|_{1}\big)^{3}
  + \frac{1}{8} \mathfrak{h}\big|_{1} \!\tr\!\Big[\big(\mathfrak{h}\big|_{1}\big)^{2}\Big]
  - \frac{1}{6} \!\tr\!\big[(\mathfrak{h}\big|_{1})^{3}\big]\,.
\end{aligned}\label{H_currents}
\end{equation}

Defining $\mathbf{j}^{\mu\nu} \equiv  |-\det\tilde{\mathfrak{g}}|^{-\frac{1}{2}} \mathfrak{j}^{\mu\nu}$ we can thus expand in powers of $G$,
\begin{equation}
  \frac{1}{\sqrt{-\tilde{\mathfrak{g}}}} \mathfrak{j}^{\mu\nu}
  = 
  \sum_{n=1}^{\infty}\mathbf{j}^{\mu\nu}\big|_{n} G^{n}\,,
\label{}\end{equation}
so that the coefficients $\mathbf{j}^{\mu\nu}\big|_{n} $ can be written as
\begin{equation}
\begin{aligned}
  \mathbf{j}^{\mu\nu}\big|_{1}
  &= 
  \mathfrak{j}^{\mu\nu}\big|_{1} \,,
  \\
  \mathbf{j}^{\mu\nu}\big|_{n}
  &= 
    \mathfrak{j}^{\mu\nu}\big|_{n} 
  + H* \mathfrak{j}^{\mu\nu}\big|_{n}\,, \qquad n>1\,.
\end{aligned}\label{}
\end{equation}
With the help of these definitions we can write the hierarchy of $G$-expanded Einstein equations very compactly as
\begin{equation}
  \Box \mathfrak{h}^{\mu\nu}\big|_{n} 
  = 
    \tau^{\mu\nu}\big|_{n}
  - 2 \mathbf{j}^{\mu\nu}\big|_{n} \,,
\label{Expansion_Einstein_eq}\end{equation}
where $\tau^{\mu\nu}$ is the energy-momentum pseudotensor defined as
\begin{equation}
\begin{aligned}
  \tau^{\mu\nu}\big|_{n}
  =
    W^{\mu\nu}\big|_{n} 
  + Z^{(\mu}{}_{\kappa} \eta^{\nu) \kappa}\big|_{n}
  - Z^{(\mu}{}_{\kappa} * \mathfrak{h}^{\nu) \kappa}\big|_{n}
  + 2d^{\mu\nu}\big|_{n}
\end{aligned}\label{}
\end{equation}
%

\subsection{The geodesic equation}

For motion in an external fixed metric, motion of matter is well known to follow geodesics. For a two-body interacting set-up, the metric will be a dynamical variable that
changes due to the motion of both of the two involved bodies, thus apparently leaving open how to specify the trajectories. As in the worldline approach to post-Minkowskian
perturbation theory, simplicity is fortunately restored when perturbing around flat Minkowski space: at any given order in perturbation theory the next order will be determined
by the metric computed up to the previous order, evaluated at the pertinent space-time point.  We thus need to focus on the geodesic equation at every fixed order in $G$,
\begin{equation}
  \ddot{X}^{\mu}_{\alpha} +\Gamma_{\nu\rho}^{\mu} \dot{X}^{\nu}_{\alpha} \dot{X}^{\rho}_{\alpha} =0 \,.
\label{}\end{equation}
Again, as with worldline effective field theory, it is convenient to adopt the equivalent expression in terms of lowered indices, obtained by considering the simpler equation obeyed by
$X_{\mu}$ with the index lowered by the conventional metric $g_{\mu\nu}$, 
\begin{equation}
  \frac{\mathrm{d}}{\mathrm{d}\tau} \bigg[g_{\mu\nu}(X_{\alpha})\dot{X}^{\nu}_{\alpha}\bigg]
  = \frac{1}{2} \partial_{\mu} g_{\gamma\rho}( X_{\alpha}) \dot{X}^{\gamma}_{\alpha} \dot{X}^{\rho}_{\alpha}\,.
\label{}\end{equation}
Similar to the case of the gothic metric density, we may treat the metric and its inverse as two independent fields, $g^{\mu\nu}$ and $\tilde{g}_{\mu\nu}$, and then impose the following constraint
\begin{equation}
  g^{\mu\nu} \tilde{g}_{\nu\rho} = \delta^{\mu}{}_{\rho} \,.
\label{}\end{equation}
Thus far, we have considered only the perturbative expansion of the gothic metric $\mathfrak{g}^{\mu\nu}$ and $\tilde{\mathfrak{g}}_{\mu\nu}$. However, in order to analyze the geodesic equation as well we are naturally led to also work explicitly with the standard metric tensor $g_{\mu\nu}$ and its inverse $\tilde{g}^{\mu\nu}$, as well as their perturbative expansions which we will denote by $\tilde{h}_{\mu\nu}$ and $h_{\mu\nu}$, respectively. To establish the connection, we must relate these two metric perturbations to our field variables $\mathfrak{h}$ and $\tilde{\mathfrak{h}}$. Consistent with our choice of keeping the classical equations of motion in $D=4$ dimensions, we have the relations
\begin{equation}
\begin{aligned}
  g^{\mu\nu} &= \sqrt{-\mathfrak{g}} \mathfrak{g}^{\mu\nu} = \eta^{\mu\nu} - h^{\mu\nu}\,,
  \\
  \tilde{g}_{\mu\nu} &= \frac{1}{\sqrt{-\mathfrak{g}}}\tilde{\mathfrak{g}}_{\mu\nu} = \eta_{\mu\nu} + \tilde{h}_{\mu\nu}\,.
\end{aligned}\label{}
\end{equation}
Thus $\tilde{h}_{\mu\nu}\big|_{n}$ is related to $\tilde{\mathfrak{h}}_{\mu\nu}\big|_{n}$ as follows:
\begin{equation}
  \tilde{h}_{\mu\nu}\big|_{n}
  =
  \tilde{\mathfrak{h}}_{\mu\nu}\big|_{n}
  + H * \tilde{\mathfrak{h}}_{\mu\nu}\big|_{n}
  + H\big|_{n} \eta_{\mu\nu} ~.
\label{metric_fluctuations}\end{equation}
With these variables, the hierarchy of geodesic equations expanded in powers of $G$ can be written
\begin{equation}
\begin{aligned}
  &\eta_{\rho\sigma} \ddot{X}^{\sigma}_{\alpha}(\tau)\big|_{n} 
    + \frac{d}{d\tau} \bigg(\tilde{h}_{\rho\sigma}(X_{\alpha})\big|_{n} v^{\sigma}_{\alpha}
      + \Big(\tilde{h}_{\rho\sigma}(X_{\alpha}) *\dot{X}^{\sigma}_{\alpha}\Big)\Big|_{n}
  \bigg) 
  \\
  &=
  \frac{1}{2} \partial_{\rho} \tilde{h}_{\mu\nu}(X_{\alpha})\big|_{n} v^{\mu}_{\alpha} v^{\nu}_{\alpha}
  + \Big(\partial_{\rho} \tilde{h}_{\mu\nu} (X_{\alpha})*\dot{X}^{\mu}_{\alpha}v^{\nu}_{\alpha}\Big) \Big|_{n}
  + \frac{1}{2} \Big(
    \partial_{\rho} \tilde{h}_{\mu\nu}(X_{\alpha}) * \dot{X}^{\mu}_{\alpha} * \dot{X}^{\nu}_{\alpha}
  \Big)\Big|_{n}\,,
\end{aligned}\label{perturbedGeodesicEq}
\end{equation}
%

\section{Definitions of Currents}

We now proceed to solve the combination of the perturbative Einstein equations of motion and the perturbative geodesic equations simultaneously using Green function methods. The natural approach would be to go to Fourier space, find the involved Green functions, perform the related momentum integrals and thus finally revert to space-time variables. However, 
in the case of the two-body problem we find it computationally simpler to introduce a doubled formalism where there are two distinct Fourier transforms, one for each of the two bodies. We will motivate this as follows. Consider the leading-order problem where, as discussed in the previous section, the matter sources are naturally expressed in Fourier space by the introduction of Fourier-space representation of the delta-function. Performing the $\tau$-integration, this leads to
\begin{equation}
  \mathfrak{j}^{\mu\nu}\big|_{1}
  =
  8 \pi \sum_{\alpha=1}^{2} m_{\alpha} v^{\mu}_{\alpha}v^{\nu}_{\alpha} 
  \int_{\ell} e^{i\ell \cdot (x - b_{\alpha})} \hat{\delta} (\ell\cdot v_{\alpha})\,.
\label{1PM_source}\end{equation}
At this leading order, the Einstein equation is identical to Poisson's equation, 
\beq
\Box \mathfrak{h}^{\mu\nu}\big|_{1} = - 2 \mathfrak{j}^{\mu\nu}\big|_{1} ~.
\eeq
Solving this equation, we introduce the Fourier transform of $\mathfrak{h}^{\mu\nu}\big|_{1}$ through
\begin{equation}
  \mathfrak{h}^{\mu\nu}\big|_{1}
  =
  \int_{\ell} e^{i \ell \cdot x}\,
  \mathfrak{J}^{\mu\nu}\big|_{1}^{\ell}\,.
\label{}\end{equation}
Substituting this, we find that to this order
\begin{equation}
  \mathfrak{J}^{\mu\nu}\big|_{1}^{\ell}
  =
  \frac{16\pi}{\ell^{2}} \Big(
  	m_{1} v^{\mu}_{1}v^{\nu}_{1} \hat{\delta} (\ell\cdot v_{1}) e^{-i\ell \cdot b_{1}}
  + m_{2} v^{\mu}_{2}v^{\nu}_{2} \hat{\delta} (\ell\cdot v_{2}) e^{-i\ell \cdot b_{2}}
  \Big)\,,
\label{}\end{equation}
Unlike ordinary Fourier transforms, this procedure does not eliminate all phase factors appearing in the external sources of eq. \eqref{1PM_source} by a shift of the Fourier momentum $\ell$. When one uses this to iterate and thus compute higher-order solutions, these residual phase factors remain, making the integrals of Fourier-type rather than standard loop integrals. 

To overcome this computational issue, we introduce two distinct Fourier transforms, one for each particle. A systematic prescription for this involves introducing separate coordinate variables for each particle and performing independent Fourier transforms with respect to each coordinate. We thus define a new pair of coordinates $x^{\mu}_{1}$ and $x^{\mu}_{2}$ as follows:
\begin{equation}
\begin{aligned}
 1 : \quad x \to x^{\mu}_{1} = x^{\mu} - b^{\mu}_{1}\,,
  \\  
 2 : \quad x \to x^{\mu}_{2} = x^{\mu} - b^{\mu}_{2}\,,
\end{aligned}\label{}
\end{equation}
and we denote their corresponding Fourier duals as follows: 
\begin{equation}
  x_{1} \iff \ell_{1}\,,
  \qquad
  x_{2} \iff \ell_{2}\,.
\label{}\end{equation}
Correspondingly, the metric perturbations $\mathfrak{h}^{\mu\nu}(x)\big|_{n}$ in position space should be replaced by
\begin{equation}
  \mathfrak{h}^{\mu\nu}(x)\big|_{n}
  \to
  \mathfrak{h}^{\mu\nu}\big|^{x_{1},x_{2}}_{n}
  \equiv
  \mathfrak{h}^{\mu\nu}(x_{1},x_{2})\big|_{n} \,,
\label{}\end{equation}
and the Fourier transforms on this doubled set of coordinates are given by
\begin{equation}
\begin{aligned}
  \mathfrak{h}^{\mu\nu}\big|^{x_{1},x_{2}}_{n}
  &=
  \int \frac{\mathrm{d}^{D}\ell_{1}}{(2\pi)^{D}} e^{i\ell_{1}\cdot x_{1}} \mathfrak{J}^{\mu\nu}\big|_{n}^{\ell_{1},0}
  +
  \int \frac{\mathrm{d}^{D}\ell_{2}}{(2\pi)^{D}} e^{i\ell_{2}\cdot x_{2}} \mathfrak{J}^{\mu\nu}\big|_{n}^{0,\ell_{2}}
  \\&\quad
  +
  \int \frac{\mathrm{d}^{D}\ell_{1}}{(2\pi)^{D}} \frac{\mathrm{d}^{D}\ell_{2}}{(2\pi)^{D}} e^{i\ell_{1}\cdot x_{1}+i\ell_{2}\cdot x_{2}}\, \mathfrak{J}^{\mu\nu}\big|_{n}^{\ell_{1},\ell_{2}} \,.
\end{aligned}\label{doubleFourier_h}
\end{equation}
We will refer to the Fourier dual of the metric perturbations as their \emph{currents}.

For a concise expression, it is convenient to introduce a formal vector notation $L_{\alpha}=\{0,\ell_{\alpha}\}$, which combines a loop momentum $\ell_{\alpha}$ and 
what we can call a zero mode in one go. We thus denote an integral over $L_{\alpha}$ by
\begin{equation}
  \int_{L_{\alpha}} f(L_{\alpha})
  =
  f(0)
  +
  \int_{\ell_{\alpha}} f(\ell_\alpha)\,.
\label{}\end{equation}
With the help of this notation the Fourier transforms in eq. \eqref{doubleFourier_h} can be written as one single term 
\begin{equation}
  \mathfrak{h}^{\mu\nu}\big|^{x_{1},x_{2}}_{n}
  =
  \int_{L_{1},L_{2}} e^{i L_{1}\cdot x_{1}+i L_{2}\cdot x_{2}}\, 
  \mathfrak{J}^{\mu\nu}\big|_{n}^{L_{1},L_{2}}\,,
\label{}\end{equation}
where by construction we take $\mathfrak{J}^{\mu\nu}\big|_{n}^{0,0} = 0$ and we define
\begin{equation}
  \int_{\ell_{1},0} = \int_{\ell_{1}}\,,
  \qquad
  \int_{0,\ell_{2}} = \int_{\ell_{2}}\,.
\label{}\end{equation}
Similarly the double Fourier expansion for $\tilde{\mathfrak{h}}_{\mu\nu}(x_{1},x_{2})$ is given by
\begin{equation}
\begin{aligned}
  \tilde{\mathfrak{h}}_{\mu\nu} \big|^{x_{1},x_{2}}_{n}
  &= 
  \int_{L_{1},L_{2}} e^{iL_{1}\cdot x_{1}+iL_{2}\cdot x_{2}} \tilde{\mathfrak{J}}_{\mu\nu}\big|_{n}^{L_{1},L_{2}}  \,,
  \qquad
  \tilde{\frak{J}}_{\mu\nu}\big|_{n}^{0,0}=0\,.
\end{aligned}\label{}
\end{equation}
The Fourier transform for derivatives acting on a field such as $\partial_{\rho} \frak{h}^{\mu\nu}\big|^{n}_{x_{1},x_{2}}$ is found by acting with the derivative operator on both $x_{1}$ and $x_{2}$:
\begin{equation}
\begin{aligned}
  \partial_{\rho} \frak{h}^{\mu\nu}\big|_{n}^{x_{1},x_{2}}
  &=
    \int_{\ell_{1}} e^{i \ell_{1} \cdot x_{1}} i \ell^{\rho}_{1} \,\mathfrak{J}^{\mu\nu}\big|_{n}^{\ell_{1},0} 
  + \int_{\ell_{2}} e^{i \ell_{2} \cdot x_{2}} i \ell^{\rho}_{2} \,\mathfrak{J}^{\mu\nu}\big|_{n}^{0,\ell_{2}}
  \\&\quad
  + \int_{\ell_{1},\ell_{2}} e^{i \ell_{1}\cdot x_{1} + i \ell_{2} \cdot x_{2}} i \ell_{12}^{\rho}\, \frak{J}^{\mu\nu}\big|_{n}^{\ell_{1},\ell_{2}}\,,
  \\
  &=
  \int_{L_{1},L_{2}} e^{iL_{1}\cdot x_{1}+iL_{2}\cdot x_{2}} i L^{\rho}_{12}\, \frak{J}^{\mu\nu}\big|_{n}^{L_{1},L_{2}}\,,
\end{aligned}\label{}
\end{equation}
where $\ell_{12} = \ell_{1}+\ell_{2}$ and $L_{12} = L_{1}+L_{2}$.

We now consider the Fourier expansion for the matter currents. One may deduce the structure of these currents for the trajectories $X^{\mu}_{\alpha}$ from the geodesic equations
\begin{equation}
\begin{aligned}
  \ddot X^{\rho}_{\alpha}(\tau)
  &=
  \frac{1}{2} \partial_{\rho} \tilde{h}_{\mu\nu}(X_{1},X_{2}) v^{\mu}_{\alpha} v^{\nu}_{\alpha} + \cdots\,,
  \\
  &= \frac{1}{2} \int_{L_{1},L_{2}} i L_{12}^{\rho}\, e^{iL_{1} \cdot (X_{\alpha}-b_{1}) + iL_{2} \cdot (X_{\alpha}-b_{2})} \tilde{J}_{\mu\nu}\big|^{L_{1},L_{2}} + \cdots\,.
\end{aligned}\label{}
\end{equation}
This implies that the perturbative expansion for the matter currents, $X^{\mu}_{\alpha}\big|_{n}$, should have the following form:
\begin{equation}
\begin{aligned}
  X^{\mu}_{\alpha}(\tau)\big|_{1}
  &= 
    \int_{\ell_{1}} X^{\mu}_{\alpha} \big|_{1}^{\ell_{1},0}\, e^{i\ell_{1}\cdot (b_{\alpha} - b_{1} + v_{\alpha} \tau)}
  + \int_{\ell_{2}} X^{\mu}_{\alpha} \big|_{1}^{0,\ell_{2}}\, e^{i\ell_{2}\cdot (b_{\alpha} - b_{2} + v_{\alpha} \tau)}\,,
  \\
  X^{\mu}_{\alpha}(\tau)\big|_{n}
  &= 
    \int_{\ell_{1}} X^{\mu}_{\alpha} \big|_{n}^{\ell_{1},0}\, e^{i\ell_{1}\cdot (b_{\alpha} - b_{1} + v_{\alpha} \tau)} 
  + \int_{\ell_{2}} X^{\mu}_{\alpha} \big|_{n}^{0,\ell_{2}}\, e^{i\ell_{2}\cdot (b_{\alpha} - b_{2} + v_{\alpha}\tau)}
  \\&\quad
  + \int_{\ell_{1},\ell_{2}} X^{\mu}_{\alpha}\big|_{n}^{\ell_{1},\ell_{2}} \, e^{i\ell_{1}\cdot (b_{\alpha} - b_{1} + v_{\alpha} \tau) +i\ell_{2}\cdot (b_{\alpha} - b_{2} + v_{\alpha} \tau)}\,,
  \qquad n\geq2
\end{aligned}\label{}
\end{equation}
In total, we can thus summarize the perturbative expansion by
\begin{equation}
  \quad X^{\mu}_{\alpha}(\tau)\big|_{n}
  =
  \int_{L_{1},L_{2}} X^{\mu}_{\alpha}\big|_{n}^{L_{1},L_{2}} \, 
  e^{i\left(L_{12}\cdot (b_{\alpha}+v_{\alpha}\tau)-L_{1}\cdot b_{1}-L_{2}\cdot b_{2}\right)}  ~.
\label{X_current_expansion}\end{equation}
%

\subsection{A convolution bracket}

Our aim is to establish a simple algebraic scheme by means of which the post-Minkowskian expansion becomes automatized and following simple iterative equations can be
solved without recourse to diagrammatic expansions and other visual aids. However, since these recursion relations in terms of currents involve many indices and momenta, it can be difficult to follow their overall structure. For a more compact and transparent representation, we therefore introduce a bracket notation that generalizes the notion of convolution in Fourier transforms. 

To this end, let us consider an arbitrary field depending on doubled coordinates, $f^{A}(x_{1},x_{2})$, where $A,B,\cdots$ are formal labels that can be any type of indices. The $n$th order contribution to the product of these functions can be written as $\big(f^{A} * f^{B}\big)\big|_{n}^{x_{1},x_{2}}$ using the product defined in eq. \eqref{convolution_product}. If we denote the current corresponding to $f^{A}\big|_{n}^{x_{1},x_{2}}$ by $\mathfrak{f}^{A}\big|_{n}^{L_{1},L_{2}}$, the current for the product of the two fields is represented by the convolution of each momentum, which we can write by means of a bracket notation as $\big[\mathfrak{f}^{A}*\mathfrak{f}^{B}\big]_{n}^{L_{1},L_{2}}$. Obviously, the convolution bracket is symmetric and linear 
\begin{equation}
  \big[c_{A} \mathfrak{f}^{A} + c_{B} \mathfrak{f}^{B} * \mathfrak{f}^{C}\big]
  =
    c_{A} \big[ \mathfrak{f}^{A} * \mathfrak{f}^{C}\big] 
  + c_{B} \big[ \mathfrak{f}^{B} * \mathfrak{f}^{C}\big]\,.
\label{}\end{equation}
Explicitly, for the three types of different combinations it reads
\begin{naligned}
  &\big[\frak{f}^{A}*\mathfrak{f}^{B}\big]_{n}^{\ell_{1},0} 
  = 
  \sum_{m=1}^{n-1} \int_{k_1}
  \frak{f}^{A}\big|_{n-m}^{\ell_{1}-k_{1},0}\, \frak{f}^{B}\big|_{m}^{k_{1},0}\,,
  \\&
  \big[\frak{f}^{A}*\mathfrak{f}^{B}\big]_{n}^{\ell_{1},\ell_{2}}
  =
  \sum_{m=1}^{n-1} \Bigg[
    \frak{f}^{A}\big|_{n-m}^{\ell_{1},0} \, \frak{f}^{B}\big|_{m}^{0,\ell_{2}}
  + \frak{f}^{A}\big|_{n-m}^{0,\ell_{2}} \, \frak{f}^{B}\big|_{m}^{\ell_{1},0}
  +\int_{k_{1},k_{2}} 
  \frak{f}^{A}\big|_{n-m}^{\ell_{1}-k_{1},\ell_{2}-k_{2}} \frak{f}^{B}\big|_{m}^{k_{1},k_{2}}
  \\&\quad
  + \int_{k} \Big(
  	  \frak{f}^{A}\big|_{n-m}^{\ell_{1}-k,\ell_{2}} \frak{f}^{B}\big|_{m}^{k,0}
  	+ \frak{f}^{A}\big|_{n-m}^{\ell_{1},\ell_{2}-k} \frak{f}^{B}\big|_{m}^{0,k}
  	+ \frak{f}^{A}\big|_{n-m}^{\ell_{1}-k,0} \frak{f}^{B}\big|_{m}^{k,\ell_{2}}
  	+ \frak{f}^{A}\big|_{n-m}^{0,\ell_{2}-k} \frak{f}^{B}\big|_{m}^{\ell_{1},k}
  	\Big)\Bigg] \,,
  \\
  &\big[\frak{f}^{A}*\mathfrak{f}^{B}\big]_{n}^{0,\ell_{2}}
  =
  \sum_{m=1}^{n-1}\int_{k_{2}} \frak{f}^{A}\big|_{n-m}^{0,\ell_{2}-k_{2}} \frak{f}^{B}\big|_{m}^{0,k_{2}} \,.
\label{example_product_currents}
\end{naligned}
We also require that a convolution bracket of a current with a constant $c$ is a simple product between them,
\begin{equation}
  \big[\,\mathfrak{f}^{A}* c \big]_{n}^{L_{1},L_{2}}
  =
  c \, \mathfrak{f}^{A}\big|_{n}^{L_{1},L_{2}}\,,
  \qquad
  \big[\, \hat{\ell}^{\rho}\, \mathfrak{f}^{A}* c \big]_{n}^{L_{1},L_{2}}
  =
  c L^{\rho}_{12} \,\mathfrak{f}^{A}\big|_{n}^{L_{1},L_{2}}\,.
\label{}\end{equation}

It is not difficult to generalize this to include an operator $\hat{\ell}^{\rho}$ inside the convolution that includes momentum,
\begin{equation}
\begin{aligned}
  \big[\hat{\ell}^{\rho}\, \frak{f}^A*\frak{f}^B\big]_{n}^{\ell_{1},0} 
  &=
  \sum_{m=1}^{n-1} \int_{k} (\ell_{1}-k)^{\rho} \frak{f}^A\big|_{n-m}^{\ell_{1}-k,0}\, \frak{f}^B\big|_{m}^{k,0}\,,
  \\
  \big[\frak{f}^A*\hat{\ell}^{\rho}\, \frak{f}^B\big]_{n}^{\ell_{1},0} 
  &=
  \sum_{m=1}^{n-1} \int_{k} k^{\rho} \frak{f}^A\big|_{n-m}^{\ell_{1}-k,0}\, \mathfrak{f}^B\big|_{m}^{k,0}\,.
\end{aligned}\label{}
\end{equation}
We also note that, similar to the conventional convolution, the bracket has the associativity property,
\begin{equation}
  \big[\frak{f}^A*[\frak{f}^B*\frak{f}^C]\big]^{L_{1},L_{2}}
  = \big[[\frak{f}^A*\frak{f}^B]*\frak{f}^C\big]^{L_{1},L_{2}}
  = \big[\frak{f}^A*\frak{f}^B*\frak{f}^C\big]^{L_{1},L_{2}} \,.
\label{}\end{equation}
Finally, we also define what can be considered an exponentiation of a current  by means of the operation $\mathcal{E}$:
\begin{equation}
\begin{aligned}
  \mathcal{E}\big[\mathfrak{f}^{A}\big]_{n}^{L_{1},L_{2}}
  &\equiv
  \mathfrak{f}^{A}\big|_{n}^{L_{1},L_{2}}
  + i\big[
  	 \hat{\ell}_{\kappa} \mathfrak{f}^{A} * X^{\kappa}_{\alpha}
  	\big]_{n}^{L_{1},L_{2}}
  - \frac{1}{2}\big[
      \hat{\ell}_{\kappa}\hat{\ell}_{\lambda}\mathfrak{f}^{A}* X^{\kappa}_{\alpha}*
    	X^{\lambda}_{\alpha}
    \big]_{n}^{L_{1},L_{2}}
  + \cdots
  \\&\quad 
  +\frac{i^{n-1}}{(n-1)!} \big[
      \hat{\ell}_{\sigma_{1}}\hat{\ell}_{\sigma_{2}} \cdots \hat{\ell}_{\sigma_{n-1}} \mathfrak{f}^{A} * X^{\sigma_{1}}_{\alpha} * \cdots * X^{\sigma_{n-1}}_{\alpha}
    \big]_{n}^{L_{1},L_{2}} \,
\end{aligned}\label{E_operator}
\end{equation}
and this naturally generalizes to  
\begin{equation}
\begin{aligned}
  &\mathcal{E}\big[\mathfrak{f}^{A}*\mathfrak{f}^{B_{1}} * \cdots *\mathfrak{f}^{B_{m}}\big]_{n}^{L_{1},L_{2}}
  \\
  &\equiv
  \big[\mathfrak{f}^{A}*\mathfrak{f}^{B_{1}} * \cdots *\mathfrak{f}^{B_{m}}\big]_{n}^{L_{1},L_{2}}
  + i\big[
  	 \hat{\ell}_{\kappa} \mathfrak{f}^{A} *\mathfrak{f}^{B_{1}} * \cdots *\mathfrak{f}^{B_{m}}* X^{\kappa}_{\alpha} 
  	\big]_{n}^{L_{1},L_{2}}
  \\&\quad
  - \frac{1}{2}\big[
      \hat{\ell}_{\kappa}\hat{\ell}_{\lambda}\mathfrak{f}^{A}*\mathfrak{f}^{B_{1}} * \cdots *\mathfrak{f}^{B_{m}}* X^{\kappa}_{\alpha}*
    	X^{\lambda}_{\alpha}
    \big]_{n}^{L_{1},L_{2}}
  + \cdots
  \\&\quad 
  +\frac{i^{n-m-1}}{(n-m-1)!} \big[
      \hat{\ell}_{\sigma_{1}}\hat{\ell}_{\sigma_{2}} \cdots \hat{\ell}_{\sigma_{n-m-1}} \mathfrak{f}^{A}* \mathfrak{f}^{B_{1}} * \cdots *\mathfrak{f}^{B_{m}}* X^{\sigma_{1}}_{\alpha} * \cdots * X^{\sigma_{n-m-1}}_{\alpha}
    \big]_{n}^{L_{1},L_{2}} \,.
\end{aligned}\label{}
\end{equation}
%

\section{Derivation of the Recursion Relations}\label{Section:4}

We now have all the algebraic tools needed to set up the recursion relations in a manner that will allow us to perform post-Minkowskian calculations in a systematic manner. 
First, we will
employ the momentum-space expressions that we introduced in the previous section to derive recursion relations for both the graviton currents and the matter currents. These follow from the Einstein equations of motion and the geodesic equations, respectively. For readability we will here focus on outlining the overall structure of the recursive formalism, stressing the iterative way in which the equations can be solved, and how they can be algebraically extended to any order in the expansion. Explicit computations will be provided
in the subsequent sections.

\subsection{Recursion relations for the graviton currents}
We have already derived the PM expansion of the equations of the graviton field in eq. \eqref{Expansion_Einstein_eq}. Using the notation introduced in the previous section we can write the solution in terms of the currents as
\begin{equation}
\begin{aligned}
  \mathfrak{J}^{\mu\nu}\big|_{1}^{L_{1},L_{2}}
  &=
    \frac{2}{(L_{12})^{2}} \mathbf{j}^{\mu\nu}\big|_{1}^{L_{1},L_{2}}\,.
  \\
  \mathfrak{J}^{\mu\nu}\big|_{n}^{L_{1},L_{2}}
  &=
    \frac{1}{(L_{12})^{2}} \Big(
  	- \tau^{\mu\nu}\big|_{n}^{L_{1},L_{2}}
  	+ 2\mathbf{j}^{\mu\nu}\big|_{n}^{L_{1},L_{2}}
  \Big)\,.
\end{aligned}\label{LLgraviton_current}
\end{equation}
Here the currents of the energy-momentum pseudotensor are represented by 
\begin{equation}
\begin{aligned}
  \tau^{\mu\nu}\big|_{n}^{L_{1},L_{2}}
  =
  \Big(W^{\mu\nu}
    + Z^{(\mu}{}_{\kappa} \eta^{\nu) \kappa}+2d^{\mu\nu}\Big)\Big|_{n}^{L_{1},L_{2}}
    - \Big[Z^{(\mu}{}_{\kappa} * \mathfrak{J}^{\nu) \kappa}\Big]_{n}^{L_{1},L_{2}}
\end{aligned}\label{EMpseudotensor}
\end{equation}
where, from the equations of motion,
\begin{equation}
\begin{aligned}
  W^{\mu\nu}\big|_{n}^{L_{1},L_{2}}
  &= 
    \big[ 
      \hat{\ell}_{\lambda} \frak{J}^{\kappa\mu} * \hat{\ell}_{\kappa} \frak{J}^{\lambda\nu}
    \big]_{n}^{L_{1},L_{2}}
  - \big[ 
      \frak{J}^{\kappa\lambda} * \hat{\ell}_{\kappa}\hat{\ell}_{\lambda} \frak{J}^{\mu\nu}
    \big]_{n}^{L_{1},L_{2}}
\,,
  \\
  Z^{\mu}{}_{\nu}\big|^{n}_{L_{1},L_{2}}
  &=
    \Big(Z^{\kappa\mu}{}_{\kappa\nu}
  {-} 2 Z^{\mu\kappa}{}_{\kappa\nu}
  + \frac{1}{2} Z^{\mu\kappa}{}_{\nu\kappa}\Big) \Big|_{n}^{L_{1},L_{2}}
  + \frac{\delta^{\mu}{}_{\nu}}{4} \Big(
  	  2Z^{\kappa\lambda}{}_{\lambda\kappa}
  	- Z^{\kappa\lambda}{}_{\kappa\lambda}
  	\Big)\Big|_{n}^{L_{1},L_{2}} \,,
  \\
  d^{\mu\nu} \big|^{n}_{L_{1},L_{2}}
  &= 
    2 \big[d^{\mu}*d^{\nu}\big]_{n}^{L_{1},L_{2}}
  - \Big[\big(\eta^{\mu\nu}-\frak{J}^{\mu\nu}\big)*\big(\eta_{\kappa\lambda}+\tilde{\frak{J}}_{\kappa\lambda}\big)*d^{\kappa} * d^{\lambda}\Big]_{n}^{L_{1},L_{2}} \,.
\end{aligned}\label{recursion_subcurrents_WZ}
\end{equation}
and
\begin{equation}
\begin{aligned}
  Z^{\mu\nu}{}_{\rho\sigma}\big|_{n}^{L_{1},L_{2}}
  &=
  \Big[ \big(\eta^{\mu\kappa}-\frak{J}^{\mu\kappa}\big)*\hat{\ell}_{\kappa} \frak{J}^{\nu \lambda} * \hat{\ell}_{\rho} \tilde{\frak{J}}_{\lambda \sigma}\, \Big]_{n}^{L_{1},L_{2}} \,,
  \\
  d^{\mu}\big|_{n}^{L_{1},L_{2}}
  &= 
   \frac{i}{4} \Big[\big(\eta^{\mu\nu}-\frak{J}^{\mu\nu}\big) * \big(\eta_{\kappa\lambda}+\tilde{\frak{J}}_{\kappa\lambda}\big) * \hat{\ell}_{\nu} \frak{J}^{\kappa\lambda}\, \Big]_{n}^{L_{1},L_{2}} \,.
\end{aligned}\label{recursion_subcurrents_d}
\end{equation}
We also need the solution for the graviton current associated with the metric perturbations $\tilde{h}_{\mu\nu}|_{n}$ in order to solve the geodesic equations. This follows from relation  \eqref{metric_fluctuations}, which yields
\begin{equation}
  \tilde{J}_{\mu\nu}\big|_{n}^{L_{1},L_{2}}
  = 
  \tilde{\mathfrak{J}}_{\mu\nu}\big|_{n}^{L_{1},L_{2}}
  + \big[H * \tilde{\mathfrak{J}}_{\mu\nu}\big]_{n}^{L_{1},L_{2}}
  + \eta_{\mu\nu} H\big|_{n}^{L_{1},L_{2}}\,.
\label{}\end{equation}

We next consider the external source current $\mathbf{j}^{\mu\nu}$ associated with $\mathfrak{j}^{\mu\nu}$
\begin{equation}
\begin{aligned}
  \mathbf{j}^{\mu\nu}\big|_{1}^{L_{1},L_{2}}
  &= 
  \mathfrak{j}^{\mu\nu}\big|_{1}^{L_{1},L_{2}} \,,
  \\
  \mathbf{j}^{\mu\nu}\big|_{n}^{L_{1},L_{2}}
  &= 
    \mathfrak{j}^{\mu\nu}\big|_{n}^{L_{1},L_{2}}
  + H* \mathfrak{j}^{\mu\nu}\big|_{n}^{L_{1},L_{2}} \,, \qquad n>1\,.
\end{aligned}\label{fat_j}
\end{equation}
We remind the reader of the series expansions for the matter source $\mathfrak{j}^{\mu\nu}$ in \eqref{Expansion_j_density}
\begin{equation}
\begin{aligned}
  \mathfrak{j}^{\mu\nu}\big|_{1}^{x_{1},x_{2}}
  &= 
  8\pi \sum_{\alpha=1}^{2} m_{\alpha} v^{\mu}_{\alpha}v^{\nu}_{\alpha} \int_{\ell} e^{i\ell \cdot x_{\alpha}} \hat{\delta} (\ell\cdot v_{\alpha})\,,
  \\
  \mathfrak{j}^{\mu\nu}\big|_{n}^{x_{1},x_{2}}
  &= 
  8\pi \sum_{\alpha=1}^{2} m_{\alpha}
  \int_{\tau} \int_{\ell} 
  e^{i \ell\cdot (x_{\alpha}-v_{\alpha}\tau)} \zeta_{\alpha}^{\mu\nu}(\ell)\big|_{n-1}\,, \quad n>1\,,
\label{external_source_current}\end{aligned}
\end{equation}
where $\int_{\tau} = \int_{-\infty}^{\infty} \mathrm{d} \tau$.
This may appear to be quite involved but it simply expresses the projection on the given fixed order $n$ of the expansion in $G$. It can readily be programmed
so as to provide explicit expressions to any order $n$.
We obtain the Fourier transform for doubled cooordinates of $\zeta_{\alpha}^{\mu\nu}(\ell)\big|_{n}$ by substituting the expansion of $X^{\mu}_{\alpha}$,
\begin{equation}
\begin{aligned}
  \zeta_{\alpha}^{\mu\nu}(\ell)\big|_{n}
  = 
  \int_{L_{1},L_{2}} \zeta_{\alpha}^{\mu\nu}(\ell)\big|_{n}^{L_{1},L_{2}}
  e^{ i L_{1}\cdot (b_{\alpha 1}+v_{\alpha}\tau) + i L_{2}\cdot (b_{\alpha 2}+v_{\alpha}\tau)}\,,
  \quad n\geq1\,,
\end{aligned}\label{}
\end{equation}
where $b_{\alpha\beta} = b_{\alpha}-b_{\beta}$ is the relative impact parameter, and 
\begin{equation}
\begin{aligned}
  \zeta_{\alpha}^{\mu\nu}(\ell)\big|_{n}^{L_{1},L_{2}}
  &= 
  \Big[ 
      e^{-i\ell\cdot X_{\alpha}*} 
    * \Big(v^{\mu}_{\alpha} + i(\hat{\ell}\cdot v_{\alpha}) X_{\alpha}^{\mu}\Big) 
    * \left(v^{\nu}_{\alpha}+i(\hat{\ell}\cdot v_{\alpha}) X_{\alpha}^{\nu}\right)
  \Big]_{n}^{L_{1},L_{2}} \,.
\end{aligned}\label{nPMZeta}
\end{equation}
Here the exponentiation of the convolution operation is defined in the obvious way by
\begin{equation}
\begin{aligned}
  \big[ e^{-i\ell\cdot X_{\alpha}*}\big]_{n}^{L_{1},L_{2}}
  &=
  1 - i\ell\cdot X_{\alpha}\big|_{n}^{L_{1},L_{2}}
  - \frac{1}{2!}\ell_{\rho}\ell_{\sigma}
  	\big[X^{\rho}_{\alpha}*X^{\sigma}_{\alpha}\big]_{n}^{L_{1},L_{2}}
  \\&\quad
  + \frac{i}{3!} \ell_{\rho_{1}} \ell_{\rho_{2}} \ell_{\rho_{3}} 
  	\big[X^{\rho_{1}}_{\alpha}*X^{\rho_{2}}_{\alpha}*X^{\rho_{3}}_{\alpha}\big]_{n}^{L_{1},L_{2}} +\cdots\,.
\end{aligned}\label{}
\end{equation}
Using the compact notation for doubled Fourier transforms the matter currents $\mathfrak{j}^{\mu\nu}(x)$ are then expanded as follows:
\begin{equation}
\begin{split}
  \mathfrak{j}^{\mu\nu}\big|_{n}
  &=
  8\pi \sum_{\alpha=1}^{2} m_{\alpha} 
  \int_{\tau} 
  \int_{\ell,L_{1},L_{2}}
  \zeta_{\alpha}^{\mu\nu}(\ell)\big|_{n-1}^{L_{1},L_{2}}
  e^{i \ell\cdot (x_{\alpha}-v_{\alpha}\tau) + i L_{1}\cdot (b_{\alpha 1}+v_{\alpha}\tau) + i L_{2}\cdot (b_{\alpha 2}+v_{\alpha}\tau)}\,.
\end{split}\label{}
\end{equation}
Shifting $\ell \to \ell+L_{1}+L_{2}$, this reduces to 
\begin{equation}
\begin{aligned}
  \mathfrak{j}^{\mu\nu}\big|_{n}
  &=
    8\pi \sum_{\alpha=1}^{2} m_{\alpha} \int_{\ell,L_{1},L_{2}} \hat{\delta}(\ell \cdot v_{\alpha}) 
    \zeta_{\alpha}^{\mu\nu}(\ell+L_{12})\big|_{n-1}^{L_{1},L_{2}}
  	e^{i \ell\cdot x_{\alpha} + i L_{1}\cdot x_{1} + i L_{2}\cdot x_{2} }\,.
\end{aligned}\label{}
\end{equation}
Since the matter currents satisfy $X^{\mu}_{1}\big|_{n}^{\ell_{1},0}=0$ and $X^{\mu}_{2}\big|_{n}^{0,\ell_{2}}=0$, it is straightforward to show that
\begin{equation}
  \zeta_{1}^{\mu\nu}(\ell)\big|_{n-1}^{\ell_{1},0}=0\,,
  \qquad
  \zeta_{2}^{\mu\nu}(\ell)\big|_{n-1}^{0,\ell_2}=0\,.
\label{}\end{equation}
%
After shifting and relabeling momentum, we can read off the currents for the external sources:
\begin{equation}
\begin{aligned}
  \mathfrak{j}^{\mu\nu}\big|_{n}^{\ell_{1},0}
  &=0\,,
  \\
  \mathfrak{j}^{\mu\nu}\big|_{n}^{0,\ell_{2}}
  &=0\,,
  \\
  \mathfrak{j}^{\mu\nu}\big|_{n}^{\ell_{1},\ell_{2}}
  &=
    8\pi \Big[
      m_{1}\hat{\delta}(\ell_{1} \cdot v_{1}) \zeta_{1}^{\mu\nu}(\ell_{12})\big|_{n-1}^{0,\ell_{2}}
  	+ m_{2}\hat{\delta}(\ell_{2} \cdot v_{2}) \zeta_{2}^{\mu\nu}(\ell_{12})\big|_{n-1}^{\ell_{1},0}
  \Big]
  \\&\quad
  + 8\pi \int_{k} \bigg[
      m_{1}\hat{\delta}(k \cdot v_{1}) \zeta_{1}^{\mu\nu} (\ell_{12}) \big|_{n-1}^{\ell_{1}-k,\ell_{2}}
  	+ m_{2}\hat{\delta}(k \cdot v_{2}) \zeta_{2}^{\mu\nu} (\ell_{12}) \big|_{n-1}^{\ell_{1},\ell_{2}-k}\bigg]\,,
\end{aligned}\label{currents_external_source}
\end{equation}
and these expressions are valid for $n\ge 2$.

\subsection{Matter Currents and momentum kick} \label{Sec:4.2}

We are finally ready to derive the recursions for the $n$th order matter current $X^{\mu}_{\alpha}\big|_{n}^{L_{1},L_{2}}$. We obtain these by substituting the post-Minkowskian expansion of each field into the perturbed geodesic equation \eqref{perturbedGeodesicEq},
\begin{equation}
\begin{aligned}
  \frac{d}{d\tau} \bigg(\eta_{\rho\sigma}\dot{X}^{\sigma}_{\alpha}\big|^n 
  + \Big(\tilde{h}_{\rho\sigma}\big(X_{\alpha}\big) *\dot{X}^{\sigma}_{\alpha}\Big)\Big|_{n}
  \bigg) 
  = 
  \frac{1}{2} \Big(
  \partial_{\rho} \tilde{h}_{\mu\nu}\big(X_{\alpha}\big) * \dot{X}^{\mu}_{\alpha} * \dot{X}^{\nu}_{\alpha}
  \Big)\Big|_{n}\,.
\end{aligned}\label{def_goedesic}
\end{equation}
Here the Fourier expansion of the perturbation of metric tensor, $i.e.$, $\tilde{h}_{\mu\nu}(X_{\alpha})\big|^{n}$ is given by
\begin{equation}
\begin{aligned}
  \tilde{h}_{\mu\nu}(X_{\alpha})
  &=
  \int_{L_{1},L_{2}} e^{i L_{1} \cdot (X_{\alpha}-b_{1}) + i L_{2} \cdot (X_{\alpha}-b_{2})} \tilde{J}_{\mu\nu}\big|^{L_{1},L_{2}}\,,
\end{aligned}\label{}
\end{equation}
and by substituting the Fourier expansion of $X_{\alpha}$, we obtain
\begin{equation}
\begin{split}
  \tilde{h}_{\mu\nu}(X_{\alpha})\big|_{n}
  &\!=\!
  \int_{L_{1},L_{2}} \!
  e^{i L_{1} \cdot \bar{X}_{\alpha,1} + i L_{2} \cdot \bar{X}_{\alpha,2}} \mathcal{E}\big[\tilde{J}_{\mu\nu}\big]_{n}^{L_{1},L_{2}}
\end{split}\label{}
\end{equation}
where the $\mathcal{E}$ operator is defined in \eqref{E_operator} and
\beq
  \bar{X}^{\mu}_{\alpha} ~\equiv~ X^{\mu}_{\alpha}\big|_{0}\,,
  \qquad
  \bar{X}^{\mu}_{\alpha,\beta} ~\equiv~ X^{\mu}_{\alpha}|^{0}-b^{\mu}_{\beta}
\eeq
This is the general all-order expression, truncated to the desired order in $G$.

These relations give us the post-Minkowskian expansion of the equations of motion for the matter fields to any order, as determined by the geodesic equation in perturbation theory. As an example, the recursion relation for the matter currents up to third post-Minkowskian order is given by
\begin{naligned}
  X^{\rho}_{\alpha}\big|_{n}^{L_{1},L_{2}}
  =
   \frac{i}{L_{12}\cdot v_{\alpha}}
    \mathcal{E}\big[\, \tilde{J}^{\rho}{}_{\sigma} * \mathcal{X}^{\sigma}_{\alpha}\big]_{n}^{L_{1},L_{2}}
  - \frac{i}{2(L_{12}\cdot v_{\alpha})^{2}} \mathcal{E}\big[ \hat{\ell}^{\rho} \tilde{J}_{\mu\nu}
    * \mathcal{X}^\mu_{\alpha} * \mathcal{X}^{\nu}_{\alpha}
    \big]_{n}^{L_{1},L_{2}}\,,
\label{Recursion_worldline_current}\end{naligned}
where
\begin{equation}
  \mathcal{X}^{\rho}_{\alpha}
  \equiv
  v^{\rho}_{\alpha}+i(\hat{\ell} \cdot v_{\alpha}) X^{\rho}_{\alpha}\,.
\label{}\end{equation}
The above expansion to third post-Minkowskian order suffices for the explicit recursive computations we will present in this paper.

The momentum kick of each participating matter source is defined perturbatively by
\begin{equation}
  \Delta P^{\rho}_{\alpha}\big|_{n}
  ~\equiv~
  m_{\alpha} \int_{-\infty}^{\infty}\mathrm{d}\tau\, \ddot{X}^{\rho}_{\alpha}(\tau)\big|_{n}\,.
\end{equation}
Using the perturbative expansion of the matter current \eqref{X_current_expansion} and evaluating the $\tau$ integration, the momentum kick for the particle 1 is 
then computed from the pure and mixed modes,
\begin{equation}
\begin{split}
  \Delta P^{\rho}_{1}\big|_{n}
  &=
  - m_{1} \int_{-\infty}^{\infty}\mathrm{d}\tau 
  \int_{L_{1},L_{2}}
  e^{iL_{12}\cdot v_{1} \tau + iL_{2}\cdot b} 
  (L_{12} \cdot v_{1})^{2} X^{\rho}_{1}\big|_{n}^{L_{1},L_{2}}
  \,,
  \\
  &=
  - m_{1} \int_{\ell} e^{i \ell \cdot b} \bigg[
  	  \hat{\delta}(\ell\cdot v_{1}) (\ell\cdot v_{1})^{2} X^{\rho}_{1} \big|_{n}^{0,\ell}
  	+ \int_{k} \hat{\delta}\big(k\cdot v_{1}) (k\cdot v_{1})^{2} X^{\rho}_{1} \big|_{n}^{k-\ell,\ell} 
  \bigg]\,,
\end{split}\label{}
\end{equation}
where $b \equiv b_{1}-b_{2}$.

Using the geodesic equation \eqref{def_goedesic}, momentum kick can also be written as
\begin{equation}
\begin{aligned}
  \eta_{\rho\sigma}\Delta P^{\sigma}_{\alpha}\big|_{n} 
  &=
  - m_{\alpha} \int_{-\infty}^{\infty}\mathrm{d}\tau \, \frac{\mathrm{d}}{\mathrm{d}\tau}
  \Big(\tilde{h}_{\rho\sigma}\big(X_{\alpha}\big) *\dot{X}^{\sigma}_{\alpha}\Big)\Big|_{n}
  \\&\quad
  + \frac{m_{\alpha}}{2} \int_{-\infty}^{\infty}\mathrm{d}\tau \,
  \partial_{\rho} \tilde{h}_{\mu\nu}\big(X_{\alpha}\big)* \dot{X}^{\mu}_{\alpha} *\dot{X}^{\nu}_{\alpha}
  \Big|_{n}
  \\
\end{aligned}\label{Recursion_worldline_current}
\end{equation}
We find that the first term on the right-hand side does not contribute to $\Delta P^{\mu}_{\alpha}$ up to 3PM order due to the appearance of a delta-function arising upon 
integration over $\tau$. Then the momentum kick for particle 1 is simply given by the second line of \eqref{Recursion_worldline_current} only, $viz.$,
\begin{equation}
\begin{aligned}
&-\frac{2i}{m_{1}}\Delta P^{\rho}_{1}\big|_{n} 
  \\
  &=
  \int_{\ell} e^{i\ell\cdot b} \bigg[ 
    \hat{\delta} \big( \ell\cdot v_{1}\big)
    \mathcal{E}\Big[ 
        \hat{\ell}^{\rho} \tilde{J}_{\mu\nu}
      * \mathcal{X}^{\mu}_{1}
      * \mathcal{X}^{\nu}_{1}
    \Big]_{n}^{0,\ell}
  + \int_{k}\hat{\delta} \big( k\cdot v_{1}\big)
  \mathcal{E}\Big[ 
      \hat{\ell}^{\rho} \tilde{J}_{\mu\nu}
    * \mathcal{X}^{\mu}_{1}
    * \mathcal{X}^{\nu}_{1}
    \Big]_{n}^{k-\ell,\ell} 
  \bigg] \,.
\end{aligned}\label{momentum_kick_1}
\end{equation}

We can conveniently introduce the integrand of the 2-dimensional Fourier transform for the momentum kick
\begin{equation}
  \Delta P^{\rho}_{1} \big|_{n}
  =
  \int_{\ell} e^{i\ell\cdot b} \hat{\delta} \big( \ell\cdot v_{1}\big) \hat{\delta} \big( \ell\cdot v_{2}\big)
  \Big(
      \widetilde{\Delta P}{}^{\rho}_{1} (\ell)\big|_{n}^{(A)}
    + \widetilde{\Delta P}{}^{\rho}_{1} (\ell)\big|_{n}^{(B)}
  \Big)\,,
\label{Fourier_integrand}\end{equation}
where $\widetilde{\Delta P}{}^{\rho}_{1} (\ell)\big|^{n}_{A,B}$ are the integrands under the orthogonality condition $\ell\cdot v_{1} = \ell\cdot v_{2} = 0$,
imposed by the delta-functions,
\begin{equation}
\begin{aligned}
  \widetilde{\Delta P}{}^{\rho}_{1} (\ell)\big|_{n}^{(A)}
  &=
  \frac{m_{1}}{2}
  \mathcal{E}\Big[ 
      i\hat{\ell}^{\rho} \tilde{J}_{\mu\nu}
    * \mathcal{X}^{\mu}_{1}
    * \mathcal{X}^{\nu}_{1}
  \Big]_{n}^{0,\ell}\Big|_{\ell\cdot v_{1} \to 0\atop \ell\cdot v_{2} \to 0}
  \\
  \widetilde{\Delta P}{}^{\rho}_{1} (\ell)\big|_{n}^{(B)}
  &= 
  \frac{m_{1}}{2} \int_{k}\hat{\delta} \big( k\cdot v_{1}\big)
  \mathcal{E}\Big[ 
      i\hat{\ell}^{\rho} \tilde{J}_{\mu\nu}
    * \mathcal{X}^{\mu}_{1}
    * \mathcal{X}^{\nu}_{1}
    \Big]_{n}^{k-\ell,\ell}\Big|_{\ell\cdot v_{1} \to 0\atop \ell\cdot v_{2} \to 0}\,.
\end{aligned}\label{Fourier_integrand2}
\end{equation}
%


\section{Iteration of Loop Integrals}\label{Sec:5}

We now turn to a most crucial aspect: the systematic evaluation of loop 
integrals. Here, we propose a novel approach for what in quantum field 
theory correspond to multi-loop integrals, constructed by iteration of 
one-loop integrals. A key observation underlying this method is that the 
integrals under consideration essentially share the same structure as 
the two-loop four-point functions familiar from quantum field theory in
both the amplitude-based approach and the worldline approaches. In 
comparison with the iterative procedure in the graviton sector alone the 
crucial difference arises from the delta functions present in the 
numerators (the velocity-cuts in the amplitude formulation 
\cite{Bjerrum-Bohr:2021din}), which, as we shall see, play a crucial 
role when carrying out the iterative procedure.

At first non-trivial order in $G$, which is the second post-Minkowskian 
order to which Westpfahl computed \cite{Westpfahl:1985tsl}, the momentum 
integrations are straightforward. As we shall demonstrate in the next 
section, the computation to third post-Minkowskian order involves three 
distinct 2-loop integral families, depending on the types of delta 
functions appearing in the numerator of the momentum integrands. Let us denote them by the labels family I, family II and radiation, defined as follows:
\begin{equation}
\begin{aligned}
  J^{\rm I,\pm}_{n_{1},n_{2},n_{3},\cdots,n_{9}}
  &=
  \frac{(-1)^{n_{8}-1} (-1)^{n_{9}-1}}{(n_{8}-1)!(n_{9}-1)!} \int_{k_{1},k_{2}}\frac{\hat{\delta}^{(n_{8-1})} \left(k_{1} \cdot v_{1}\right) \hat{\delta}^{(n_{9-1})}\left(k_{2} \cdot v_{1}\right)}{
  D_{1}^{n_{1}} D_{2}^{n_{2}} D_{3}^{n_{3}} \cdots D^{n_{6}}_{6,\ord{2}} D^{n_{7}}_{7,\ord{2}}}\,,
  \\
  J^{\rm II,\pm}_{n_{1},n_{2},n_{3},\cdots,n_{9}}
  &=
   \frac{(-1)^{n_{8}-1} (-1)^{n_{9}-1}}{(n_{8}-1)!(n_{9}-1)!} \int_{k_{1},k_{2}}\frac{\hat{\delta}^{(n_{8-1})} \left(k_{1} \cdot v_{1}\right) \hat{\delta}^{(n_{9-1})} \left(k_{2} \cdot v_{2}\right)}{D_{1}^{n_{1}} D_{2}^{n_{2}} D_{3}^{n_{3}} \cdots D^{n_{6}}_{6,\ord{2}} D^{n_{7}}_{7,\ord{1}}}\,.
  \\
  M_{n_{2},n_{3},n_{5},n_{7_{-}},n_{7_{+}}}
  &=
  \int_{k_{1},k_{2}} 
  \frac{\hat{\delta} (k_{1} \cdot v_{1}) \hat{\delta} (k_{2} \cdot v_{2})}{ D_{2}^{n_{2}} D_{3}^{n_{3}} D_{5}^{n_{5}} D^{n_{7_{-}}}_{7_{-}} D^{n_{7_{+}}}_{7_{+}} }
\end{aligned}
\label{integral_Families}\end{equation}
where $\hat{\delta}^{(n)}(x)$ is $n$-th derivative of a delta-function,
\begin{equation}
   \hat{\delta}^{(n)}(x) = \frac{d^{n}}{d x^{n}}\hat{\delta}(x)\,,
\label{def_delta_n}\end{equation}
and the propagators are defined by
\begin{equation}
\begin{aligned}
&D_{1} = k_{1}^{2}\,, \quad
D_{2} = k_{2}^{2}\,, \quad
D_{3} = \big(\ell-k_{1}-k_{2}\big)^{2}-\text{sgn}(\ell^{0}-k_{1}^{0}-k_{2}^{0})i 0_{+}\,, \quad
\\&
D_{4} = \big(\ell-k_{1}\big)^{2}\,, \quad
D_{5} = \big(\ell-k_{2}\big)^{2}\,, \quad
D_{6,(i)} = k_{1}\cdot v_{i}-i0_{+}\,,\quad 
D_{7,(i)} = k_{2}\cdot v_{i}\mp i 0_{+}\,,
\end{aligned}\label{}
\end{equation}
where $D_{1}, D_{2}, D_{4}, D_{5}$ are postive definite due to the delta functions and $D_{3}$, $D_{6}$ and $D_{7}$ are the retarded propagators. Here $\ell$ is the Fourier-transformed variable associated with either particle 1 or 2 satisfying the orthogonality condition $\ell\cdot v_{1} = 0$ and $\ell\cdot v_{2} = 0$.

Not surprisingly, the $i\epsilon$-prescriptions are crucial here, and they can be summarized by the following prescription: we impose causality by requiring all Green functions
to be {\em retarded}. For some Green functions we do not need to specify the contours of integration around the poles because they never approach those, and for others, such
as those of linear propagators, the $i\epsilon$-prescription is not sensitive to the sign of the $0$-component of momentum. But when Green functions for the metric field can 
approach the singularity we impose the retarded Green function prescription, as required by causality. This matches precisely the worldline formalism when observables
there are computed using the {\em in-in} formalism \cite{Jakobsen:2022psy,Kalin:2022hph}. 

In our mostly-plus metric convention, the four-velocities $v^{\mu}_{1}$ and $v^{\mu}_{2}$ for 
particles 1 and 2 satisfy
\begin{equation}
   v_{1}^{2} = -1\,,
   \qquad
   v_{2}^{2} = -1\,,
   \qquad
   v_{1}\cdot v_{2} = - \gamma\,, \quad \gamma > 1\,.
\label{}\end{equation}
Let us parametrize an arbitrary $D$-dimensional vector $k^{\mu}$ as
\begin{equation}
   k^{\mu}
   =
   (k^{0}, k^{i}_{\perp},k^{z})
   =
   (k^{0}, \vec{k}_{\perp},k^{z})\,,
\label{}\end{equation}
and chose as reference frame the rest frame of particle 1 with particle 
2 moving initially along the $z$-direction,
\begin{equation}
   v_{1}^{\mu} = (1, \vec{0})\,,
   \qquad
   v_{2}^{\mu} = \big(\gamma, \vec{0}_{\perp}, \sqrt{\gamma^{2}-1}\big)\,,
\label{v_parametrization}\end{equation}
Eventually, when we compute the momentum kick the momentum-space 
variable $\ell^{\mu}$ will have to satisfy two delta-function 
constraints corresponding to $\ell \cdot v_{1} = \ell \cdot v_{2} = 0$. 
In the chosen rest frame of particle of particle 1 above we can thus 
parametrize $\ell^{\mu}$ as
\begin{equation}
   \ell^{\mu} = (0,\vec{\ell}_{\perp},0)\,.
\label{}\end{equation}
for both indices 1 and 2.

\subsection{Family-I Integrals} \label{Sec:5.1}

We first consider the type-I integral family, characterized by the 
presence of delta-functions $\hat{\delta}(k_{1}\cdot v_{1}) 
\hat{\delta}(k_{2}\cdot v_{1})$. By employing LiteRed \cite{Lee:2013mka}, we can identify the three master integrals in family $\rm I$, under the condition $n_{4}=n_{5}=0$, which is the only relevant case for our actual computation,
\begin{equation}
\begin{aligned}
   f^{\rm I}_{1} = J^{\rm I,+}_{1,1,1,0,0,0,0,1,1}\,,
   \qquad
   f^{\rm I}_{2} = J^{\rm I,+}_{1,1,1,0,0,1,0,1,1}\,,
   \qquad
   f^{\rm I,\pm}_{3} = J^{\rm I,\pm}_{1,1,1,0,0,1,1,1,1}\,.
\end{aligned}\label{Fam_IMI}
\end{equation}
Since these master integrals satisfy $n_{8}=n_{9} =1$, the numerator of 
the integrand reduces to delta functions only. Then, using the 
parametrizations of $v^{\mu}_{1,2}$ in \eqref{v_parametrization}, the $\delta$--functions reduce as follows:
\begin{equation}
   \hat{\delta}(k_{1}\cdot v_{1})
   \to
   \hat{\delta} (k_{1}^{0})\,,
   \qquad
   \hat{\delta}(k_{2}\cdot v_{1})
   \to
   \hat{\delta} (k_{2}^{0})\,.
\label{}\end{equation}
After integrating out the delta functions, the loop momenta are $ 
D-1$-dimensional vectors,
\begin{equation}
\begin{aligned}
   k_{1}^{\mu} =
   \begin{pmatrix}
   0 \\ \vec{k}_{1}
   \end{pmatrix}\,,
   \qquad
   k_{2}^{\mu} =
   \begin{pmatrix}
   0 \\ \vec{k}_{2}
   \end{pmatrix}\,.
\end{aligned}\label{}
\end{equation}
and the corresponding loop integrands are simplified as
\begin{equation}
   J^{\rm I,\pm}_{n_{1},n_{2},n_{3},n_{4},n_{5}, n_{6},n_{7}}
   =
   \int_{\vec{k}_{1},\vec{k}_{2}}
   \frac{1}{\mathbf{D}_{1}^{n_{1}} \mathbf{D}_{2}^{n_{2}} 
\mathbf{D}_{3}^{n_{3}} \cdots \mathbf{D}^{n_{6}}_{6} 
\mathbf{D}^{n_{7}}_{7}}\,,
\label{}\end{equation}
where the propagators are
\begin{equation}
\begin{aligned}
&\mathbf{D}_{1} = \big|\vec{k}_{1}\big|^{2}\,, \qquad
\mathbf{D}_{2} = \big|\vec{k}_{2}\big|^{2}\,, \qquad
\mathbf{D}_{3} = \big|\vec{\ell}-\vec{k}_{1}-\vec{k}_{2}\big|^{2}\,, \qquad
\mathbf{D}_{4} = \big|\vec{\ell}-\vec{k}_{1}\big|^{2}\,,
\\
&\mathbf{D}_{5} = \big|\vec{\ell}-\vec{k}_{2}\big|^{2}\,, \qquad
\mathbf{D}_{6} = \vec{k}_{1}\cdot \vec{v}_{2}-i0_+ \,, \qquad
\mathbf{D}_{7} = \pm \vec{k}_{2}\cdot \vec{v}_{2}-i0_+ \,.
\end{aligned}\label{}
\end{equation}

Let us now evaluate the master integrals \eqref{Fam_IMI} by using 
iterations of one-loop integrals. First, $f^{\rm I}_{1}$ can be 
reorganized as double bubble one-loop integrals, and we can easily 
evaluate using the bubble integral formula \eqref{bubble_integral}
\begin{equation}
\begin{aligned}
   f^{\rm I}_{1}
   &=
   \int_{\vec{k}_{1}} \frac{1}{|\vec{k}_{1}|^{2}}\int_{k_{2}} 
\frac{1}{|\vec{k}_{2}|^{2}|\vec{\ell}-\vec{k}_{1}-\vec{k}_{2}|^{2}}\,,
   \\
   &=
   (4 \pi)^{-3+2 \epsilon} 
\frac{\Gamma^{3}\left(\frac{1}{2}-\epsilon\right) \Gamma(2 
\epsilon)}{\Gamma\left(\frac{3}{2}-3 \epsilon\right)} |\vec{\ell}\ 
|^{-4\epsilon} \,.
\end{aligned}\label{}
\end{equation}
Similarly, $f^{\rm I}_{2}$ can be rearranged as bubble--triangle 
integrals, and we can evaluate using \eqref{triangle_integral} and 
\eqref{bubble_integral}
\begin{equation}
\begin{aligned}
   f^{\rm I}_{2}
   &=
   \int_{\vec{k}_{1}} \frac{1}{|\vec{k}_{1}|^{2} (\vec{k}_{1}\cdot \vec{v}_{2}-i0_+)}
   \int_{\vec{k}_{2}} 
\frac{1}{|\vec{k}_{2}|^{2}|\vec{\ell}-\vec{k}_{1}-\vec{k}_{2}|^{2}}\,,
   \\&
   =
   \frac{i 16^{3 \epsilon -1} \pi ^{2 \epsilon -1} \Gamma (-4 \epsilon ) 
\Gamma \left(2 \epsilon +\frac{1}{2}\right) }{\Gamma \left(\frac{1}{2}-3 
\epsilon \right) \Gamma (1-\epsilon ) \sqrt{\gamma ^2-1}} |\vec{\ell}\ 
|^{-1-4\epsilon}\,.
\end{aligned}\label{}
\end{equation}
Note that the order of integration is crucial. The one-loop triangle 
formula given in \eqref{triangle_integral} assumes that the external 
momentum $\ell$ satisfies the condition $\ell\cdot v_{1} = 0$ and $\ell\cdot 
v_{2} = 0$. However, if we perform the triangle integral first in the 
above expression,
\begin{equation}
   f^{\rm I}_{2}
   =
   \int_{k_{2}} \frac{1}{|\vec{k}_{2}|^{2}}
   \int_{k_{1}} \frac{1}{|\vec{k}_{1}|^{2} 
|\vec{\ell}-\vec{k}_{1}-\vec{k}_{2}|^{2} (\vec{k}_{1}\cdot \vec{v}_{2}-i0_+)}\,,
\label{}\end{equation}
we encounter a one-loop triangle integral with respect to $k_{1}$ in which the momentum $\vec{\ell}-\vec{k}_{2 }$ does not satisfy the delta-function condition, $i.e.$, $(\vec{\ell}-\vec{k}_{2})\cdot v_{\alpha} \neq 0$. 
Thus, we would require a new, what we can call off-shell version of the triangle 
integral. We avoid it by evaluating the triangle 
integral subsequently.

Finally, let us consider $f^{\rm I}_{3}$. As before, we can reorganize 
the integrand as a double triangle integral. Now it is not possible to 
use the one-loop results due to the absence of a corresponding off-shell 
formula, just as described above. Instead, we employ the trick described below 
that replaces the linear propagators with delta functions. We first 
enlarge the momentum integrals to three variables $\vec{k}_i $, for $ i 
= 1, 2, 3$ by means of a delta-function constraint:
\begin{equation}
\begin{aligned}
   f^{\rm I,\pm}_{3}
   =
   \int_{\vec{k}_{1},\vec{k}_{2},\vec{k}_{3}}
\frac{\hat{\delta}(\vec{k}_{1}+\vec{k}_{2}+\vec{k}_{3}-\vec{\ell})}{|\vec{k}_{1}|^{2} 
|\vec{k}_{2}|^{2} |\vec{k}_{3}|^{2}
   (\vec{k}_{1}\cdot \vec{v}_{2}-i0_+)(\vec{k}_{2}\cdot 
\vec{v}_{2}\mp i0_+)}\,.
\end{aligned}\label{fI3}
\end{equation}
We now need to be careful with the signs of $i\epsilon$-terms.
Let us define $X^{\pm}_{1,2} \equiv \vec{k}_{i}\cdot \vec{v}_{2} \pm i 0_+$. When $X^{\pm}_{\alpha}$ appear in the numerator, we can 
clearly ignore the sign in front of $i\epsilon$, so we can use just 
$X_{\alpha}$ itself. Using the on-shell condition on $\ell$, we have the 
following relation due to the $\delta$-function in \eqref{fI3}
\begin{equation}
   X^{\pm}_{1} + X^{\pm}_{2}+ X^{\pm}_{3} = 0\\.
\label{}\end{equation}
Then, following the explanations above, we note that we have the following identities:
\begin{equation}
\begin{aligned}
   &\frac{1}{X^{-}_{1}X^{-}_{2}} = \frac{X_{3}}{X^{-}_{1} X^{-}_{2} 
X^{+}_{3}} = \frac{-X_{1}-X_{2}}{X^{-}_{1} X^{-}_{2} X^{+}_{3}} = - 
\frac{1}{X_{2}^{-}X_{3}^{+}} - \frac{1}{X^{-}_{1}X_{3}^{+}}\,,
   \\
   &\frac{1}{X^{-}_{1} X^{+}_{2}} + \text{perm}(1,2,3) = 4\pi^{2} 
\delta(X_{1}) \delta(X_{2})\delta(X_{3}) \,.
\end{aligned}\label{}
\end{equation}
Since the loop momenta are just dummy variables, we can exploit the 
permutation symmetry among these loop momenta $k_{1} \leftrightarrow  
k_{2} \leftrightarrow k_{3}$ to get the symmetrized form
\begin{equation}
\begin{aligned}
   f^{\rm I,+}_{3}
   &=
   \int_{\vec{k}_{1},\vec{k}_{2},\vec{k}_{3}}
   \frac{1}{6} 
\frac{\hat{\delta}(\vec{k}_{1}+\vec{k}_{2}+\vec{k}_{3}-\vec{\ell})}{|\vec{k}_{1}|^{2} 
|\vec{k}_{2}|^{2} |\vec{k}_{3}|^{2}}
   \bigg(\frac{1}{
   X_{1}^{-} X_{2}^{-}}
   + \text{perm}(k_{1},k_{2},k_{3})
   \bigg)\,,
   \\
   &=
   \int_{\vec{k}_{1},\vec{k}_{2},\vec{k}_{3}}
   \frac{1}{3} 
\frac{\hat{\delta}(\vec{k}_{1}+\vec{k}_{2}+\vec{k}_{3}-\vec{\ell})}{|\vec{k}_{1}|^{2} 
|\vec{k}_{2}|^{2} |\vec{k}_{3}|^{2}}
   \bigg(- \frac{1}{
   X_{1}^{-} X_{2}^{+}}
   - \text{perm}(k_{1},k_{2},k_{3})
   \bigg)\,,
   \\
   &=
   -\frac{1}{3} \int_{\vec{z}} e^{-i \vec{\ell}\cdot \vec{z}}
   \int_{\vec{k}_{1}} \frac{\hat{\delta}(\vec{k}_{1}\cdot \vec{v}_{2}) 
e^{i \vec{k}_{1}\cdot z}}{|\vec{k}_{1}|^{2}}
   \int_{\vec{k}_{2}} \frac{\hat{\delta}(\vec{k}_{2}\cdot \vec{v}_{2}) 
e^{i \vec{k}_{2}\cdot z}}{|\vec{k}_{2}|^{2}}
   \int_{\vec{k}_{3}}
   \frac{e^{i \vec{k}_{3}\cdot z}}{|\vec{k}_{3}|^{2}} \,,
\end{aligned}\label{}
\end{equation}
where we have introduced a Fourier-integral representation of the 
delta-function.
The second equality above implies $f^{\rm I,+}_{3} =-2 f^{\rm 
I,-}_{3}$\,. Note how the integrations have factorized due to the 
introduction of the spurious new integration variable $\vec{k}_3$.

The $k_{1}$ and $k_{2}$ integrals can now be evaluated as
\begin{equation}
\begin{aligned}
   \int_{\vec{k}_{1}} \frac{\hat{\delta}(\sqrt{\gamma^{2}-1}k^{z}_{1}) 
e^{i \vec{k}_{1}\cdot \vec{z}_{\perp}}}{|\vec{k}_{1}|^{2}}
   =
   \frac{1}{\sqrt{\gamma^{2}-1}}
   \int_{\vec{k}^{^{\perp}}_{1}} \frac{e^{i \vec{k}^{\perp}_{1}\cdot 
\vec{z}}}{|\vec{k}^{\perp}_{1}|^{2}}
   = \frac{\Gamma(-\epsilon)}{4\pi^{1-\epsilon} \sqrt{\gamma^{2}-1}} 
|\vec{z}_{\perp}|^{2\epsilon}\,,
\end{aligned}\label{}
\end{equation}
while the $k_{3}$ integral is trivial,
\begin{equation}
   \int_{\vec{k}_{3}} \frac{e^{i \vec{k}_{3}\cdot 
\vec{z}}}{|\vec{k}_{3}|^{2}}
   =
   \frac{\Gamma\big(\tfrac{1-2\epsilon}{2}\big) 
}{4\pi^{\frac{3-2\epsilon}{2}}}
   |\vec{z}|^{2\epsilon-1}\,.
\label{}\end{equation}
Combining these results, we can evaluate the last $\vec{z}$ integral and 
thus finally get $f^{\rm I,+}_{3}$,
\begin{equation}
\begin{aligned}
   f^{\rm I,+}_{3}
   &=
   - \frac{\Gamma\big(\tfrac{1-2\epsilon}{2}\big) 
\Gamma(-\epsilon)^{2}}{3\times64 \pi^{\frac{7}{2}-3\epsilon}(\gamma^{2}-1)}
   \int_{\vec{z}_{\perp},z_{z}} e^{-i \vec{\ell}_{\perp}\cdot 
\vec{z}_{\perp}}
   |\vec{z}_{\perp}|^{4\epsilon}|\vec{z}|^{2\epsilon-1}\,,
   \\
   &=
   - \frac{ \Gamma(-\epsilon)^{3}}{3\times64 
\pi^{3-3\epsilon}(\gamma^{2}-1)}
   \int_{\vec{z}_{\perp}} e^{-i \vec{\ell}_{\perp}\cdot \vec{z}_{\perp}}
   |\vec{z}_{\perp}|^{6\epsilon}\,,
   \\
   &=
   - \frac{ \Gamma(-\epsilon)^{3}\Gamma(1+2\epsilon)}{3\times 
(4\pi)^{2-2\epsilon}\Gamma(-3\epsilon)(\gamma^{2}-1)} 
|\vec{\ell}\,|^{-2-4\epsilon}\,.
\end{aligned}\label{f_I_3}
\end{equation}
%

\subsection{Family-II Integrals}\label{Sec:5.2}
We next consider the second family of integrals in which the 
$\delta$-functions in the integrand are given by 
$\hat{\delta}(k_{1}\cdot v_{1})$ and $\hat{\delta}(k_{2}\cdot v_{2})$. 
Using again LiteRed \cite{Lee:2013mka}, we derive the master integrals 
for each sector. There are a total of 16 master integrals in this 
family. We may classify them according to whether 
$n_{6}+n_{7}+n_{8}+n_{9}$ is even or odd.

The master integrals for the even sector, $\{f^{\rm II,e}_{1},\cdots, 
f^{\rm II,e}_{10}\}$, are
\begin{equation}
\begin{aligned}
   f^{\rm II,e}_{1} = J^{\rm II, \pm}_{0,1,1,0,1,0,0,1,1} \,, \qquad
   f^{\rm II,e}_{2} = J^{\rm II, \pm}_{1,1,0,1,1,0,0,1,1} \,, \\
   f^{\rm II,e}_{3} = J^{\rm II, \pm}_{1,1,1,0,0,0,0,1,1} \,, \qquad
   f^{\rm II,e}_{4} = J^{\rm II, \pm}_{2,1,1,0,0,0,0,1,1} \,, \\
   f^{\rm II,e}_{5} = J^{\rm II, \pm}_{1,1,2,0,0,0,0,1,1} \,, \qquad
   f^{\rm II,e}_{6} = J^{\rm II, \pm}_{1,1,1,1,1,0,0,1,1} \,, \\
   f^{\rm II,e}_{7} = J^{\rm II, \pm}_{1,1,2,1,1,0,0,1,1} \,, \qquad
   f^{\rm II,e}_{8} = J^{\rm II, \pm}_{1,1,1,0,0,1,1,1,1} \,, \\
   f^{\rm II,e}_{9} = J^{\rm II, \pm}_{0,1,1,0,1,2,0,1,1} \,, \qquad
   f^{\rm II,e}_{10} = J^{\rm II, \pm}_{1,1,0,1,1,1,1,1,1} \,.
\end{aligned}\label{master_Integrals_II_even}
\end{equation}
while for the odd sector, $\{f^{\rm II,o}_{1},\cdots, f^{\rm 
II,o}_{6}\}$, they are
\begin{equation}
\begin{aligned}
   f^{\rm II,o}_{1} = J^{\rm II, \pm}_{0,1,1,0,1,0,0,2,1} \,, \qquad
   f^{\rm II,o}_{2} = J^{\rm II, \pm}_{0,0,1,1,1,0,0,2,1} \,, \\
   f^{\rm II,o}_{3} = J^{\rm II, \pm}_{1,1,1,1,1,0,0,2,1} \,, \qquad
   f^{\rm II,o}_{4} = J^{\rm II, \pm}_{0,0,1,1,1,0,1,1,1} \,, \\
   f^{\rm II,o}_{5} = J^{\rm II, \pm}_{0,1,1,0,1,1,0,1,1} \,, \qquad
   f^{\rm II,o}_{6} = J^{\rm II, \pm}_{1,1,0,1,1,0,1,1,1} \,.
\end{aligned}\label{master_Integrals_II_odd}
\end{equation}
Note that $f^{\rm II,e}_{9}$ and $f^{\rm II,e}_{10}$ in the even sector 
and $f^{\rm II,o}_{5}$ and $f^{\rm II,o}_{6}$ do not arise in the loop 
integrand. So we will ignore them.

In the integral families of \eqref{integral_Families} we note again that 
when $n_{8} = 1$ and $n_{9}=1$ the numerators just correspond to delta 
functions. They can be handled using a method that we will describe 
below. On the other hand, when $n_{8}$ or $n_{9}$ are positive integers 
these terms represent derivatives of $\delta$-functions, as defined in 
eq.\eqref{def_delta_n}. These derivatives can be recast as derivatives 
with respect to the loop momenta,
\begin{equation}
   \hat{\delta}^{\ord{n+1}}(k \cdot v_{\alpha})
   =
   \frac{(-1)^{n}}{n!} \bigg( - v^{\mu}_{\alpha}\frac{\partial}{\partial 
k_{\mu}}\bigg)^{n} \hat{\delta}(k \cdot v_{\alpha})\,.
\label{}\end{equation}
Through application of integration by parts, we can subsequently 
eliminate these derivatives. Thus it suffices to evaluate only the 
configurations where $n_8 = n_9 = 1$, as all higher-order cases can be 
systematically reduced to this form through the integration by parts 
procedure.

Among the master integrals in family II, the non-trivial values of $n_8$ 
or $n_9$ arise only in the odd sector \eqref{master_Integrals_II_odd}, 
and these can be recast through
integrations by parts as follows:
\begin{equation}
\begin{aligned}
   f^{\rm II,o}_{1} &= 2J^{\rm II, \pm}_{0,1,2,0,1,0,-1,1,1} \,, &\qquad
   f^{\rm II,o}_{2} &= 2J^{\rm II, \pm}_{1,1,2,0,0,0,-1,1,1} \,, \\
   f^{\rm II,o}_{3} &= 2J^{\rm II, \pm}_{1,1,2,1,1,0,-1,1,1} \,, &\qquad
   f^{\rm II,o}_{4} &= J^{\rm II, \pm}_{0,0,1,1,1,0,1,1,1} \,, \\
   f^{\rm II,o}_{5} &= J^{\rm II, \pm}_{0,1,1,0,1,1,0,1,1} \,, &\qquad
   f^{\rm II,o}_{6} &= J^{\rm II, \pm}_{1,1,0,1,1,0,1,1,1} \,.
\end{aligned}\label{master_Integrals_II_odd2}
\end{equation}

Let us finally consider $n_{8} = 1$ and $n_{9}=1$ case, in which the 
numerator contains only ordinary delta-function constraints.
In the parametrization of $v_{\alpha}$ in eq. \eqref{v_parametrization}, 
the $\delta$--functions are reduced to

\begin{equation}
\begin{aligned}
   \hat{\delta}(k_{1}\cdot v_{1})
   & \to
   \hat{\delta} (k_{1}^{0})\,,
   \\
   \hat{\delta}(k_{2}\cdot v_{2})
   & \to
   \frac{1}{\gamma}\hat{\delta} (k_{2}^{0}- 
\sqrt{1-\gamma^{-2}}k_{2}^{z})\,.
\end{aligned}\label{}
\end{equation}
We can now integrate out $k^{0}_{i}$ by means of the $\delta$-functions, 
and thus parametrize the loop momenta as follows:
\begin{equation}
\begin{aligned}
   k_{1}^{\mu}
=
\begin{pmatrix}
0 \\ \vec{k}_{1}
\end{pmatrix}
=
\begin{pmatrix}
0 \\ \vec{k}^{\perp}_{1} \\ k^{z}_{1}
\end{pmatrix}
\,,
\qquad
k_{2}^{\mu}
=
\begin{pmatrix}
\sqrt{1-\gamma^{-2}} k^{z}_{2}
\\
\vec{k}_{2}
\end{pmatrix}
=
\begin{pmatrix}
\sqrt{1-\gamma^{-2}} k^{z}_{2}
\\
\vec{k}^{\perp}_{2} \\ k^{z}_{2}
\end{pmatrix} \,.
\end{aligned}\label{}
\end{equation}
Thus, after integration over $k_{1}^{0}\,, k_{2}^{0}$, the denominators 
in $J^{\rm II,\pm}_{n_{1},n_{2},n_{3},n_{4},n_{5}, n_{6},n_{7}}$ in eq. 
\eqref{integral_Families} become
\begin{equation}
\begin{aligned}
k_{1}^{2}
& \to
\big|\vec{k_{1}}\big|^{2}\,,
   \\
   k_{2}^{2}
   &\to
   \big|\vec{k}^{\perp}_{2}\big|^{2} + \gamma^{-2}\big(k^{z}_{2}\big)^{2}\,,
   \\
   (\ell-k_{1}-k_{2})^{2}
   &\to
   |\vec{\ell}-\vec{k}_{1}|^{2} -2 (\vec{\ell}-\vec{k}_{1})\cdot 
\vec{k}_{2} +|\vec{k}^{\perp}_{2}|^{2} + \gamma^{-2}\big(k^{z}_{2}\big)^{2}\,,
   \\
   (\ell-k_{1})^{2}
   &\to
   \big|\vec{\ell}-\vec{k}_{1}\big|^{2}\,,
   \\
   (\ell-k_{2})^{2}
   &\to
   \big|\vec{\ell}^{\perp}-\vec{k}^{\perp}_{2}\big|^{2}
   +\gamma^{-2}\big(k^{z}_{2}\big)^{2}\,,
   \\
   k_{1}\cdot v_{2}
   &\to
   \sqrt{\gamma^{2}-1}\, k_{1}^{z}\,,
   \\
   k_{2}\cdot v_{1}
   &\to
   -\gamma^{-1} \sqrt{\gamma^{2}-1}\, k_{2}^{z}\,,
\end{aligned}\label{}
\end{equation}
If we rescale $k^{z}_{2} \to \gamma k^{z}_{2}$ the denominators 
involving $k_{2}$ simplify considerably due to
\begin{equation}
\begin{aligned}
&\mathrm{d}^{D-1} k_{2}\to \gamma \mathrm{d}^{D-1} k_{2}\,,
\\
&\big|\vec{k}^{\perp}_{2}\big|^{2} +
\gamma^{-2}\big(k^{z}_{2}\big)^{2}
\to
\big|\vec{k}_{2}\big|^{2}\,,
\\
&|\vec{\ell}-\vec{k}_{1}|^{2}
-2 (\ell-\vec{k}_{1})\cdot \vec{k}_{2} +|\vec{k}^{\perp}_{2}|^{2}
+ \gamma^{-2}\big(k^{z}_{2}\big)^{2}
\to
\big|\vec{\ell}-\vec{k}_{1}-\vec{k}_{2}\big|^{2}
+2x (\vec{q}\cdot \vec{k}_{1}) (\vec{q}\cdot \vec{k}_{2})
\,,
\\
&\big|\vec{\ell}^{\perp}-\vec{k}^{\perp}_{2}\big|^{2}
+\gamma^{-2}\big(k^{z}_{2}\big)^{2}
\to
\big|\vec{\ell}-\vec{k}_{2}\big|^{2}\,,
\\
&- \gamma^{-1} \sqrt{\gamma^{2}-1}\, k_{2}^{z}
to
- \sqrt{\gamma^{2}-1}\, k_{2}^{z}
=
-  \sqrt{\gamma^{2}-1}\, (\vec{q}\cdot \vec{k}_{2})\,.
\end{aligned}
\end{equation}
where $x = \gamma-1 >0$ and $\vec{q} = (\vec{0}^{\perp},1)$.

In this way, the family-II integrals reduce, after using $\vec{q}\cdot\vec{\ell} =0$ to
\begin{equation}
   J^{\rm II,\pm}_{n_{1},n_{2},n_{3},n_{4},n_{5},n_{6},n_{7}}
   =
   \frac{1}{(\gamma^{2}-1)^{\frac{n_{6}+n_{7}}{2}}}
   \int_{\vec{k}_{1},\vec{k}_{2}}
   \frac{1}{\mathcal{D}_{1}^{n_{1}} \mathcal{D}_{2}^{n_{2}} 
\mathcal{D}_{3}^{n_{3}} \mathcal{D}_{4}^{n_{4}} \mathcal{D}_{5}^{n_{5}} 
\mathcal{D}_{6}^{n_{6}} \mathcal{D}_{7}^{n_{7}}}
\label{familyII}\end{equation}
where
\begin{equation}
  \int_{\vec{k}} = \int \frac{d^{D-1} \vec{k}}{(2\pi)^{D-1}}\,,
\label{}\end{equation}
and
\begin{equation}
\begin{aligned}
\mathcal{D}_{1} &= \big|\vec{k}_{1}\big|^{2}\,,\quad
\mathcal{D}_{2} = \big|\vec{k}_{2}\big|^{2}\,,
\\
\mathcal{D}_{3} &= \big|\vec{\ell}-\vec{k}_{1}-\vec{k}_{2}\big|^{2}+ 2x (\vec{q}\cdot \vec{k}_{1}) (\vec{q}\cdot \vec{k}_{2})-\text{sgn}(\ell^{0}-k_{1}^{0}-k_{2}^{0})i0_+ \,,
\\
\mathcal{D}_{4} &= \big|\vec{\ell} - \vec{k}_{1}\big|^{2}\,,\quad
\mathcal{D}_{5} = \big|\vec{\ell} - \vec{k}_{2}\big|^{2}\,,\quad
\mathcal{D}_{6} = \vec{k}_{1} \cdot \vec{q}-i0_+ \,,\quad
\mathcal{D}_{7} = - \vec{k}_{2} \cdot  \vec{q} \mp i0_+ \,.
\end{aligned}\label{propagator_set_without_expansion}
\end{equation}
Note that any $x$-dependence arises in the $\mathcal{D}_{3}$ propagator only. We may thus consider a small-$x$ expansion of $1/(\mathcal{D}_{3})^{n_{3}}$ according to the method of regions, and there are only two possibilities in this case: potential and radiation regions. To see this structure explicitly, we make a shift $\vec{k}_{1} \to \vec{\ell} -  \vec{k}_{1} - \vec{k}_{2}$ 
where
\begin{equation}
\begin{aligned}
  \mathcal{D}_{1} &= \big|\vec{\ell} - \vec{k}_{1} - \vec{k}_{2}\big|^{2}\,,\qquad
  \mathcal{D}_{2} = \big|\vec{k}_{2}\big|^{2}\,,
  \\
  \mathcal{D}_{3} &= \big|\vec{k}_{1}\big|^{2}-2x(\vec{q}\cdot \vec{k}_{2})^{2}- 2 x (\vec{q}\cdot \vec{k}_{1}) (\vec{q}\cdot \vec{k}_{2})-\text{sgn}(\vec{q}\cdot \vec{k}_{2})i0_+ \,,
  \\
  \mathcal{D}_{4} &= \big|\vec{k}_{1}+\vec{k}_{2}\big|^{2}\,,\qquad
  \mathcal{D}_{5} = \big|\vec{\ell} - \vec{k}_{2}\big|^{2}\,,\qquad
  \mathcal{D}_{6} = - (\vec{k}_{1}+\vec{k}_{2}) \cdot \vec{q}-i0_+ \,
  \\
  \mathcal{D}_{7} &= - \vec{k}_{2} \cdot  \vec{q} \mp i0_+ \,.
\end{aligned}\label{propagator_set_without_expansion_shifted}
\end{equation}
The only non-trivial scaling of the loop momenta under the small-$x$ expansion is as follows:
\begin{equation}
\begin{aligned}
\text{potential} : \quad &( \vec{k}_{1}\rightarrow \vec{k}_{1} , \vec{k}_{2}\rightarrow \vec{k}_{2})\,,
\\
\text{radiation} : \quad &( \vec{k}_{1}\rightarrow \sqrt{2x}\vec{k}_{1} , \vec{k}_{2}\rightarrow \vec{k}_{2})\,.
\end{aligned}\label{mor scaling}
\end{equation}
Since the rescaling of the potential region is trivial, we will use \eqref{propagator_set_without_expansion}. On the other hand, we will use the shifted propagator given by \eqref{propagator_set_without_expansion_shifted} with the rescaling \eqref{mor scaling} for the radiation region. The master integrals are the same for each region, and the family II case is represented by
\begin{equation}
\begin{aligned}
  f^{\rm II, e}_{i} &= f^{\rm II,P,e}_{i} + f^{\rm II,R,e}_{i}\,, \qquad  i = 1,2, \cdots , 10\,,
  \\
  f^{\rm II, o}_{i} &= f^{\rm II,P,o}_{i} + f^{\rm II,R,o}_{i}\,, \qquad  i = 1,2, \cdots , 6 \,.
\end{aligned}\label{}
\end{equation}
Here $f^{\rm II,P,e/o}_{i}$ and $f^{\rm II,R,e/o}_{i}$ are the master integrals in the potential and radiation regions. 

\subsubsection{The potential region}\label{Sec:Potential Region}
Let us consider the potential region by expanding the small-$x$ expansion directly
\begin{equation}
   \frac{1}{\mathcal{D}^{n_{3}}_{3}}
   = \sum_{m=0}^{\infty}
     \frac{2^{m} \Gamma(n_{3}+m)}{\Gamma(n_{3}) \Gamma(m+1)}
     \frac{\big(\vec{k_{1}}\cdot \vec{q}\big)^{m} \big(\vec{k_{2}}\cdot 
\vec{q}\big)^{m}}{\big|\vec{\ell}-\vec{k}_{1}-\vec{k}_{2}\big|^{2n_{3}+2m}}(-x)^{m}\,.
\label{}\end{equation}
It is thus convenient to define a new integral family $J^{\rm \overline{II}}_{n_{1},n_{2},n_{3},n_{4},n_{5},n_{6},n_{7}}$ in which $\mathcal{D}_{3}$ is replaced by $\tilde{\mathcal{D}}_{3} =  \big|\vec{\ell}-\vec{k}_{1}-\vec{k}_{2}\big|^{2}$,
\begin{equation}
   J^{\rm \overline{II}}_{n_{1},n_{2},n_{3},n_{4},n_{5},n_{6},n_{7}}
   =
   \int_{\vec{k}_{1},\vec{k}_{2}} 
   \frac{(-1)^{n_{7}}}{\mathcal{D}_{1}^{n_{1}} \mathcal{D}_{2}^{n_{2}} 
\tilde{\mathcal{D}}_{3}^{n_{3}} \mathcal{D}_{4}^{n_{4}} 
\mathcal{D}_{5}^{n_{5}} \mathcal{D}_{6}^{n_{6}} \mathcal{D}_{7}^{n_{7}}}\,,
\label{}\end{equation}
where the reduced propagators are
\begin{equation}
\begin{aligned}
   \mathcal{D}_{1} &= \big|\vec{k}_{1}\big|^{2}\,,\quad
   \mathcal{D}_{2} = \big|\vec{k}_{2}\big|^{2}\,,\quad
   \tilde{\mathcal{D}}_{3} = 
\big|\vec{\ell}-\vec{k}_{1}-\vec{k}_{2}\big|^{2}\,,
   \\
   \mathcal{D}_{4} &= \big|\vec{\ell} - \vec{k}_{1}\big|^{2}\,,\quad
   \mathcal{D}_{5} = \big|\vec{\ell} - \vec{k}_{2}\big|^{2}\,,\quad
   \mathcal{D}_{6} = \vec{k}_{1} \cdot \vec{q}-i0_+ \,,\quad
   \mathcal{D}_{7} = \vec{k}_{2} \cdot  \vec{q} + i0_+ \,.
\end{aligned}\label{integralFamilybarII}
\end{equation}
In this way, the family-II integrals in the potential region \eqref{familyII} are represented by 
expansions in terms of the $J_{n_{1},\cdots, n_{7}}^{\rm \overline{II}}$ 
integrals,
\begin{equation}
\begin{aligned}
   J^{\rm II,P}_{n_{1},\cdots, n_{7}}
  =
\frac{1}{(\gamma^{2}-1)^{\frac{n_{6}+n_{7}}{2}}} \sum_{m=0}^{\infty} \frac{2^{m} \Gamma(n_{3}+m) x^{m}}{\Gamma(n_{3}) \Gamma(m+1)}J^{\rm \overline{II},P}_{n_{1},n_{2},n_{3}+m,n_{4},n_{5},n_{6}-m,n_{7}-m}\,.
\end{aligned}\label{expansionFamilyII}
\end{equation}
Using this relation, we may evaluate the master integrals in 
\eqref{master_Integrals_II_even} and \eqref{master_Integrals_II_odd}.

There are a total of 8 master integrals in this integral family, which 
can also be classified according to whether $n_6+n_7$ is even or odd, 
but not all of the master integrals contribute to the evaluations of the 
master integrals for the family-II eq. \eqref{master_Integrals_II_even} 
and eq. \eqref{master_Integrals_II_odd}. The relevant master integrals 
in the even sector are
\begin{equation}
\begin{aligned}
   f^{\rm \overline{II},P,e}_{1} = J^{\rm \overline{II},P}_{1,1,1,0,0,0,0}\,,\qquad
   f^{\rm \overline{II},P,e}_{2} = J^{\rm \overline{II},P}_{1,1,0,1,1,0,0}\,,\qquad
   f^{\rm \overline{II},P,e}_{3} = J^{\rm \overline{II},P}_{0,0,1,1,1,1,1}\,,
\end{aligned}\label{}
\end{equation}
and those in the odd sector are
\begin{equation}
\begin{aligned}
   f^{\rm \overline{II},P,o}_{1} = J^{\rm \overline{II},P}_{0,0,1,1,1,0,1}\,,\qquad
   f^{\rm \overline{II},P,o}_{2} = J^{\rm \overline{II},P}_{0,1,1,0,1,1,0}\,,\qquad
   f^{\rm \overline{II},P,o}_{3} = J^{\rm \overline{II},P}_{1,1,0,1,1,0,1}\,,
\end{aligned}\label{}
\end{equation}
One remarkable property of these master integrals is that they can be 
evaluated by iterations of one-loop integrals. We refer the reader to the
Appendix for a detailed derivation of this.

We now turn to the evaluation of the master integrals 
\eqref{master_Integrals_II_even} and \eqref{master_Integrals_II_odd} 
using the $J_{n_{1},\cdots, n_{7}}^{\rm \overline{II}}$. We consider the 
even and odd sectors separately.

The even sector master integrals are
{\small%
\begin{equation}
\begin{aligned}
  f^{\rm II,P,e}_{1} &= 0\,.,
  \\
  f^{\rm II,P,e}_{2}
  &=
  \frac{(4 \pi )^{2 \epsilon -3} \left(\Gamma(\frac{1}{2}-\epsilon) \right)^4 \left(\Gamma (\epsilon +\frac{1}{2})\right)^2}{\Gamma (1-2\epsilon )^2}  |\vec{\ell}\,|^{-4 \epsilon -2}\,,
  \\
  f^{\rm II,P,e}_{3}
  &=
  (4\pi)^{2\epsilon-2}e^{-2\gamma_{\rm E}\epsilon}
  \bigg[
  \frac{1}{4\epsilon\sqrt{\gamma^{2}-1}}\cosh^{-1}\gamma + \mathcal{O}(\epsilon^{0})
  \bigg] |\vec{\ell}\, |^{-4\epsilon}\,,
  \\
  f^{\rm II,P,e}_{4}
&= -(4\pi)^{2\epsilon-2}e^{-2\gamma_{\rm E}\epsilon}
\bigg[
- \frac{\gamma}{2}
+ \epsilon\left(\gamma -\sqrt{\gamma ^2-1} \cosh^{-1}\gamma \right)
+ \mathcal{O}(\epsilon^{2})
\bigg] |\vec{\ell}\, |^{-4\epsilon}\,,
  \\
  f^{\rm II,P,e}_{5}
  &=
  -(4\pi)^{2\epsilon-2}e^{-2\gamma_{\rm E}\epsilon} \bigg[\frac{1}{2\sqrt{\gamma^{2}-1}} \cosh^{-1}\gamma+ \mathcal{O}(\epsilon^{1})\bigg] |\vec{\ell}\, |^{-4\epsilon}
  \\
  f^{\rm II,P,e}_{6}
  &=
  \frac{(4 \pi )^{2 \epsilon -2} e^{-2 \gamma_{\rm E} \epsilon}}{|\vec{\ell}\,|^{4+4\epsilon}} \left[ \frac{\cosh^{-1}\!\gamma}{2 \epsilon \sqrt{\gamma^2-1}}
  + \frac{(\cosh^{-1}\!\gamma)^2 +\text{Li}_2\big(2 \gamma 
  (\sqrt{\gamma ^2-1}{-}\gamma)+2\big)}{2 \sqrt{\gamma ^2-1}}+\mathcal{O}(\epsilon)\right]\,,
  \\
  f^{\rm II,P,e}_{7}
  &=
  \frac{(4 \pi )^{2 \epsilon -2} e^{-2 \gamma_{\rm E} \epsilon}}{|\vec{\ell}\,|^{4 \epsilon +6}} \Bigg[ \frac{\gamma (5 \epsilon +1)}{2\epsilon \left(\gamma^2-1\right)} - \frac{\left(2 \gamma ^2 \epsilon +\epsilon +1\right) \cosh^{-1}\gamma}{2 \left(\gamma ^2-1\right)^{3/2} \epsilon }
  \\&\qquad\qquad\qquad\qquad
  - \frac{ \big(\cosh^{-1}\gamma \big)^2 + \text{Li}_2\big(2 \gamma 
  (\sqrt{\gamma ^2-1}{-}\gamma)+2\big)}{2 \left(\gamma ^2-1\right)^{3/2} }+\mathcal{O}(\epsilon^{2})\bigg]\,,
  \\
  f^{\rm II,P,e}_{8}
  &=
  \frac{(4 \pi )^{2 \epsilon -2} e^{-2 \gamma_{\rm E} \epsilon}}{|\vec{\ell}|^{2+4 \epsilon}} \bigg[ - \frac{1}{2\epsilon^{2}(\gamma^{2}-1)} + \mathcal{O}(\epsilon^{0})\bigg]\,.
\end{aligned}\label{}
\end{equation}
}
Next the odd sector is 
\begin{equation}
\begin{aligned}
  f^{\rm II,P,o}_{1} &= 0\,,
  \\  
  f^{\rm II,P,o}_{2} &= 0\,,
  \\
  f^{\rm II,P,o}_{3} &= 0\,.
  \\
  f^{\rm II,P,o}_{4}
  &=
  \frac{i}{\sqrt{\gamma^{2}-1}} \frac{4^{2 \epsilon -3} \pi ^{2\epsilon -\frac{5}{2}}
   \Gamma \left(\frac{1}{2}-2 \epsilon \right) \Gamma \left(\frac{1}{2}-\epsilon \right)^2 \Gamma (-\epsilon ) \Gamma \left(2 \epsilon +\frac{1}{2}\right)}{\Gamma \left(\frac{1}{2}-3 \epsilon \right) \Gamma (1-2 \epsilon )~|\vec{\ell}\,|^{1+4 \epsilon}} \,.
\end{aligned}\label{}
\end{equation}
These results are all consistent with \cite{Jakobsen:2022fcj}. Details of the derivation of the results are presented in Appendix \ref{App:B}.

\subsubsection{Radiation region}
The radiation region is defined by the small-$x$ expansion of \eqref{propagator_set_without_expansion_shifted} after the rescaling in \eqref{mor scaling}, which is given by 
\begin{equation}
\begin{aligned}
  J^{\rm II,R,\pm}_{n_{1},n_{2},n_{3},n_{4},n_{5},n_{6},n_{7}}
  =
  \frac{(2x)^{\frac{D-1}{2}}}{(\gamma^{2}-1)^{\frac{n_{6}+n_{7}}{2}}}
  \int_{\vec{k}_{1},\vec{k}_{2}} 
  \frac{1}{\mathcal{D}_{1}^{n_{1}} \mathcal{D}_{2}^{n_{2}} \mathcal{D}_{3}^{n_{3}} \mathcal{D}_{4}^{n_{4}} \mathcal{D}_{5}^{n_{5}} \mathcal{D}_{6}^{n_{6}} \mathcal{D}_{7}^{n_{7}}}\,,
\end{aligned}\label{}
\end{equation}
where 
\begin{equation}
\begin{aligned}
  \mathcal{D}_{1} &= \big|\vec{\ell} - \vec{k}_{2}\big|^{2}-2\sqrt{2x}\vec{k}_{1}\cdot(\vec{\ell}-\vec{k}_{2})+2x\big|\vec{k}_{1}\big|^{2}\,,
  \qquad
  \mathcal{D}_{2} = \big|\vec{k}_{2}\big|^{2}\,,
  \\
  \mathcal{D}_{3} &= 2x\big|\vec{k}_{1}\big|^{2}-2x(\vec{q}\cdot \vec{k}_{2})^{2} - 2 x\sqrt{2x} (\vec{q}\cdot \vec{k}_{1}) (\vec{q}\cdot \vec{k}_{2})-\text{sgn}(\vec{q}\cdot \vec{k}_{2})i 0_{+} \,,
  \\
  \mathcal{D}_{4} &= \big|\vec{k}_{2}\big|^{2}+2\sqrt{2x}(\vec{k}_{1}\cdot\vec{k}_{2})+2x\big|\vec{k}_{1}\big|^{2}\,,
  \qquad
  \mathcal{D}_{5} = \big|\vec{\ell} - \vec{k}_{2}\big|^{2}\,,
  \qquad
  \\
  \mathcal{D}_{6} &= - \vec{k}_{2} \cdot \vec{q}-\sqrt{2x}(\vec{k}_{1} \cdot \vec{q})-i 0_{+}\,,
  \qquad
  \mathcal{D}_{7\mp} = - \vec{k}_{2} \cdot  \vec{q} \mp i 0_{+}\,.
\end{aligned}\label{}
\end{equation}
The small $x$-expansion of the propagators containing $x$ are given by
\begin{equation}
\begin{aligned}
  \frac{1}{\mathcal{D}^{n_{1}}_{1}} 
  &= \sum_{m=0}^{\infty} 
    \frac{(2x)^{\frac{m}{2}}\Gamma(n_{1}+m)}{\Gamma(n_{1}) \Gamma(m+1)}
    \frac{(2\vec{k}_{1}\cdot(\vec{\ell}-\vec{k}_{2})-\sqrt{2x}\big|\vec{k}_{1}\big|^{2})^{m}}{\mathcal{D}_{5}^{n_{1}+m}}\,,
    \\
      \frac{1}{\mathcal{D}^{n_{3}}_{3}} 
  &= \sum_{n=0}^{\infty} 
    \frac{(2x)^{\frac{n}{2}-n_{3}} \Gamma(n_{3}+n)}{\Gamma(n_{3}) \Gamma(n+1)}
    \frac{\big(\vec{k_{1}}\cdot \vec{q}\big)^{n} \big(\vec{k_{2}}\cdot \vec{q}\big)^{n}}{\tilde{\mathcal{D}}_{3} ^{n_{3}+n}}\,,
    \\
  \frac{1}{\mathcal{D}^{n_{4}}_{4}} 
  &= \sum_{r=0}^{\infty} 
    \frac{(2x)^{\frac{r}{2}} \Gamma(n_{4}+r)}{\Gamma(n_{4}) \Gamma(r+1)}
    \frac{(-2(\vec{k}_{1}\cdot\vec{k}_{2})-\sqrt{2x}\big|\vec{k}_{1}\big|^{2})^{r}}{\mathcal{D}_{2}^{n_{4}+r}}\,,
    \\
      \frac{1}{\mathcal{D}^{n_{6}}_{6}} 
  &= \sum_{s=0}^{\infty} 
    \frac{(2x)^{\frac{s}{2}} \Gamma(n_{6}+s)}{\Gamma(n_{6}) \Gamma(s+1)}
    \frac{\big(\vec{k_{1}}\cdot \vec{q}\big)^{s}}{\mathcal{D}_{7-}^{n_{6}+s}}\,,
\end{aligned}\label{}
\end{equation}
where
\begin{equation}
  \tilde{\mathcal{D}}_{3} 
  = 
  \big|\vec{k}_{1}\big|^{2}-(\vec{q}\cdot \vec{k}_{2})^{2} -\text{sgn}(\vec{q}\cdot \vec{k}_{2})i0_{+}\,.
\label{}\end{equation}

Substituting these expansions of the propagators into the above expression, we obtain
\begin{equation}
\begin{aligned}
  &J^{\rm II,R,\pm}_{n_{1},n_{2},n_{3},n_{4},n_{5},n_{6},n_{7}}
  \\
  &=\!\sum_{m,n,r,s}\!
    \frac{(2x)^{\frac{1}{2}(d+m+n+r+s)-n_{3}}\Gamma(n_{1}{+}m)\Gamma(n_{3}{+}n)\Gamma(n_{4}{+}r)\Gamma(n_{6}{+}s)}{(\gamma^{2}-1)^{\frac{n_{6}+n_{7}}{2}}\Gamma(n_{1})\Gamma(n_{3})\Gamma(n_{4})\Gamma(n_{6}) \Gamma(m{+}1)\Gamma(n{+}1)\Gamma(r{+}1)\Gamma(s{+}1)}
  \\
  &\qquad
  \times \int_{\vec{k}_{1},\vec{k}_{2}}
  \frac{C^{\rm II,R,\pm}_{m,n,r,s}}{\mathcal{D}_{2}^{n_{2}+n_{4}+r} \tilde{\mathcal{D}}_{3}^{n_{3}+n} \mathcal{D}_{5}^{n_{1}+n_{5}+m} \mathcal{D}_{7-}^{n_{6}+s} \mathcal{D}_{7\mp}^{n_{7}}}\,,
\end{aligned}\label{Radiation_region_reduction}
\end{equation}
where
\begin{equation}
    C^{\rm II,R,\pm}_{m,n,r,s}
  =
  (-1)^{m+r}\big(\vec{k}_{1} \cdot \vec{q}\, \big)^{n+s}
  \big(\vec{k}_{2} \cdot \vec{q}\, \big)^{n} 
  \Big(2\vec{k}_{1}\cdot\big(\vec{k}_{2}-\vec{\ell}\ \big)+\sqrt{2x}\big|\vec{k}_{1}\big|^{2}\Big)^{m} 
  \Big(2\vec{k}_{1} \cdot\vec{k}_{2}+\sqrt{2x}\big|\vec{k}_{1}\big|^{2}\Big)^{r}\,.
\label{}\end{equation}

Interestingly, among the four propagators, $\mathcal{D}_{2}$, $\tilde{\mathcal{D}}_{3}$, $\mathcal{D}_{5}$ and  $\mathcal{D}_{7\mp}^{n_{7}}$, only $\tilde{\mathcal{D}}_{3}$ involves $\vec{k}_{1}$, thus we can evaluate $\vec{k}_{1}$ and $\vec{k}_{2}$ integrals successively. After the IBP reduction, the $\vec{k}_{1}$ integration reduces to
\begin{equation}
\begin{aligned}
  \int_{\vec{k}_{1}}
  \frac{1}{\tilde{\mathcal{D}}_{3}}
  &=
  (-1)^{d-1}(4\pi)^{-\frac{d}{2}} e^{i\pi \frac{d}{2}} \big( \vec{k}_{2}\cdot \vec{q}\ \big)^{d-2} \Gamma\big(\tfrac{2-d}{2}\big)\,,
\end{aligned}\label{}
\end{equation}
where $d = D-1 = 3-2\epsilon$. Then the remaining the two-loop integral takes the following form
\begin{equation}
\begin{aligned}
  &\int_{\vec{k}_{2}}
  \frac{1}{ \big|\vec{k}_{2}\big|^{2n'_{1}} \big|\vec{\ell} - \vec{k}_{2}\big|^{2n'_{2}} \big(- \vec{k}_{2} \cdot \vec{q}-i0_{+}\big)^{n'_{3}} \big(- \vec{k}_{2} \cdot \vec{q}+i0_{+}\big)^{n'_{4}}}\
  \\
  &= 
  i^{n'_3 - n'_4}  (4\pi)^{\frac{2-D}{2}}
  \frac{
    \Gamma\big(\frac{d - 2n'_1 - n'_3 - n'_4}{2}\big)
    \Gamma\big(\frac{d - 2n'_2 - n'_3 - n'_4}{2}\big)
    \Gamma\big(\frac{-d + 2n'_1 + 2n'_2 + n'_3 + n'_4}{2}\big)}{
    2\Gamma(n_1)\Gamma(n_2) \Gamma(d - n'_1 - n'_2 - n'_3 - n'_4) \Gamma\big(\frac{1 + n'_3 + n'_4}{2}\big)
}
  \\&\quad\times
  \frac{\cos \big(\frac{\pi}{2}(n'_3 - n'_4)\big)}{\cos(\frac{\pi}{2}(n'_3 + n'_4))}
  \big|\vec{\ell}\big|^{d - 2n'_1 - 2n'_2 - n'_3 - n'_4}\,.
\end{aligned}\label{}
\end{equation}

In this way, we can evaluate the master integrals in the potential region using \eqref{Radiation_region_reduction} for a given $m,n,r,s$ explicitly and organize the result as an expansion in powers of $x$
\begin{equation}
\begin{aligned}
  &f^{\rm II, R, e/o}_{i}
  = x^{\alpha} \sum_{n=0}^{\infty} C^{\rm II, R, e/o}_{n, i}\, x^{n/2}
\end{aligned}\label{integral_in_x}
\end{equation}
As the potential region, we can perform the resummation and the master integrals are given by as follows:
\begin{equation}
\begin{aligned}
  \mathcal{C}^{-\epsilon}f^{\rm II,R,e}_{1}
  &=
  -\frac{\sqrt{\gamma^2 - 1}}{64\pi^2 \big|\vec{\ell}\,\big|^{4\epsilon}} \Big[(1 + 6\epsilon) - \epsilon \log\left(\tfrac{1+\gamma}{2}\right)\Big]
  + \mathcal{O}(\epsilon^2) \,,
  \\
  \mathcal{C}^{-\epsilon}f^{\rm II,R,e}_{2}
  &=0 \,,
  \\
  \mathcal{C}^{-\epsilon}f^{\rm II,R,e}_{3}
  &=\mathcal{O}(\epsilon^{0})\,,
  \\
  \mathcal{C}^{-\epsilon}f^{\rm II,R,e}_{4}
  &=
  \frac{\sqrt{\gamma^2 - 1}}{64\pi^2 |\vec{\ell}|^{2 +4\epsilon}}
  \bigg[ (1 - 2\epsilon)
    + \frac{4 \gamma \epsilon \cosh^{-1}(\gamma)}{\sqrt{\gamma^2-1}}
    - \epsilon \log\left(\tfrac{1+\gamma}{2}\right)
  \bigg]
  + \mathcal{O}(\epsilon^2) \,,
  \\
  \mathcal{C}^{-\epsilon}f^{\rm II,R,e}_{5}
  &=
  -\frac{1}{64\pi^2 \sqrt{\gamma^2 - 1}}\bigg[\frac{1}{\epsilon} -\log\left(\tfrac{1+\gamma}{2}\right)\bigg] |\vec{\ell}|^{-2 - 4\epsilon}
  + \mathcal{O}(\epsilon) \,,
\\
  \mathcal{C}^{-\epsilon}f^{\rm II,R,e}_{6}
  &=
  \frac{(\cosh^{-1}\gamma)^2}{32\pi^2 \sqrt{\gamma^2 - 1}|\vec{\ell}|^{4 +4\epsilon}}
  + \mathcal{O}(\epsilon) \,,
\\
  \mathcal{C}^{-\epsilon}f^{\rm II,R,e}_{7}
  &=
  \frac{-|\vec{\ell}|^{-6 - 4\epsilon}}{32\pi^2 \sqrt{\gamma^2 - 1}} 
  \bigg[
   \frac{2 {+} 9\epsilon}{\epsilon}
  {+} \frac{2 \gamma \cosh^{-1}\gamma}{\sqrt{\gamma^{2}-1}}
  {-} \frac{(\cosh^{-1}\gamma)^2}{(\gamma^2 - 1)}
  {-} 2\log\!\left(\tfrac{1 + \gamma}{2}\right)
  \bigg] 
  + \mathcal{O}(\epsilon) \,,
\\
  \mathcal{C}^{-\epsilon}f^{\rm II,R,e}_{8,+}
  &=
  -\frac{\cosh^{-1}\gamma}{32\pi^2 (\gamma^2 - 1)\epsilon} |\vec{\ell}|^{-2 - 4\epsilon}
  + \mathcal{O}(\epsilon^{0}) \,,
  \\
  f^{\rm II,R,e}_{8,-}
  &=
  f^{\rm II,R,e}_{8,+} \,,
\end{aligned}\label{}
\end{equation}
where the overall factor $\mathcal{C}$ is
\begin{equation}
\begin{aligned}
\mathcal{C}&=\frac{2\pi^2 e^{-2\gamma_E}}{\gamma - 1}\,.
\end{aligned}\label{}
\end{equation}
Next the odd sector is 
\begin{equation}
\begin{aligned}
  \mathcal{C}^{-\epsilon}f^{\rm II,R,o}_{1}
  &=
  -\frac{i (1+3\epsilon \log 4)-i \epsilon \log\left(\frac{1 + \gamma}{2}\right)}{32\pi|\vec{\ell}|^{1 +4\epsilon}} + \mathcal{O}(\epsilon^2) \,,
\\
  \mathcal{C}^{-\epsilon}f^{\rm II,R,o}_{2}
  &=
  -\frac{i (1+3\epsilon \log 4)+3 i \epsilon \log \left(\frac{1 + \gamma}{2}\right)}{32\pi |\vec{\ell}|^{1+ 4\epsilon}}
  + \mathcal{O}(\epsilon^2) \,,
\\
  \mathcal{C}^{-\epsilon}f^{\rm II,R,o}_{3}
  &=
  -\frac{i \epsilon \big(\gamma^2+2\gamma-3
    + 2 (1 + \gamma)\log\left(\tfrac{1 + \gamma}{2}\right)\big)}{8\pi(\gamma^2-1)|\vec{\ell}|^{5 +4\epsilon}}
  + \mathcal{O}(\epsilon^2) \,,
\\
  \mathcal{C}^{-\epsilon}f^{\rm II,R,o}_{4}
  &=
  -\frac{i\cosh^{-1}(\gamma)}{32\pi \sqrt{\gamma^2 - 1} |\vec{\ell}|^{1+ 4\epsilon}}
  + \mathcal{O}(\epsilon) \,,
\\
  f^{\rm II,R,o}_{\bar{4}}
  &=
 f^{\rm II,R,o}_{4} \,.
\end{aligned}\label{}
\end{equation}
Our result is consistent with \cite{Jakobsen:2022psy} up to linear combinations of master integrals.

\subsection{The radiation integrals}\label{sec:mushroom}
The last integral family is the radiation integrals that produce the radiation-reaction at 3PM order. They are special in that for those integrals a Green function for the metric
field can hit a singularity. Physically, we can interpret that as real radiation, although, as is well known, the 3PM is a borderline case where energy is not radiated away.
The corresponding integrals can be evaluated as nested one-loop integrals. For simplicity, we relabel the propagators and their powers as follows:
\begin{equation}
\begin{aligned}
  M_{n_{1},n_{2},n_{3},n_{4},n_{5}}
  =
  \int_{k_{1},k_{2}} 
  \frac{\hat{\delta} (k_{1} \cdot v_{1}) \hat{\delta} (k_{2} \cdot v_{2})}{ D_{1}^{n_{1}} D_{2}^{n_{2}} D_{3}^{n_{3}} D^{n_{4}}_{4} D^{n_{5}}_{5} }
\end{aligned}
\label{}\end{equation}
and the propagators are 
\begin{equation}
\begin{aligned}
  D_{1} = k_{2}^{2}\,, \qquad
  D_{2} = \big(\ell-k_{1}-k_{2} \big)^{2}
  -\text{sgn}(\ell^{0}-k_{1}^{0}-k_{2}^{0})i0_{+}\,, \qquad\qquad
  \\
  D_{3} = \big(\ell-k_{2}\big)^{2}\,, \qquad
  D_{4} = k_{2}\cdot v_{1}-i0_+ \,, \qquad
  D_{5} = k_{2}\cdot v_{1}+i0_+ \,. \qquad
\end{aligned}\label{}
\end{equation}
Integrating out the delta functions and shifting $k_{1}$ as $\ell-k_{1}-k_{2} \to k_{1}$, we have
\begin{equation} 
  M_{n_{1},n_{2},n_{3},n_{4},n_{5}}
  =
  \frac{1}{(\gamma^{2}-1)^{\frac{n_{4}+n_{5}}{2}}}
  \int_{\vec{k_{2}}}
  \frac{1}{\mathcal{D}_{1}^{n_{1}} \mathcal{D}_{3}^{n_{3}} \mathcal{D}_{4}^{n_{4}} \mathcal{D}_{5}^{n_{5}} }
  \int_{\vec{k}_{1}} 
  \frac{1}{\mathcal{D}_{2}^{n_{2}}}\,.
\label{rearanged_Mushroom}\end{equation}
where
\begin{equation}
\begin{aligned}
  \mathcal{D}_{1} = \big|\vec{k}_{2}\big|^{2}\,,\qquad
  \mathcal{D}_{2} = \big|\vec{k}_{1}\big|^{2}-2x(\vec{q}\cdot \vec{k}_{2})^{2} - 
  2 x (\vec{q}\cdot \vec{k}_{1}) (\vec{q}\cdot \vec{k}_{2})-\text{sgn}\big( \vec{k}_{2}\cdot \vec{q} \big)i0_+ \,,
  \\
  \mathcal{D}_{3} = \big| \vec{\ell}-\vec{k}_{2} \big|^{2}\,,\qquad
  \mathcal{D}_{4} = - \vec{k}_{2} \cdot \vec{q}-i0_+ \,,\qquad
  \mathcal{D}_{5} =- \vec{k}_{2} \cdot  \vec{q}+i0_+ \,. \qquad \qquad
\end{aligned}\label{}
\end{equation}
We may evaluate the integral iteratively. First $\vec{k}_{1}$ integral gives
\begin{equation} 
\begin{aligned}
  \int_{k_{1}}
  \frac{1}{\mathcal{D}_{2}^{n_{2}} }
  &=
  e^{i\frac{D-2n_{2}-1}{2} \pi} (4\pi)^{\frac{1-D}{2}}(\gamma^{2}-1)^{\frac{D-2n_{2}-1}{2}} \mathcal{D}_{4}^{D-2n_{2}-1} \frac{\Gamma\!\bigl( \frac{2n_{2}+1-D}{2}\bigr)}{\Gamma(n_{2})} \,.
\end{aligned}\label{}
\end{equation}
Substituting the result into \eqref{rearanged_Mushroom}, we have 
\begin{equation} 
\begin{aligned}
  M_{n_{1},n_{2},n_{3},n_{4},n_{5}}
  &=
  e^{i\frac{D-2n_{2}-1}{2} \pi}
  (4\pi)^{\frac{1-D}{2}} (\gamma^{2}-1)^{\frac{D-2n_{2}-n_{4}-n_{5}-1}{2}} \frac{\Gamma\!\bigl( \frac{2n_{2}+1-D}{2}\bigr)}{\Gamma(n_{2})}
  \\&\qquad\times
  \int_{\vec{k}_{2}} 
  \frac{1}{\mathcal{D}_{1}^{n_{1}} \mathcal{D}_{3}^{n_{3}} \mathcal{D}_{4}^{n_{4}+2n_{2}+1-D} \mathcal{D}_{5}^{n_{5}} } \,.
\end{aligned}\label{}
\end{equation}
The remaining $\vec{k}_{2}$ integral can be evaluated as
\begin{equation} 
\begin{aligned}
  &\int_{\vec{k}_{2}}
  \frac{1}{\mathcal{D}_{1}^{n_{1}} \mathcal{D}_{3}^{n_{3}} \mathcal{D}_{4}^{n_{4}} \mathcal{D}_{5}^{n_{5}} }
  \\
  &=
i^{n_4 - n_5} (4\pi)^{\frac{2-D}{2}} \frac{
\Gamma\!\big(\frac{D - 2 n_1 - n_4 - n_5 - 1}{2}\big)
\Gamma\!\big(\frac{D - 2 n_3 - n_4 - n_5 - 1}{2}\big)
\Gamma\!\big(\frac{- D + 2 n_1 + 2 n_3 + n_4 + n_5+1}{2}\big)
}{2\Gamma(n_1)
\Gamma(n_3)
\Gamma(D - n_1 - n_3 - n_4 - n_5 - 1)
\Gamma\!\Big(\frac{1 + n_4 + n_5}{2}\Big)}
\\
&\quad
\times
\displaystyle\frac{\cos\!\big(\tfrac{\pi}{2}(n_4-n_5)\big)}{\cos\!\big(\tfrac{\pi}{2}(n_4+n_5)\big)} |\ell|^{D - 2 n_1 - 2 n_3 - n_4 - n_5 - 1} \,.
\end{aligned}\label{}
\end{equation}
Combining everything together, we obtain what in the amplitude formalism is known as a mushroom integral (because it diagrammatically appears when a graviton line
proceeds directly from incoming to outgoing line, like a cap upon a stem),
\begin{equation} \!\!\!\!
\begin{aligned}
   M_{n_{1},n_{2},n_{3},n_{4},n_{5}}
   &=
  i^{n_4 - n_5}
  (4\pi)^{\frac{3-2D}{2}}(\gamma^2-1)^{\frac{D-2n_2-n_4-n_5-1}{2}} |\ell|^{2D-2n_1-2n_2-2n_3-n_4-n_5-2}
  \\[2mm]&\quad
  \times\frac{
  \Gamma\!\big(\frac{-D+2n_2+1}{2}\big)\,
  \Gamma\!\big(\frac{2D-2n_1-2n_2-n_4-n_5-2}{2}\big)\,
  \Gamma\!\big(\frac{2D-2n_2-2n_3-n_4-n_5-2}{2}\big)\,}
  {2\Gamma(n_1)\,\Gamma(n_2)\,\Gamma(n_3)\,\Gamma(2D-n_1-2n_2-n_3-n_4-n_5-2)}
  \\[2mm]&\quad\times
  \frac{\Gamma\!\big(\frac{-2D+2n_1+2n_2+2n_3+n_4+n_5+2}{2}\big)\sin\!\big(\frac{\pi}{2}(D-2n_2-n_4+n_5)\big)}
     {\Gamma\!\big(\frac{-D+2n_2+n_4+n_5+2}{2}\big)\sin\!\big(\frac{\pi}{2}(D-2n_2-n_4-n_5)\big)}\,.
\end{aligned}\label{}
\end{equation}
In above result, we can see a non-integer power of $(\gamma^2-1)^{\frac{D-2n_2-n_4-n_5-1}{2}}$ which is a characteristic feature of the dissipative effect.

\section{Warm-up: the first post-Minkowskian term}\label{Sec:6}

We now drive the explicit form of the 1PM currents from the recursions in Section \ref{Section:4}, which play the role of initial conditions for the iterative procedure.

\subsection{Graviton Currents}
The 1PM graviton current $\mathfrak{J}^{\mu\nu}\big|_{1}^{L_{1},L_{2}}$ is purely determined by the 1PM external source $\mathfrak{j}^{\mu\nu}|^{1}_{L_{1},L_{2}}$, corresponding to point masses moving along straight-line trajectories in a flat spacetime \eqref{external_source_current}. The corresponding 1PM source currents are 
\begin{equation}
\begin{aligned}
  \mathfrak{j}^{\mu\nu} \big|_{1}^{\ell_{1},0}
  &=
  8 \pi m_{1} \hat{\delta}(v_{1}\cdot \ell_{1}) v_{1}^{\mu} v_{1}^{\nu}\,,
  \\
  \mathfrak{j}^{\mu\nu}\big|_{1}^{0,\ell_{2}}
  &=
  8 \pi m_{2} \hat{\delta}(v_{2}\cdot \ell_{2}) v_{2}^{\mu} v_{2}^{\nu}\,,
  \\
  \mathfrak{j}^{\mu\nu} \big|_{1}^{\ell_{1},\ell_{2}}
  &= 
  0\,.
\end{aligned}\label{}
\end{equation}
Then, we have the 1PM graviton currents from the recursion at 1PM \eqref{LLgraviton_current}, which is the initial condition of the recursions
\begin{equation}
\begin{aligned}
  \mathfrak{J}^{\mu\nu}\big|_{1}^{\ell_{1},0}
  &=
  \eta^{\mu\kappa} \eta^{\nu\lambda}\tilde{\mathfrak{J}}_{\kappa\lambda}\big|_{1}^{\ell_{1},0}
  =
  \frac{16\pi m_{1}\hat{\delta}(v_{1}\cdot \ell_{1})}{\ell_{1}^{2}} v^{\mu}_{1} v^{\nu}_{1} \,,
  \\
  \mathfrak{J}^{\mu\nu} \big|_{1}^{0,\ell_{2}}
  &=
  \eta^{\mu\kappa} \eta^{\nu\lambda}\tilde{\mathfrak{J}}_{\kappa\lambda}\big|_{1}^{0,\ell_{2}}
  =
  \frac{16\pi m_{2}\hat{\delta}(v_{2}\cdot \ell_{2})}{\ell_{2}^{2}} v^{\mu}_{2} v^{\nu}_{2} \,,
  \\
  \mathfrak{J}^{\mu\nu}\big|_{1}^{\ell_{1},\ell_{2}}
  &=
  \eta^{\mu\kappa} \eta^{\nu\lambda}\tilde{\mathfrak{J}}_{\kappa\lambda}\big|_{1}^{\ell_{1},\ell_{2}}
  =
  0\,.
\end{aligned}\label{}
\end{equation}
One can easily check that the current $\mathfrak{J}^{\mu\nu}\big|_{1}^{L_{1},L_{2}}$ satisfies the Ward identity due to the delta functions in the currents
\begin{equation}
\begin{aligned}
  \ell_{1,\mu} \mathfrak{J}^{\mu\nu}\big|_{1}^{\ell_{1},0} = 0\,,
  \\
  \ell_{2,\mu} \mathfrak{J}^{\mu\nu}\big|_{1}^{0,\ell_{2}} = 0\,.
\end{aligned}\label{}
\end{equation}

We also obtain the $H\big|_{1}^{L_{1},L_{2}}$ current from \eqref{H_currents}, $H\big|_{1}^{L_{1},L_{2}} = - \frac{1}{2} \mathfrak{J}^{\kappa}{}_{\kappa}\big|_{1}^{L_{1},L_{2}}\,,$
\begin{equation}
\begin{aligned}
  H\big|_{1}^{\ell_{1},0}
  &= 
  \frac{8 \pi m_{1} \hat{\delta} (\ell_{1}\cdot v_{1})}{|\ell_{1}|^{2}}\,,
  \\
  H\big|_{1}^{0,\ell_{2}} 
  &= 
  \frac{8 \pi m_{2} \hat{\delta} (\ell_{2}\cdot v_{2})}{|\ell_{2}|^{2}}\,,
  \\
  H\big|_{1}^{\ell_{1},\ell_{2}} &= 0\,.
\end{aligned}\label{}
\end{equation}

Then the corresponding current for the usual metric tensor $\tilde{J}_{\mu\nu}\big|_{1}^{L_{1},L_{2}}$ from the following relation
\begin{equation}
  \tilde{J}_{\mu\nu}\big|_{1}^{L_{1},L_{2}}
  =
    \tilde{\mathfrak{J}}_{\mu\nu}\big|_{1}^{L_{1},L_{2}} 
  + \eta_{\mu\nu} H\big|_{1}^{L_{1},L_{2}}\,,
\label{}\end{equation}
and the explicit form is given by
\begin{equation}
\begin{aligned}
  \tilde{J}_{\mu\nu}\big|_{1}^{\ell_{1},0}
  &=
  \frac{8 \pi m_{1} \hat{\delta} (\ell_{1}\cdot v_{1}) \big( \eta_{\mu\nu} + 2 v_{1\mu}v_{1\nu}\big)}{ |\ell_{1}|^{2}}\,,
  \\
  \tilde{J}_{\mu\nu}\big|_{1}^{0,\ell_{2}}
  &=
  \frac{8 \pi m_{2} \hat{\delta} (\ell_{2}\cdot v_{2}) \big( \eta_{\mu\nu} + 2v_{2\mu}v_{2\nu}\big)}{|\ell_{2}|^{2}}\,,
  \\
  \tilde{J}_{\mu\nu}\big|_{1}^{\ell_{1},\ell_{2}}
  &=
  0\,.
\end{aligned}\label{}
\end{equation}
%


\subsection{Matter currents and momentum kick}

We now denote a substitution of the $X^{\mu}_{\alpha}$ into the $\tilde{h}_{\mu\nu}(x_{1},x_{2})$ as 
\begin{equation}
\begin{aligned}
  &\tilde{h}_{\mu\nu} \big(X_{\alpha}\big)
  \equiv
  \tilde{h}_{\mu\nu}(x_{1},x_{2})\big|_{x_{1}\to X_{\alpha}-b_{1}\atop x_{2}\to X_{\alpha}-b_{2}} \,,
  \\&\qquad
  =
    \int_{\ell_{1}} e^{i\ell_{1}\cdot X_{\alpha,1}} \tilde{J}_{\mu\nu}\big|^{\ell_{1},0}
  + \int_{\ell_{2}} e^{i\ell_{2}\cdot X_{\alpha,2}} \tilde{J}_{\mu\nu}\big|^{0,\ell_{2}} 
  + \int_{\ell_{1},\ell_{2}} e^{i\ell_{1}\cdot X_{\alpha,1}+i\ell_{2}\cdot X_{\alpha,2}} \tilde{J}_{\mu\nu}\big|^{\ell_{1},\ell_{2}} \,.
\end{aligned}\label{}
\end{equation}
Then the geodesic equation at first order in $G$ is given by
\begin{equation}
  \frac{d}{d\tau} \bigg[
  	    \eta_{\rho\sigma}\dot{X}^{\sigma}_{\alpha}\big|_{1}
  	  + \tilde{h}_{\rho\sigma}\big(X_{\alpha}\big)\big|_{1} v^{\sigma}_{\alpha}
  	\bigg]
  = 
    \frac{1}{2} \partial_{\rho} \tilde{h}_{\mu\nu}\big(X_{\alpha}\big)\big|_{1} v^{\mu}_{\alpha} v^{\nu}_{\alpha}\,.
\label{}\end{equation}
Substituting the perturbative expansion for the matter field \eqref{X_current_expansion}, we have the recursions for the matter currents
\begin{equation}
\begin{aligned}
  &	(i\ell_{1}\cdot v_{\alpha})^{2} \eta_{\rho\sigma}X_{\alpha}^{\sigma} \big|_{1}^{\ell_{1},0}
  + (i\ell_{1}\cdot v_{\alpha}) \tilde{J}_{\rho\sigma}\big|_{1}^{\ell_{1},0} v^{\sigma}_{\alpha} 
  = \frac{i}{2} \ell_{1}^{\rho}\tilde{J}_{\mu\nu}\big|_{1}^{\ell_{1},0} v^{\mu}_{\alpha} v^{\nu}_{\alpha}\,,
  \\&
  	(i\ell_{2}\cdot v_{\alpha})^{2} \eta_{\rho\sigma}X_{\alpha}^{\sigma} \big|_{1}^{0,\ell_2}
  + (i\ell_{2}\cdot v_{\alpha}) \tilde{J}_{\rho\sigma}\big|_{1}^{0,\ell_2} v^{\sigma}_{\alpha} 
  = \frac{i}{2}\ell_{2}^{\rho} \tilde{J}_{\mu\nu}\big|_{1}^{0,\ell_2} v^{\mu}_{\alpha} v^{\nu}_{\alpha}\,,
  \\
  &X_{\alpha}^{\rho}\big|_{1}^{\ell_{1},\ell_{2}} = 0\,.
\end{aligned}\label{}
\end{equation}
Further, substituting the expression $\tilde{J}_{\mu\nu}\big|^{1}_{L_{1},L_2}$, we can easily solve the recursions 
\begin{equation}
\begin{aligned}
  X^{\rho}_{1}\big|_{1}^{\ell_1,0} &= 0\,,
  \\
  X^{\rho}_{2}\big|_{1}^{\ell_1,0}
  &= - \frac{4 i \pi m_{1}}{\ell_{1}^{2}(\ell_{1}\cdot v_{2})^{2}} \hat{\delta}(v_{1}\cdot \ell_{1})
  	\Big[ (\ell_{1}\cdot v_{2})\big( 4\gamma v^{\rho}_{1} - 2v^{\rho}_{2}\big)
  		+ \ell_{1}^{\rho} \big(2\gamma^{2} -1\big)\Big] \,,
  \\
  X^{\rho}_{1}\big|_{1}^{0,\ell_2}
  &= - \frac{4 i \pi m_{2}}{\ell_{2}^{2}(\ell_{2}\cdot v_{1})^{2}} \hat{\delta}(v_{2}\cdot \ell_{2})
  	\Big[
  		  (\ell_{2}\cdot v_{1})\big( 4\gamma v^{\rho}_{2} - 2v^{\rho}_{1}\big)
  		+ \ell_{2}^{\rho} \big(2\gamma^{2} -1\big)
  	\Big] \,,
  \\
  X^{\rho}_{2}\big|_{1}^{0,\ell_2} &= 0\,,
\end{aligned}\label{Geodesic_currents_G1}
\end{equation}
where $\gamma = - v_{1}\cdot v_{2}$. Note that we do not need the $i\varepsilon$ prescription because $\ell_{1}^{2}$ and $\ell^{2}_{2}$ are strictly positive.

Then the momentum kick $\Delta P_{1,\ord{1}}$ can be obtained by
\begin{equation}
  \Delta P^{\rho}_{1}\big|_{1}
  = 
  m_{1} \int_{-\infty}^{\infty} \mathrm{d} \tau \ddot{X}^{\rho}_{1}(\tau) \big|_{1}
  = 
  - m_{1} \int_{-\infty}^{\infty} \mathrm{d} \tau \int_{\ell_{2}} \big(\ell_{2}\cdot v_{1}\big)^{2} X^{\rho}_{1}\big|_{1}^{0,\ell_{2}} e^{i\ell_{2}\cdot (b + v_{1} \tau)}\,.
\label{}\end{equation}
Substituting \eqref{Geodesic_currents_G1} and the $\tau$ integration, we have
\begin{equation}
\begin{aligned}
  \Delta P^{\rho}_{1}\big|_{1}
  &= 4i\pi m_{1}m_{2} \int_{\ell_{2}}
  \frac{\hat{\delta}(v_{1}\cdot \ell_{2})\hat{\delta}(v_{2}\cdot \ell_{2}) }{\ell_{2}^{2}}
  \ell_{2}^{\rho} \big(2\gamma^{2} - 1\big)
  e^{i\ell_{2}\cdot b}\,,
  \\
  &= 4\pi m_{1} m_{2} \big(2\gamma^{2}-1\big)  \frac{\partial}{\partial b_{\rho}}\int_{\ell_{2}}
  \frac{\hat{\delta}(v_{1}\cdot \ell_{2}) \hat{\delta}(v_{2}\cdot \ell_{2})}{\ell_{2}^{2}} e^{i\ell_{2}\cdot b} 
  \\
  &= -2m_1m_2\frac{\big(2\gamma^{2}-1\big)}{\sqrt{\gamma^2 - 1}}\frac{b^{\rho }}{|b^2|} ~,
\end{aligned}\label{}
\end{equation}
which is the known answer.

\section{The Westpfahl result}

We now turn to the first non-trivial extension of our set-up where we will re-derive the Westpfahl result for scattering at second post-Minkowskian order using our iterative formulation. This is the first time we see the recursion in action, and where we will compute the momentum kick using the iterative ingredients. As will become clear below, the pure and mixed momentum modes exhibit distinct integration structures. In the case of the graviton current, we will explicitly perform all needed momentum integrals appearing in the pure mode convolutions. In contrast, for the mixed modes, we will keep the expressions at the integrand level. Similarly, we leave the matter currents at the integrand level, and the remaining integrations will only be carried out during the final computation of the momentum kick.

\subsection{Graviton currents}
The recursion for the graviton current $\mathfrak{J}^{\mu\nu}\big|^{2}_{L_{1},L_{2}}$ is given by 
\begin{equation}
\begin{aligned}
  \mathfrak{J}^{\mu\nu}\big|_{2}^{L_{1},L_{2}}
  &=
    \frac{1}{(L_{12})^{2}} \bigg(
  	- \tau^{\mu\nu}\big|_{2}^{L_{1},L_{2}}
  	+ 2 \mathbf{j}^{\mu\nu}\big|_{2}^{L_{1},L_{2}}
  \bigg)\,,
\end{aligned}\label{2PMMathfrakJ}
\end{equation}
and we first derive the 2PM currents for the matter source $\mathbf{j}^{\mu \nu}\big|_{2}^{L_{1},L_{2}}$ using the relation 
\begin{equation}
  \mathbf{j}^{\mu\nu}\big|_{2}^{L_{1},L_{2}}
  = 
    \mathfrak{j}^{\mu\nu}\big|_{2}^{L_{1},L_{2}} 
  + \big(H* \mathfrak{j}^{\mu\nu}\big)\big|_{2}^{L_{1},L_{2}}\,.
\label{}\end{equation}
We can obtain the external source $\mathfrak{j}^{\mu\nu}\big|_{2}^{L_{1},L_{2}}$ by substituting $\zeta^{\mu\nu}_{\alpha}\big|_{2}^{L_{1},L_{2}}$ into eq. \eqref{external_source_current} defined in \eqref{nPMZeta}
\begin{equation}
\begin{aligned}
  \mathfrak{j}^{\mu\nu}\big|_{2}^{\ell_{1},0}
  &= 0\,,
  \qquad
  \mathfrak{j}^{\mu\nu} \big|_{2}^{0,\ell_{2}}
  = 0\,,
  \\
  \mathfrak{j}^{\mu\nu}\big|_{2}^{\ell_{1},\ell_{2}}
  &= 
  8i\pi m_{1} \hat{\delta}\big(\ell_{1}\cdot v_{1}\big) \Big[
  	  2 \big(\ell_{2}\cdot v_{1} \big) v^{(\mu}_{1} X^{\nu)}_{1}\big|_{1}^{0,\ell_{2}}
  	- \big(\ell_{12}\cdot X_{1}\big|_{1}^{0,\ell_{2}}\big) v^{\mu}_{1} v^{\nu}_{1}
  \Big]
  \\&\quad
  +  8i\pi m_{2} \hat{\delta}\big(\ell_{2}\cdot v_{2}\big) \Big[
  	  2 \big(\ell_{1}\cdot v_{2} \big) v^{(\mu}_{2} X^{\nu)}_{2}\big|_{1}^{\ell_{1},0}
  	- \big(\ell_{12}\cdot X_{2}\big|_{1}^{\ell_{1},0}\big) v^{\mu}_{2} v^{\nu}_{2}\Big] \,.
\end{aligned}\label{}
\end{equation}
Using the 1PM matter currents $X^{\mu}_{\alpha}\big|_{1}^{L_{1},L_{2}}$ derived in the previous section, we have the explicit form of the 2PM currents of the external source $\mathfrak{j}^{\mu\nu}\big|_{2}^{L_{1},L_{2}}$
\begin{equation}
\begin{aligned}
  \!\!\!\!\mathfrak{j}^{\mu\nu}\big|_{2}^{\ell_{1},\ell_{2}}
  &\!=
  \frac{2\hat{\mu}_{1,2}}{\ell_{2}^{2}} \Bigg[
    4\gamma v_{1}^{(\mu} v_{2}^{\nu)}
  + \ell_{2}^{(\mu}v^{\nu)}_{1} \frac{2\gamma^{2}{-}1}{\ell_{2}{\cdot} v_{1}}
  - v^{\mu}_{1} v^{\nu}_{1} \bigg[
  	1 {+} \frac{2 \gamma (\ell_{1}{\cdot} v_{2})}{\ell_{2}\cdot v_{1}}
  	{+} (\ell_{12}\cdot \ell_{2}) \frac{2\gamma^{2}-1}{2(\ell_{2}{\cdot} v_{1})^{2}}
  	\bigg] \Bigg]
  \\&\quad
  + (1\leftrightarrow 2 ) \,,
\end{aligned}\label{source_G2}
\end{equation}
where 
\begin{equation}
  \hat{\mu}_{\alpha, \beta}=32 \pi^{2} m_{\alpha} m_{\beta} \hat{\delta}\left(\ell_{1} \cdot v_{\alpha}\right) \hat{\delta}\left(\ell_{2} \cdot v_{\beta}\right)
  \,.
\label{}\end{equation}
The corresponding matter current $\mathbf{j}^{\mu\nu}|^{2}_{L_{1},L_{2}}$ is given by
\begin{equation}
\begin{aligned}
  \mathbf{j}^{\mu\nu}\big|_{2}^{L_{1},L_{2}}
  &= 
    \mathfrak{j}^{\mu\nu}\big|_{2}^{L_{1},L_{2}}
  + \big[H*\mathfrak{j}^{\mu\nu}\big]_{2}^{L_{1},L_{2}} \,.
\end{aligned}\label{}
\end{equation}
Evaluating the convolution product $\big[H*\mathfrak{j}^{\mu\nu}\big]^{2}_{L_{1},L_{2}}$, one finds that the following are trivial in dimensional regularization because they are scale-free  integrals:
\begin{equation}
\begin{aligned}
  \big[\mathfrak{J}^{\mu\nu} *\mathfrak{j}^{\kappa}{}_{\kappa} \big]_{2}^{\ell_{1},0}
  =
  -\int_{k}\frac{128 \pi^2 m_{1}^{2} v_{1}^{\mu} v_{1}^{\nu} \hat{\delta} (k \cdot v_{1}) \hat{\delta} (\ell_{1}\cdot v_{1})}{(\ell_{1}-k)^2} = 0\,,
  \\
  \big[\mathfrak{J}^{\mu\nu} *\mathfrak{j}^{\kappa}{}_{\kappa}\big]_{2}^{0,\ell_{2}} 
  =
  - \int_{k} \frac{128 \pi ^2 m_{2}^2 v_{2}^{\mu} v_{2}^{\nu} \hat{\delta} (k\cdot v_{2}) \hat{\delta} (\ell_{2}\cdot v_{2})}{(\ell_{2}-k)^2} = 0\,.
\end{aligned}\label{}
\end{equation}
Thus the pure mode sectors of $\mathbf{j}^{\mu\nu}\big|_{2}^{\ell_{1},0}$ and $\mathbf{j}^{\mu\nu}\big|_{2}^{0,\ell_{2}}$ also vanish, and only the mixed mode $\mathbf{j}^{\mu\nu}\big|_{2}^{\ell_{1},\ell_{2}}$ survives,
\begin{equation}
\begin{aligned}
  \mathbf{j}^{\mu\nu}\big|_{2}^{\ell_{1},\ell_{2}}
  &=
  \frac{2\hat{\mu}_{1,2}}{\ell_{2}^{2}} \Bigg[
    4\gamma v_{1}^{(\mu} v_{2}^{\nu)}
  + \ell_{2}^{(\mu}v^{\nu)}_{1} \frac{2\gamma^{2}{-}1}{\ell_{2}\cdot v_{1}}
  - v^{\mu}_{1} v^{\nu}_{1} \bigg[
  	  \frac{2 \gamma (\ell_{1}{\cdot} v_{2})}{\ell_{2}\cdot v_{1}}
  	+ (\ell_{12}{\cdot} \ell_{2}) \frac{2\gamma^{2}{-}1}{2(\ell_{2}{\cdot} v_{1})^{2}}
  	\bigg] \Bigg]
  \\&\quad
  + (1\leftrightarrow 2 ) \,.
\end{aligned}\label{source_G2}
\end{equation}

Next, we solve the recursions for the energy-momentum pseudotensor $\tau^{\mu\nu}|^{2}_{L_{1},L_{2}}$ and the three subcurrents $W^{\mu\nu}\big|_{2}^{L_{1},L_{2}}\,, Z^{\mu}{}_{\nu}\big|_{2}^{L_{1},L_{2}}$ and $d^{\mu\nu}\big|_{2}^{L_{1},L_{2}}$ derived in \eqref{EMpseudotensor} and \eqref{recursion_subcurrents_WZ}. The explicit form of these currents is provided in Appendix \ref{Appendix:2PM}. 

The 2PM graviton currents $\mathfrak{J}^{\mu\nu} \big|_{2}^{L_{1},L_{2}}$ follow from eq. \eqref{2PMMathfrakJ}, and substituting the subcurrents listed above, we have
\begin{equation}\!\!\!\!
\begin{aligned}
  &\mathfrak{J}^{\mu\nu}\big|_{2}^{\ell,0}
  \\
  &=
  \frac{m_{1}^{2} 16^{\epsilon} \pi^{\epsilon +2}
   \hat{\delta}(\ell\cdot v_{1})}{\Gamma(2{-}\epsilon) \cos(\pi \epsilon) |\ell|^{2\epsilon+3}}
  \Big[|\ell|^{2} 
    \big((15{-}16 \epsilon) v_{1}^{\mu}v_{1}^{\nu}
      -\tilde{\eta}^{\mu\nu}
      +2 (1{-}\epsilon) \eta^{\mu\nu}
    \big)
    -(1{-}2 \epsilon) \ell^{\mu}\ell^{\nu}
  \Big]\,,
  \\
  &\mathfrak{J}^{\mu\nu}\big|_{2}^{\ell_{1},\ell_{2}} 
  \\
  &= 
    \frac{\hat{\mu}_{12}}{|\ell_{1}|^2 |\ell_{2}|^2 |\ell_{12}|^2} \bigg[
      4 \ell_2^{(\mu} v_1^{\nu)} \left(\frac{(2 \gamma^2-1)\ell_{1}^{2}}{ \ell_{2}\cdot v_{1} -i\varepsilon} 
    - 4 \gamma (\ell_{1}\cdot v_2)\right) 
    - 2\left(2\gamma^2-1\right) \ell_1^{(\mu} \ell_2^{\nu)}
  \\&\qquad\qquad\qquad\quad
    + 8v_1^{\mu} v_1^{\nu} \left( 
  	    (\ell_{1}\cdot v_{2}) ^2 
  	  - \frac{(2\gamma^2-1) |\ell_{1}|^{2}(\ell_{12}\cdot\ell_{2}) }{4(\ell_{2}\cdot v_{1} -i\varepsilon)^2}
  	  - \frac{\gamma |\ell_{1}|^{2} ( \ell_{1}\cdot v_{2}) }{ \ell_{2}\cdot v_{1}-i\varepsilon} \right)
  \\&\qquad\qquad\qquad\quad
  + 8 v_1^{(\mu} v_2^{\nu)} \Big(
  	\gamma  ( \ell_{1}\cdot\ell_{2}) -( \ell_{1}\cdot v_{2})  ( \ell_{2}\cdot v_{1}) +\gamma \big(\ell_{1}^{2}+\ell_{2}^{2}\big)
  \Big)
  \\&\qquad\qquad\qquad\quad
  + \eta^{\mu\nu} \Big(\left(2 \gamma^2-1\right) ( \ell_{1}\cdot\ell_{2}) + 4 \gamma (\ell_{1}\cdot v_{2})  ( \ell_{2}\cdot v_{1}) \Big) 
  + (1\leftrightarrow 2)
  \bigg] \,.
\end{aligned}\label{J_G2}
\end{equation}
and $\mathfrak{J}^{\mu\nu} \big|_{2}^{0,\ell_{2}}$ is defined by $1 \leftrightarrow 2$ exchanges from $\mathfrak{J}^{\mu\nu} \big|_{2}^{\ell_{1},0}$.
Finally, recursions for the quantities $\tilde{\mathfrak{J}}^{\mu\nu} \big|_{2}^{L_{1},L_{2}}$ and $H \big|_{2}^{L_{1},L_{2}}$ follow from
\begin{equation}
\begin{aligned}
  \tilde{\mathfrak{J}}_{\mu \nu}\big|_2^{L_{1},L_{2}}
  &= 
    \eta_{\mu\kappa}\eta_{\nu\lambda} \mathfrak{J}^{\kappa\lambda} \big|_2^{L_{1},L_{2}}
  + \big[\tilde{\mathfrak{J}}_{\mu\lambda} * \eta_{\nu\kappa}\mathfrak{J}^{\kappa\lambda}\big]_{2}^{L_{1},L_{2}}\,,
  \\
  H\big|_2^{L_{1},L_{2}}
  &=
  - \frac{1}{2} \mathfrak{J}^{\kappa}{}_{\kappa}\big|_{2}^{L_{1},L_{2}}
  + \frac{1}{8} \big[\mathfrak{J}^{\kappa}{}_{\kappa}*\mathfrak{J}^{\lambda}{}_{\lambda}\big]_{2}^{L_{1},L_{2}}
  - \frac{1}{4} \big[\tilde{\mathfrak{J}}_{\kappa\lambda}*\mathfrak{J}^{\kappa\lambda}\big]_{2}^{L_{1},L_{2}}\,.
\end{aligned}\label{}
\end{equation}
In addition,  the current corresponding to the standard metric perturbation $\tilde{J}_{\mu\nu}\big|^{L_{1},L_{2}}_{2}$ follows from
\begin{equation}
  \tilde{J}_{\mu\nu}\big|_{2}^{L_{1},L_{2}}
  = 
  	\tilde{\mathfrak{J}}_{\mu\nu}\big|_{2}^{L_{1},L_{2}}
  + \big[H* \tilde{\mathfrak{J}}_{\mu\nu} \big]_{2}^{L_{1},L_{2}}
  + \eta_{\mu\nu} H\big|_{2}^{L_{1},L_{2}}\,.
\label{}\end{equation}
%

\subsection{Matter currents and momentum kick}

The recursion for the 2PM worldline current  $X^{\rho}_{\alpha}\big|_2^{L_{1},L_{2}}$ is derived from eq. \eqref{Recursion_worldline_current} for $n=2$ case,
\begin{equation}\!\!\!
\begin{aligned}  
&(L_{12}\cdot v_{\alpha})^{2} X^{\rho}_{\alpha}\big|_{2}^{L_{1},L_{2}}
  \\
  &=
  -\frac{v^{\mu}_{\alpha} v_{\alpha}^{\nu}}{2} \bigg(
  	  iL_{12}^{\rho} \tilde{J}_{\mu\nu} \big|_{2}^{L_{1},L_{2}}
    - \Big[\hat{\ell}_{\rho} \hat{\ell}_{\sigma} \tilde{J}_{\mu\nu} * X^{\sigma}_{\alpha}\Big]_{2}^{L_{1},L_{2}}
  \bigg) 
  + \Big[\hat{\ell}^{\rho} \tilde{J}_{\mu\nu} * \big(\hat{\ell}{\cdot} v_{\alpha}\big)X^{\mu}_{\alpha} v_{\alpha}^{\nu} \Big]_{2}^{L_{1},L_{2}}
  \\&~
  + L_{12} \cdot v_{\alpha} \bigg(
  	  \tilde{J}^{\rho}{}_{\sigma}\big|_{2}^{L_{1},L_{2}} v^{\sigma}_{\alpha}
  	+ i\Big[\hat{\ell}_{\kappa} \tilde{J}^{\rho}{}_{\sigma} v^{\sigma}_{\alpha} * X^{\kappa}_{\alpha}\Big]_{2}^{L_{1},L_{2}}
  	+ i\Big[\tilde{J}^{\rho}{}_{\sigma} * \big(\hat{\ell}\cdot v_{\alpha}\big)X^{\sigma}_{\alpha}\Big]_{2}^{L_{1},L_{2}}
  \bigg) \,.
\end{aligned}\label{G2_geodesic_eq}
\end{equation}
The full expressions of $X^{\rho}_{1}\big|_{2}^{0,\ell_{2}}$ and $X^{\rho}_{1}\big|_{2}^{\ell_{1},\ell_{2}}$ are long, and given in Appendix \ref{Appendix:2PM}.

Using the matter currents, we can now finally derive the momentum kick from eq.  \eqref{momentum_kick_1}
\begin{equation}
\begin{aligned}
  \Delta P^{\rho}_{1}\big|_{2}
  &=
    m_{1} \int_{-\infty}^{\infty} \mathrm{d} \tau \ddot{X}^{\rho}_{1} (\tau)\big|_{2}\,,
  \\
  &=
  - m_{1} \int_{\ell} e^{i \ell \cdot b} \bigg[
  	  \hat{\delta}(\ell\cdot v_{1}) (\ell\cdot v_{1})^{2} X^{\rho}_{1} \big|_{2}^{0,\ell}
  	+ \int_{k} \hat{\delta}\big(k\cdot v_{1}) (k\cdot v_{1})^{2} X^{\rho}_{1} \big|_{2}^{k-\ell,\ell} 
  \bigg]\,.
\end{aligned}\label{}
\end{equation}
Due to the delta functions, only the double-pole term with respect to $\ell_{2}\cdot v_{1}$ and $k\cdot v_{1}$ contributes to the momentum kick for for $X^{\rho}_{1}\big|_{2}^{0,\ell}$ and $X^{\rho}_{1}\big|_{2}^{k-\ell,\ell}$, respectively. The double pole for terms for $\ell \cdot v_{1}$ of $X^{\rho}_{1}\big|_{2}^{0,\ell}$ are
\begin{equation}\!\!
\begin{aligned}
  &-\frac{1}{m_{2}^{2}}(\ell\cdot v_{1})^{2}X^{\rho}_{1}\big|_{2}^{0,\ell}
  \\
  &=
  \frac{\pi^{2+\epsilon}\hat{\delta} (\ell\cdot v_2)}{2^{1-4 \epsilon}|\ell|^{2\epsilon}}
  \Bigg[
    \frac{i \left(8 (1{-}2\gamma^{2})^2 \epsilon ^2 
    -4 (3\gamma^2{-}1) (4 \gamma^2{-}3) \epsilon +3 (5 \gamma^2{-}1)(\gamma^2{-}1)\right)\ell^{\rho}}{(\gamma^2-1) \Gamma(2-\epsilon ) \cos(\pi \epsilon ) |\ell|}
  \\&\qquad\qquad\qquad\quad
  + \frac{4 \sqrt{\pi} (1-2\gamma^{2})^2 \csc(\pi\epsilon ) w_{2}}{\left(\gamma^2-1\right)^{\frac{1}{2}} \Gamma(\frac{1}{2}-\epsilon)}
  \Bigg] \,.
\end{aligned}\label{X_2PM_recursion}
\end{equation}
where
\begin{equation}
  w_{1}^{\mu} \equiv \frac{\gamma v_{1}^{\mu}-v_{2}^{\mu}}{\gamma^{2}-1}\,, 
  \qquad 
  w_{2}^{\mu} \equiv \frac{\gamma v_{2}^{\mu}-v_{1}^{\mu}}{\gamma^{2}-1}\,.
\label{}\end{equation}
Next we evaluate  the $X^{\rho}_{1}\big|_{2}^{k-\ell,\ell}$ part, finding
\begin{equation}
\begin{aligned}
  &-\frac{1}{m_{1}m_{2}}\int_{k} \hat{\delta}\big(k\cdot v_{1}) (k\cdot v_{1})^{2} X^{\rho}_{1} \big|_{2}^{k-\ell,\ell}
  \\
  &=
  \frac{\hat{\delta}(\ell\cdot v_{1})\hat{\delta}(\ell\cdot v_{2})}{\pi^{-2-\epsilon}2^{1-4 \epsilon}|\ell|^{2\epsilon}} 
  \Bigg[
    \frac{i \left(8 (1{-}2\gamma^{2})^2 \epsilon^2 
    -4 (3\gamma^2{-}1) (4 \gamma^2{-}3) \epsilon +3 (5\gamma^2{-}1)(\gamma^2{-}1)\right)\ell^{\rho}}{(1{-}\gamma^2) \Gamma(2-\epsilon) \cos(\pi \epsilon) |\ell|}
  \\&\qquad\qquad\qquad\qquad
  -\frac{4 \sqrt{\pi } (1-2\gamma^{2})^2 \csc(\pi\epsilon ) w_{1}}{\left(\gamma ^2-1\right)^{\frac{1}{2}} \Gamma (\frac{1}{2}-\epsilon)}
  \Bigg] \,.
\end{aligned}\label{}
\end{equation}

Combining these results and expanding with respect to the dimensional regularization parameter $\epsilon$, the momentum kick reduces to
\begin{equation}
\begin{aligned}
  &\Delta P^{\rho}_{1} \big|_{2} 
  = 
  m_{1}m_{2} \int_{\ell} \hat{\delta} (\ell\cdot v_{1}) \hat{\delta}(\ell\cdot v_{2}) e^{i \ell b} \Big( 
      m_{1}\mathcal{P}_{1}^{\rho}\big|_{2}^{(A)} 
    + m_{2}\mathcal{P}_{1}^{\rho}\big|_{2}^{(B)}\Big)\,,
\end{aligned}\label{}
\end{equation}
where
\begin{equation}
\begin{aligned}
  \mathcal{P}_{1}^{\rho}\big|_{2}^{(A)} 
  &= 
  -\frac{2\pi (1-2 \gamma ^2)^2 w^{\rho}_{1}}{(\gamma^{2}-1)^{\frac{1}{2}} \epsilon}
  - \frac{3i\pi^{2} (5 \gamma ^2-1) \ell^{\rho}}{2|\ell|} 
  +\frac{4\pi (1-2 \gamma^2)^2 \log|\ell| w_{1}^{\rho}}{(\gamma ^2-1)^{\frac{1}{2}}}\,,
  \\
  \mathcal{P}_{1}^{\rho}\big|_{2}^{(B)}
  &=
   \frac{2\pi (1-2 \gamma ^2)^2 w^{\rho}_{2}}{(\gamma^{2}-1)^{\frac{1}{2}} \epsilon}
  - \frac{3i\pi^{2} (5 \gamma ^2-1) \ell^{\rho}}{2|\ell|} 
  -\frac{4\pi (1-2 \gamma^2)^2 \log|\ell| w_{2}^{\rho}}{(\gamma^2-1)^{\frac{1}{2}}}\,,
\end{aligned}\label{}
\end{equation}
Finally performing the Fourier transform, we have
\begin{equation}
\begin{aligned}
  \Delta P_{1}^{\rho}\big|_{2}
  &= 
  \frac{m_{1}m_{2} }{|b|^{2}} \bigg[
      \frac{3\pi (m_{1} +m_{2})(5\gamma^{2}-1)b^{\rho}}{4\sqrt{\gamma^{2}-1}\ |b|}
    - \frac{2 (1-2\gamma^{2})^{2}(m_{1} w_{1}^{\rho}-m_{2} w_{2}^{\rho})}{(\gamma^{2}-1)}
  \bigg]\,,
\end{aligned}\label{}
\end{equation}
which is the standard result.


\section{Third post-Minkowskian order: radiation back-reaction}
We finally show how our formalism can be taken to higher orders in a systematic manner. We explicitly write down the 3PM recursion relations, and thus derive the corresponding graviton and matter currents. Solving the recursion relations, we derive all needed integrands and finally compute the momentum kick by evaluating the double momentum integrals. As expected, this is the first order in which the 
choice of $i\epsilon$-prescription becomes crucial. For the lower orders we were cavalier about those $i\epsilon$-prescriptions because the momentum integrals did not probe the poles of the Green functions. Starting at third order, the choice of causal (retarded) Green functions automatically ensures full inclusion of all radiative effects (and associated radiation back-reaction). This is precisely as expected from worldline computations using the $in-in$ formalism that again leads to the $i\epsilon$-prescription of retarded Green functions \cite{Jakobsen:2022psy,Kalin:2022hph}. For an exposition of how this is equivalent to the amplitude-based KMOC formalism (which can be performed entirely with Feynman propagators), see ref.
\cite{Damgaard:2023vnx}.

\subsection{Recursions for the graviton current}
The starting point is again the recursion for the graviton current, now at 3PM order,
\begin{equation}
  \mathfrak{J}^{\mu\nu}\big|_{3}^{L_{1},L_{2}}
  =
    \frac{1}{(L_{12})^{2}} \Big(
  - \tau^{\mu\nu}\big|_{3}^{L_{1},L_{2}}
  + 2 \mathbf{j}^{\mu\nu}\big|_{3}^{L_{1},L_{2}}
  \Big)\,,
\label{}\end{equation}
We first consider the external source term $\mathbf{j}^{\mu\nu}|^{3}_{L_{1},L_{2}}$ defined in \eqref{currents_external_source}. It is straightforward to show that $\mathbf{j}^{\mu\nu}|_{3}^{\ell_{1},0}$ and $\mathbf{j}^{\mu\nu}|_{3}^{0,\ell_{2}}$ vanish, then the only non-trivial contribution $\mathbf{j}^{\mu\nu}|_{3}^{\ell_{1},\ell_{2}}$, which is represented by $\mathfrak{j}^{\mu\nu}|_{3}^{\ell_{1},\ell_{2}}\!$ as in eq. \eqref{fat_j},
\begin{equation}
\begin{split}
  \frac{1}{8\pi m_{1}} \mathfrak{j}^{\mu\nu}\big|_{3}^{\ell_{1},\ell_{2}}
  &=
    \hat{\delta}(\ell_{1} \cdot v_{1}) \zeta_{1}^{\mu\nu}(\ell_{12})\big|_{2}^{0,\ell_{2}}
  + \int_{k}
      \hat{\delta}(k \cdot v_{1}) \zeta_{1}^{\mu\nu} (\ell_{12}) \big|_{2}^{\ell_{1}-k,\ell_{2}} + (1\leftrightarrow 2)\,,
\end{split}\label{}
\end{equation}
where $\zeta_{1}^{\mu\nu}(\ell_{12}) \big|_{2}^{L_{1},L_{2}}$ is defined in eq. \eqref{nPMZeta}. The energy-momentum pseudotensor $\tau^{\mu\nu}\big|_{3}^{L_{1},L_{2}}$ is determined by
\begin{equation}
\begin{aligned}
  \tau^{\mu\nu}\big|_{3}^{L_{1},L_{2}}
  &= 
    W^{\mu\nu}\big|_{3}^{L_{1},L_{2}}
  + \eta^{\kappa(\mu}Z^{\nu)}{}_{\kappa}\big|_{3}^{L_{1},L_{2}}
  - \big[Z^{(\mu}{}_{\kappa} * \mathfrak{J}^{\nu)\kappa}\big]_{3}^{L_{1},L_{2}}
  + 2d^{\mu\nu}\big|_{3}^{L_{1},L_{2}}\,.
\end{aligned}\label{}
\end{equation}
where the subcurrents $W^{\mu\nu}$, $Z^{\mu}{}_{\nu}$ and $d^{\mu\nu}$ can be found from the recursions derived in eqs. \eqref{recursion_subcurrents_WZ} and \eqref{recursion_subcurrents_d}.

Solving the recursions, we obtain $\tilde{J}_{\mu\nu} \big|_{3}^{L_{1},L_{2}}$ and its constituent subcurrents, which are divided into zero-mode and mixed-mode currents with distinct structures. The zero-mode currents can be integrated order by order in the post-Minkowskian expansion using the one-loop bubble or triangle integral formulas. On the other hand, mixed-mode currents have a more complicated structure and we have not found any simple way to perform the integration by iterated one-loop integrals. In such cases, rather than treating the two-loop integral iteratively, we perform an IBP reduction and employ the master integral results of Section \ref{Sec:5}.

\subsection{Momentum kick}
As we have used multiple times, the momentum kick $\Delta P^{\rho}_{1} \big|_{3}$ is obtained by integrating $\ddot{X}^{\rho}_{1}(\tau)\big|_{3}$. All the recursion relations required to determine the worldline currents and the momentum kick at arbitrary $n$-PM order are already obtained in Section~\ref{Sec:4.2}. To compute the 3PM momentum kick, which is our goal here, it is sufficient to apply \eqref{momentum_kick_1} for $n=3$. This requires the 3PM graviton current, but not the 3PM worldline current: only the worldline currents up to 2PM enter the computation. 

The 3PM momentum kick then splits into three sectors according to the powers of masses:
\begin{equation}
  \Delta P^{\rho}_{1} \big|_{3}
  =
    \Delta P^{\rho}_{1} \big|_{3}^{m_{1}^{3}m_{2}}
  + \Delta P^{\rho}_{1} \big|_{3}^{m_{1}^{2}m_{2}^{2}}
  + \Delta P^{\rho}_{1} \big|_{3}^{m_{1}m_{2}^{3}}\,.
\label{}\end{equation}

Solving the recursion, we can check that $m_{1}m_{2}^{3}$ terms arise only from 
\begin{equation}
\begin{aligned}
  \Delta P^{\rho}_{1} \big|_{3}^{m_{1} m_{2}^{3}}
  &=
  m_{1} m_{2}^{3}\int_{\ell} e^{i\ell\cdot b}
  \hat{\delta} \big( \ell\cdot v_{1}\big) \hat{\delta} \big( \ell\cdot v_{2}\big)
  \widetilde{\Delta P}{}^{\rho}_{1} (\ell)\big|_{3}^{(A)}
\end{aligned}\label{m1m2Cubic}
\end{equation}
where $\widetilde{\Delta P}{}^{\rho}_{1} (\ell)\big|_{3}^{(A)}$ is defined in \eqref{Fourier_integrand2}.

On the other hand, $m_{1}^{3}m_{2}$ and $m_{1}^{2}m_{2}^{2}$ sectors arise from $\widetilde{\Delta P}{}^{\rho}_{1} (\ell)\big|_{3}^{(B)}$ together
\begin{equation}
\begin{aligned}
  \Delta P^{\rho}_{1} \big|_{3}^{m_{1}^{3} m_{2}^{1}}
  &=
  m_{1}^{3} m_{2} \int_{\ell} e^{i\ell\cdot b}
  \hat{\delta} \big( \ell\cdot v_{1}\big) \hat{\delta} \big( \ell\cdot v_{2}\big)
  \widetilde{\Delta P}{}^{\rho}_{1} (\ell)\big|_{3}^{(B)|m_{1}^{3} m_{2}^{1}} \,, 
  \\
  \Delta P^{\rho}_{1} \big|_{3}^{m_{1}^{2} m_{2}^{2}}
  &=
  m_{1}^{2} m_{2}^{2} \int_{\ell} e^{i\ell\cdot b}
  \hat{\delta} \big( \ell\cdot v_{1}\big) \hat{\delta} \big( \ell\cdot v_{2}\big)
  \widetilde{\Delta P}{}^{\rho}_{1} (\ell)\big|_{3}^{(B)|m_{1}^{2} m_{2}^{2}}\,.
\end{aligned}\label{}
\end{equation}
One can show that all the two-loop integrals arising from $\Delta P^{\rho}_{1} \big|_{3}^{m_{1}^{1} m_{2}^{3}}$ and $\Delta P^{\rho}_{1} \big|_{3}^{m_{1}^{3} m_{2}^{1}}$ belong to Family-I, while all those from $\Delta P^{\rho}_{1} \big|_{3}^{m_{1}^{2} m_{2}^{2}}$ belong to Family-II. 

We may decompose the integrand of the Fourier transform into 3 parts according to the convolution bracket structure, 0 bracket, single bracket and double bracket, which is denoted by $[\bullet *\bullet *\bullet ]$, such as
\begin{equation}
  - \frac{m_1}{2}\mathcal{E}\Big[ 
        i\hat{\ell}^{\rho} \tilde{J}_{\mu\nu}
      * \mathcal{X}^{\mu}_{1}
      * \mathcal{X}^{\nu}_{1}
    \Big]_{3}^{L_{1},L_{2}} 
    \equiv 
    \mathcal{P}^{\rho}_{0} \big|_{L_{1},L_{2}}
    +
    \mathcal{P}^{\rho}_{1}\big|_{L_{1},L_{2}}
    +
    \mathcal{P}^{\rho}_{2}\big|_{L_{1},L_{2}} \,,
\label{}\end{equation}
where
\begin{equation}
\begin{aligned}
  \mathcal{P}^{\rho}_{0} \big|_{L_{1},L_{2}} 
  &= - \frac{m_1}{2}
  i L_{12}^{\rho}\tilde{J}_{\mu\nu} \big|_{3}^{L_{1},L_{2}} \, v^{\mu}_{1} v^{\nu}_{1} \,,
  \\
  \mathcal{P}^{\rho}_{1} \big|_{L_{1},L_{2}} 
  &=  
  \frac{m_1}{2} \bigg(\Big[\hat{\ell}^{\rho} \hat{\ell}_{\sigma} \tilde{J}_{\mu\nu} * X^{\sigma}_{1}
  	\Big]_{3}^{L_{1},L_{2}} v^{\mu}_{1} v^{\nu}_{1}
  + 2\Big[ \hat{\ell}^{\rho} \tilde{J}_{\mu\nu} * (\hat{\ell}\cdot v_{1})X^{\mu}_{1} 
  \Big]_{3}^{L_{1},L_{2}} v^{\nu}_{1} \bigg)\,,
  \\
  \mathcal{P}^{\rho}_{2} \big|_{L_{1},L_{2}} 
  &=
  \frac{m_1}{2} i \bigg(\frac{1}{2} \Big[
    \hat{\ell}^{\rho} \hat{\ell}_{\kappa} \hat{\ell}_{\lambda} \tilde{J}^{\mu\nu} 
   	* X^{\kappa}_{1} 
   	* X^{\lambda}_{1}
  \Big]_{3}^{L_{1},L_{2}}  v^{\mu}_{1} v^{\nu}_{1}
  \\&\qquad\qquad
  + \Big[\hat{\ell}^{\rho} \hat{\ell}_{\kappa} \tilde{J}_{\mu\nu}
  	* X^{\kappa}_{1} 
  	* (\hat{\ell}\cdot v_{1})X^{\mu}_{1} 
  \Big]_{3}^{L_{1},L_{2}}v_{1}^{\nu}
  \\&\qquad\qquad
  + \Big[\hat{\ell}^{\rho} \tilde{J}_{\mu\nu} * (\hat{\ell}\cdot v_{1}) X^{\mu}_{1} * (\hat{\ell}\cdot v_{1}) X^{\nu}_{1}\Big]_{3}^{L_{1},L_{2}} \bigg)\,.
\end{aligned}\label{momentum_kick_integrand}
\end{equation}
As we saw in the previous section, the zero-mode currents such as $[0,\ell]$ or $[\ell,0]$ appearing in the 2PM graviton and worldline currents and in the 3PM graviton current are already integrated quantities, and it is their integrated values that enter the convolution brackets. Depending on whether a term carries zero, single, and double brackets, it may require no loop integration, a one-loop integration, or a two-loop integration. Thus it is useful to organize the integrand according to its bracket structure.

From the dimensional analysis, the $n$-PM momentum kick should behave as
\begin{equation}
  \Delta P_{1}^{\mu}\big|^{n} \sim m F_{n}(\gamma)\left(\frac{G m}{|b|}\right)^{n} \propto \frac{G^{n}}{|b|^{n}}\,.
\label{}\end{equation}
Then the transverse and longitudinal parts of $\Delta P^{\mu}_{1}\big|^{n}$ are proportional to
\begin{equation}
\begin{aligned}
  \Delta P^{\mu}_{1}\big|^{n}_{\perp} \sim b^{\mu}/|b|^{n+1}\,,
  \qquad
  \Delta P^{\mu}_{1}\big|^{n}_{\parallel} \sim \frac{a v^{\mu}_{1}+b v^{\mu}_{2}}{|b|^{n}}\,.
\end{aligned}\label{}
\end{equation}
Now let us consider the 3PM order, $n=3$. The Fourier transform of $b^{\mu}/|b|^{4}$ and $1/|b|^{3}$ are proportional to $\log|\ell|\ell^{\mu}$ and $|\ell|$ respectively. We also ignore the analytic terms, which is given by $1, |\ell|^{2}, |\ell|^{4}, \cdots$, since their Fourier transforms become local terms, $(\partial_{b}^{2})^{n}\delta^{2}(|b|)$. Thus we will focus on the non-analytic pieces, $\log|\ell|\ell^{\mu}$ and $|\ell|$, which precisely give rise to the classical long-range effect after performing the Fourier integral.

\subsubsection{$m_{1} m_{2}^3$ terms}

We first consider $m_{1} m_{2}^3$ sector $\Delta P^{\rho}_{1} \big|_{3}^{m_{1} m_{2}^{3}}$. The integrand for the Fourier integral in this sector is given by $\widetilde{\Delta P}{}^{\rho}_{1} (\ell)\big|_{3}^{(A)}$ defined in \eqref{Fourier_integrand}, which is divided into the three parts
\begin{equation}
  \widetilde{\Delta P}{}^{\rho}_{1} (\ell)\big|_{3}^{(A)}
  =
  m_{1}m_{2}^{3}\Big(
  \overline{\mathcal{P}}^{\rho}_{0}\big|_{0,\ell}
  +
  \overline{\mathcal{P}}^{\rho}_{1}\big|_{0,\ell}
  +
  \overline{\mathcal{P}}^{\rho}_{2}\big|_{0,\ell}\Big) \,.
\label{def_m1m2^3_Fourier_integrand}\end{equation}
Here $\overline{\mathcal{P}}^{\rho}_{i}\big|_{0,\ell}$ is the integrand without mass term, $\big(\mathcal{P}^{\rho}_{i}\big|_{0,\ell}\big)_{m_{1},m_{2}\to 1}$, where $i = 0,1,2$. 

We first derive $\overline{\mathcal{P}}^{\rho}_{0}\big|_{0,\ell}$ given by the 3PM graviton current $\tilde{J}_{\mu\nu}\big|_{3}^{0,\ell}$ only. We obtained the graviton current by evaluating bubble-bubble integrals summarized in \eqref{3PM_tJ_zero_mode}. Using the result, we have
\begin{equation}
\begin{aligned}
  \overline{\mathcal{P}}_{0}\big|_{{0,\ell}}
  &=
  -\frac{i 2^{6 \epsilon -1} \pi^{2 \epsilon} \csc(\pi \epsilon) \sec^2(\pi  \epsilon ) }{\Gamma(\frac{3}{2}-3 \epsilon) \Gamma(\epsilon +\frac{1}{2})} |\ell|^{-4 \epsilon } \ell^{\mu}
  \\& \quad \times
  \Bigg(\frac{ \Gamma(-2 \epsilon ) \mathcal{N}^{m_{1} m_{2}^{3}}_{0,1}}{\Gamma (1-\epsilon ) \Gamma (-\epsilon )}
  -\frac{ \sec (\pi \epsilon) \mathcal{N}^{m_{1} m_{2}^{3}}_{0,2}}{3 \Gamma(1-\epsilon) \Gamma(\epsilon +\frac{3}{2})}
  +\frac{ \sec (\pi \epsilon) \mathcal{N}^{m_{1} m_{2}^{3}}_{0,3} }{3 \Gamma(2-\epsilon) \Gamma(\epsilon +\frac{3}{2})} \Bigg) \,,
\end{aligned}\label{}
\end{equation}
where
\begin{equation}
\begin{aligned}
  \mathcal{N}^{m_{1} m_{2}^{3}}_{0,1}
  &=
  \pi^3 2^{2 \epsilon +1} \,,
  \\
  \mathcal{N}^{m_{1} m_{2}^{3}}_{0,2}
  &=
  4 \pi^{7/2} \gamma ^2 (5 \epsilon +3) \,,
  \\
  \mathcal{N}^{m_{1} m_{2}^{3}}_{0,3}
  &=
  \pi^{7/2} \big(10 \gamma ^2-1+(10 \gamma ^2-3) \epsilon +(2-18 \gamma ^2) \epsilon ^2 \big) \,.
\end{aligned}\label{}
\end{equation}
Expanding in $\epsilon$ and keeping only the terms relevant for the classical contribution,
\begin{equation}
\begin{aligned}
  \overline{\mathcal{P}}_{0}\big|_{{0,\ell}}
  = 
  - \frac{4}{3} i \pi (4\gamma^2-1) \log|\ell| \ell^{\mu} 
  + \mathcal{O}(\epsilon) \,.
\end{aligned}\label{m1m2^3_0Bracket}
\end{equation}

Next, $\overline{\mathcal{P}}^{\rho}_{1}|_{0,\ell}$ is given by a single bracket between 1PM and 2PM $\tilde{J}_{\mu\nu}\big|_{1,2}^{0,\ell}$ and $X^{\mu}_{1}\big|_{2,1}^{0,\ell}$. Substituting the currents summarized in Section \ref{Sec:6} and Appendix \ref{Appendix:2PM} into the convolution bracket, we obtain one-loop triangle integrals as follows:
\begin{equation}
\begin{aligned}
  \overline{\mathcal{P}}_{1}\big|_{{0,\ell}}
  &=
  i 2^{4 \epsilon } \pi ^{\epsilon +3} \sec(\pi\epsilon)
  \Big(\tfrac{32 \gamma^2}{\Gamma (1-\epsilon )}+\tfrac{6 \gamma^2-1+2(1-2\gamma^2) \epsilon}{\Gamma(2-\epsilon)}\Big) 
  \hat{I}^{\mu}_{\frac{1}{2}+\epsilon,1,0}
  \\&\quad
  +\frac{i 2^{4\epsilon} \pi^{\epsilon+3} \sec(\pi\epsilon)}{\Gamma(2-\epsilon)}(2\epsilon-1) 
  \Big((2 \gamma ^2-1)|\ell|^{2} \hat{I}^{\mu}_{\frac{3}{2}+\epsilon,1,0} 
    + 4 \hat{I}^{\mu}_{\frac{3}{2}+\epsilon,1,-2}\Big)
  \\&\quad
  +\frac{i  2^{4 \epsilon } \pi ^{\epsilon +3} \sec (\pi  \epsilon )  }{\Gamma (2-\epsilon )} (2 \gamma ^2-1) (\gamma ^2+3-4 \epsilon) 
  \Big( |\ell|^{2} \hat{I}^{\mu}_{\frac{1}{2}+\epsilon,1,2} - \hat{I}^{\mu}_{-\frac{1}{2}+\epsilon,1,2} \Big) \,.
\end{aligned}\label{}
\end{equation}
where $\hat{I}^{\mu}_{n_{1},n_{2},n_{3}} =  2 I^{\mu}_{n_{1},n_{2},n_{3}}- \ell^{\mu} I_{n_{1},n_{2},n_{3}}$ and
\begin{equation}
\begin{aligned}
  I_{n_{1},n_{2},n_{3}}
  &=
  \int_{k}\frac{ \hat{\delta}(k\cdot v_{1})}{ |k|^{2n_{1}} |\ell-k|^{2n_{2}} \big( k\cdot v_{2} - i \epsilon \big)^{n_{3}} } \,,
  \\
  I^{\mu}_{n_{1},n_{2},n_{3}}
  &=
  \int_{k}\frac{ \hat{\delta}(k\cdot v_{1}) k^{\mu}}{ |k|^{2n_{1}} |\ell-k|^{2n_{2}} \big( k\cdot v_{2} - i \epsilon \big)^{n_{3}} } \,.
\end{aligned}\label{}
\end{equation}
Using the one-loop integral formula, we obtain 
\begin{equation}
\begin{aligned}
  \overline{\mathcal{P}}^{\mu}_{1}\big|_{{0,\ell}}
  &=
  \frac{i  4^{3 \epsilon -2} \pi ^{2 \epsilon +\frac{7}{2}} \csc (\pi  \epsilon ) \sec ^3(\pi  \epsilon )  \mathcal{N}_{1,1}  }{(\gamma ^2-1) \Gamma (\frac{5}{2}-3 \epsilon ) \Gamma (2-\epsilon ) \Gamma (\epsilon +\frac{3}{2})^2} |\ell|^{-4 \epsilon } \ell^{\mu}
  \\&\quad
  -\frac{ 2^{6 \epsilon +1} \pi ^{2 \epsilon +3}  \sin (\pi  \epsilon ) \csc (4 \pi  \epsilon ) \mathcal{N}_{1,2} }{\sqrt{\gamma ^2-1} (\epsilon -1) \Gamma (\frac{3}{2}-3 \epsilon ) \Gamma (\epsilon +\frac{1}{2})}  |\ell|^{1-4 \epsilon } w_{2}^{\mu}
\end{aligned}\label{}
\end{equation}
where
\begin{equation}
\begin{aligned}
  \mathcal{N}_{1,1}^{m_{1} m_{2}^{3}}
  &=
  \frac{4\epsilon^2-1}{2}\Big[\
  2(4+\epsilon-9\epsilon^2)(2\gamma^{2}{-}1)^2 + (6\epsilon-1)(12\epsilon^2-5\epsilon{-}5)(2\gamma^{2}{-}1) 
  \\&\qquad\qquad\quad
  + 12\epsilon^2-7\epsilon-7 \Big] \,,
  \\
  \mathcal{N}_{1,2}^{m_{1} m_{2}^{3}}
  &=
  (2\gamma^{2}-1)(\gamma^2+3 -4\epsilon)\,.
\end{aligned}\label{}
\end{equation}
Expanding in $\epsilon$ and keeping only the terms relevant to the momentum kick, we have
\begin{equation}
\begin{aligned}
  \overline{\mathcal{P}}^{\mu}_{1}\big|_{{0,\ell}}
  &=
    \frac{4 i \pi (16 \gamma^4 - 11\gamma^2 -2) }{ 3(\gamma^2-1) } \log|\ell| \ell^{\mu}
  + \frac{ \pi^2 (2\gamma^{2}-1)(\gamma^{2}+3)}{\sqrt{\gamma^2-1}} |\ell| w^{\mu}_{2} \,.
\end{aligned}\label{m1m2^3_1Bracket}
\end{equation}

We finally consider the two bracket part $\mathcal{P}^{\mu}_{2}\big|_{{0,\ell}}$ consisting of the 1PM currents. After IBP reduction, two loop integrand is written in terms of the master integrals of Family-I
\begin{equation}
  \overline{\mathcal{P}}^{\mu}_{2}\big|_{{0,\ell}}
  =
    \frac{32 i \pi^3 \mathcal{C}^{m_{1} m_{2}^{3}}_{2,1} \ell^{\mu}}{3 (\gamma ^2-1)^2} f_{1}^{\rm I}
  + \frac{512 i \pi^3 \mathcal{C}^{m_{1} m_{2}^{3}}_{2,2} |\ell|^{2} w_{2}^{\mu}}{(\gamma ^2-1) (6 \epsilon -1)} f_{2}^{\rm I}
  + \frac{16 i \pi^3 \mathcal{C}^{m_{1} m_{2}^{3}}_{2,3} |\ell|^{2} \ell^{\mu}}{3 (\gamma ^2-1) (3\epsilon-1)} f_{3,+}^{\rm I}\,,
\label{}\end{equation}
where
\begin{equation}
\begin{aligned}
  \mathcal{C}^{m_{1} m_{2}^{3}}_{2,1}
  &=
  (36\epsilon^2-36\epsilon+11) (2\gamma^2{-}1)^3
  -6(2\gamma^2{-}1)^2 +12(3\epsilon-1)(2\gamma^2{-}1) +6 \,,
  \\
  \mathcal{C}^{m_{1} m_{2}^{3}}_{2,2}
  &=(2 \gamma ^2-1) \epsilon  (-2 \gamma ^2 (\gamma ^2-1)+(1-2 \gamma ^2)^2 \epsilon ) \,,
  \\
  \mathcal{C}_{2,3}^{m_{1} m_{2}^{3}}
  &=(1-2\gamma^2)^3  \epsilon \,.
\end{aligned}\label{}
\end{equation}
By substituting the values of the master integrals
\begin{equation}
\begin{aligned}
  &\overline{\mathcal{P}}^{\mu}_{2}\big|_{{0,\ell}}
  \\
  &=
  \frac{-i\pi ^{2 \epsilon } }{3 (\gamma^2{-}1)^2}
  \bigg[\frac{ \Gamma (1{-}\epsilon )^2 \Gamma (\epsilon{+}\frac{1}{2})\mathcal{N}^{m_{1} m_{2}^{3}}_{2,1}}{\Gamma (2-3\epsilon) \sin(\pi\epsilon)}
  -\frac{\Gamma(\frac{1}{2}{-}\epsilon )^2 \Gamma(-3 \epsilon) \Gamma (\epsilon +1) \mathcal{N}^{m_{1} m_{2}^{3}}_{2,2} }{\Gamma (3-6 \epsilon)\cos(\pi\epsilon )} \bigg] \frac{\ell^{\mu}}{|\ell|^{4 \epsilon}}
  \\&\quad
+\frac{ 2^{6\epsilon+4} \pi ^{2 \epsilon +3}  \csc(4\pi \epsilon) \sin(\pi \epsilon) \mathcal{N}_{2,3} }{(\gamma ^2-1)^{3/2} \Gamma (\frac{3}{2}-3 \epsilon ) \Gamma (\epsilon +\frac{1}{2})} |\ell|^{1-4 \epsilon } w_{2}^{\mu}\,,
\end{aligned}\label{}
\end{equation}
where
\begin{equation}
\begin{aligned}
  \mathcal{N}_{2,1}&= \pi ^{3/2} (2 \gamma ^2-1)^3 2^{6\epsilon } \,,
  \quad
  \mathcal{N}_{2,2}
  = 
  3(3\epsilon{-}1)\, \mathcal{C}^{m_{1} m_{2}^{3}}_{2,1} \,,
  \quad
  \mathcal{N}_{2,3}
  =
  \frac{1}{\epsilon}\mathcal{C}^{m_{1} m_{2}^{3}}_{2,2} \,.
\end{aligned}\label{}
\end{equation}
Expanding in $\epsilon$ and keeping only the terms relevant for the classical contribution,
\begin{equation}
\begin{aligned}
  \overline{\mathcal{P}}^{\mu}_{2}\big|_{{0,\ell}}
  &=
  - \frac{2i \pi (1+18\gamma^2-44\gamma^4+24\gamma^6 ) }{ (\gamma^2{-}1)^2 } \log|\ell| \ell^{\mu}
  - \frac{16 \pi ^2  \gamma^2 (2\gamma^2-1) }{\sqrt{\gamma ^2-1}} |\ell| w_{2}^{\mu} \,.
\end{aligned}\label{m1m2^3_2Bracket}
\end{equation}

Collecting the 0,1 and 2 bracket parts in \eqref{m1m2^3_0Bracket}, \eqref{m1m2^3_1Bracket} and \eqref{m1m2^3_2Bracket} respectively, we obtain $\widetilde{\Delta P}{}^{\mu}_{1} (\ell) \big|_{3}^{(A)}
$ using \eqref{def_m1m2^3_Fourier_integrand}
\begin{equation}\!\!\!\!\!
\begin{aligned}
  &\widetilde{\Delta P}{}^{\mu}_{1} (\ell) \big|_{3}^{(A)}
  \\
  &=
  -m_{1} m_{2}^{3}\bigg[\frac{2i \pi (16 \gamma^2 (\gamma ^2-1)^2-1) }{ (\gamma ^2-1)^2 } \log|\ell| \ell^{\mu}
  + \frac{3 \pi^2 (5\gamma^{2}-1)(2\gamma^{2}-1)}{\sqrt{\gamma ^2-1}} |\ell| w_{2}^{\mu}\bigg] \,.
\end{aligned}\label{}
\end{equation}
The above Fourier integral provides the $m_{1} m_{2}^{3}$ sector of the 3PM momentum kick
\begin{equation}
\begin{aligned}
  \Delta P^{\mu}_{1} \big|_{3}^{m_{1} m_{2}^{3}}
  &=
  \int_{\ell_{2}} e^{i\ell_{2}\cdot b} \hat{\delta}(\ell_{2} \cdot v_{1}) \hat{\delta} (\ell_{2}\cdot v_{2}) \widetilde{\Delta P}{}^{\mu}_{1} (\ell) \big|_{3}^{(A)}
  \\
  &=
  \frac{2 m_{1} m_{2}^3 \big(1-16 \gamma ^2 (\gamma^2{-}1)^2\big)  }{(\gamma^2-1)^{5/2} |b|^4} b^{\mu}
  +\frac{3 \pi m_{1} m_{2}^3 (5\gamma^{2}-1)(2\gamma^{2}-1)}{2 (\gamma ^2-1) |b|^{3}} w_{2}^{\mu} \,.
\end{aligned}\label{}
\end{equation}

\subsubsection{$m_{1}^{3} m_{2}$ terms}
We next present consider $m_{1}^{3} m_{2}$ sector $\Delta P^{\rho}_{1} \big|_{3}^{m_{1}^{3} m_{2}}$, The corresponding the loop integrand of the 2-dimensional Fourier transform is given by $\widetilde{\Delta P}{}^{\rho}_{1} (\ell)\big|^{(B)|m_{1}^{3} m_{2}}_{3}$, which is decomposed into the 
\begin{equation}
  \widetilde{\Delta P}{}^{\rho}_{1} (\ell)\big|^{(B)|m_{1}^{3} m_{2}}_{3} 
  =
  m_{1}^{3} m_{2}\int_{k} \hat{\delta}(k\cdot v_{1})\Big(\overline{\mathcal{P}}_{0}^{\rho}\big|^{m_{1}^{3} m_{2}}_{k-\ell,\ell}
  +
  \overline{\mathcal{P}}_{1}^{\rho}\big|^{m_{1}^{3} m_{2}}_{k-\ell,\ell}
  +
  \overline{\mathcal{P}}_{2}^{\rho}\big|^{m_{1}^{3} m_{2}}_{k-\ell,\ell}\Big)\,,
\label{}\end{equation}
where $\overline{\mathcal{P}}_{i}^{\rho}\big|^{m_{1}^{3} m_{2}}_{k-\ell,\ell}$ are the mass removed from the terms in $\overline{\mathcal{P}}_{i}^{\rho}\big|_{k-\ell,\ell}$ that have $m_{1}^{3} m_{2}$, 
\begin{equation}
  \overline{\mathcal{P}}_{i}^{\rho}\big|^{m_{1}^{3} m_{2}}_{k-\ell,\ell}
  =
  \Big(\mathcal{P}_{i}^{\rho}\big|_{k-\ell,\ell}\big|^{m_{1}^{3} m_{2}}\Big)\Big|_{m_{1},m_{2}\to 1}\,.
\label{}\end{equation}
Since $\tilde{J}_{\mu\nu}\big|_{3}^{\ell_{1},\ell_{2}}$ already contains a loop integral in the definition\footnote{One can easily check that there is an extra loop integral $\int_{k}$ in the mixed mode case $k-\ell,\ell$, which is the second term of \eqref{momentum_kick_1}}, it is convenient to combine $\mathcal{P}_{0}\big|^{m_{1}^3 m_{2}}_{k-\ell,\ell}$ and $\mathcal{P}_{1}\big|^{m_{1}^3 m_{2}}_{k-\ell,\ell}$ and performing the remaining one-loop integral together, which is given by
\begin{naligned}
  &\int_{k}\hat{\delta} \big( k\cdot v_{1}\big)\Big(\overline{\mathcal{P}}_{0}\big|^{m_{1}^3 m_{2}}_{k-\ell,\ell}+\overline{\mathcal{P}}_{1}\big|^{m_{1}^3 m_{2}}_{k-\ell,\ell}\Big)
  \\
  &=
  \frac{i 2^{4 \epsilon} \pi ^{\epsilon +3} \sec (\pi  \epsilon ) }{\Gamma (2-\epsilon ) \ell_{2}^{2}} 
  \bigg[
     4(17 - 18 \epsilon ) I_{1,\epsilon +\frac{1}{2},-2}^{\mu}
    +\big(50 \gamma ^2+5-(44 \gamma ^2+10) \epsilon\big) I_{1,\epsilon -\frac{1}{2},0}^{\mu}
  \\&\qquad\qquad\qquad\qquad\quad
  -\big( 6 \gamma ^2+13 + (4 \gamma ^2-18) \epsilon \big) I_{1,\epsilon +\frac{1}{2},0}^{\mu}
  \\&\qquad\qquad\qquad\qquad\quad
  + (2 \gamma ^2-1) (\gamma ^2+3-4 \epsilon) \Big(I_{1,\epsilon -\frac{1}{2},2}^{\mu} -|\ell|^{2} I_{1,\epsilon +\frac{1}{2},2}^{\mu}\Big)\bigg] \,.
\end{naligned}
Substituting the one-loop formula, we have 
\begin{equation}
\begin{split}
  \int_{k}\hat{\delta} \big( k\cdot v_{1}\big)
  \Big(
     \overline{\mathcal{P}}_{0}\big|^{m_{1}^3 m_{2}}_{k-\ell,\ell}
    +\overline{\mathcal{P}}_{1}\big|^{m_{1}^3 m_{2}}_{k-\ell,\ell}
  \Big)
  &=
  \frac{i 2^{6 \epsilon -3} \pi ^{2 \epsilon +\frac{7}{2}} \csc (2 \pi  \epsilon ) \sec ^2(\pi  \epsilon ) \mathcal{N}^{^{m_{1}^3 m_{2}}}_{1,1} }{(\gamma^2-1) \Gamma (\frac{9}{2}-3 \epsilon ) \Gamma(2-\epsilon ) \Gamma (\epsilon +\frac{1}{2})^2} \frac{\ell_{2}^{\mu}}{|\ell|^{4 \epsilon }}
  \\&\!\!\!
  +\frac{ 2^{6\epsilon} \pi^{2 \epsilon +3} \sin(\pi\epsilon) \csc(4\pi\epsilon) \mathcal{N}^{m_{1}^3 m_{2}}_{1,2}}{\sqrt{\gamma^2-1} (\epsilon -1) \Gamma (\frac{3}{2}-3 \epsilon ) \Gamma (\epsilon +\frac{1}{2})} \frac{w_{1}^{\mu}}{|\ell|^{4 \epsilon -1}}\,.
\end{split}\label{m1^3m2_1bracket}
\end{equation}
where
\begin{naligned}
  \mathcal{N}^{m_{1}^3 m_{2}}_{1,1}
  &=
  \frac{2\epsilon-1}{2}\Big[
  (4-11\epsilon+20\epsilon^2-12\epsilon^3) (2\gamma^2-1)^2
  +(\epsilon-1) ( 24\epsilon^2-58\epsilon+31)
  \\&\qquad\qquad
  + \left(27-498\epsilon+1346\epsilon^2-1308\epsilon^3+432\epsilon^4 \right)(2\gamma^2-1) \Big]
  \\
  \mathcal{N}^{m_{1}^3 m_{2}}_{1,2}
  &=(2 \gamma ^2-1) (\gamma ^2+3-4 \epsilon) \,.
\end{naligned}
Expanding \eqref{m1^3m2_1bracket} in $\epsilon$ and keeping only the terms relevant to the long-range interaction, we have
\begin{equation}
  \frac{4 i \pi ( 27+8\gamma^2 ) }{105 } \log|\ell|\ell^{\mu}
  +\frac{ \pi ^2 (3-5\gamma^2-2\gamma^4) }{2\sqrt{\gamma ^2-1}}|\ell| w_{1}^{\mu}\,.
\label{m1^3m2_1bracket_result}\end{equation}

We next derive the double bracket part, which is given by the double bracket with the 1PM graviton current and two 1PM worldline currents. Substituteing the 1PM currents to the definition in \eqref{momentum_kick_integrand}, we obtain two-loop integrals which belong to the Family-I after IBP reduction
\begin{equation}\!\!\!\!\!\!\!
\begin{aligned}
  &\int_{k}\hat{\delta} \big( k\cdot v_{1}\big)
  \mathcal{P}^{\rho}_{2}\big|^{m_{1}^3 m_{2}}_{k-\ell,\ell}
  =
  \frac{16 i \pi ^3 \mathcal{C}^{m_{1}^3 m_{2}}_{2,1} \ell^{\mu}}{3 (\gamma^2-1)^2 (\epsilon -1) (6 \epsilon-7) (6 \epsilon -5)} f_{1}^{\rm I}
  \\&\qquad\qquad\qquad\quad
  +\frac{32 i \pi ^3 (2\gamma^{2}-1) \mathcal{C}^{m_{1}^3 m_{2}}_{2,2} |\ell|^{2} w_{1}^{\mu}}{(\gamma ^2-1) (\epsilon -1) (6 \epsilon -1) } f_{2}^{\rm I} 
  -\frac{32 i \pi^3 \epsilon (2 \gamma ^2-1)^3 |\ell|^{2} \ell^{\mu}}{6 (\gamma ^2-1) (3 \epsilon -1)} f_{3,+}^{\rm I}\,,
\end{aligned}\label{m13m2_2bracket}
\end{equation}
where
\begin{equation}
\begin{aligned}
  \mathcal{C}^{m_{1}^3 m_{2}}_{2,1}
  &=
  \left(2592\epsilon^5 -10368\epsilon^4 +15996\epsilon^3 -11828\epsilon^2 +4173\epsilon -564 \right) \left(2\gamma^2-1\right)^3
  \\&\qquad
  + \left( 864\epsilon^4 -2880\epsilon^3 +3534\epsilon^2 -1917\epsilon +397 \right)
  \left(2\gamma^2-1\right)^2
  \\&\qquad
  + \left( 1728\epsilon^4 -5724\epsilon^3 +6720\epsilon^2 -3201\epsilon +478 \right)
  \left(2\gamma^2-1\right)
  \\
  &\qquad
  + (\epsilon-1) \left(240\epsilon^2-490\epsilon+241\right)\,,
  \\
  \mathcal{C}_{2,2}^{m_{1}^3 m_{2}}&=\epsilon  \Big(31 \gamma ^4 -34 \gamma ^2 + 3 -4 \big(24 \gamma ^4-25 \gamma ^2+5\big) \epsilon +16 \big(2 \gamma^2-1\big)^2 \epsilon ^2 \Big) \,.
\end{aligned}\label{}
\end{equation}
Sbstituting the famil-I master integrals presented in Section \ref{Sec:5.1}, we have
\begin{equation}
\begin{aligned}
  &\frac{i \pi ^{2 \epsilon } \sin (\pi  \epsilon ) }{(\gamma ^2-1)^2}
  \Bigg[ \frac{ \csc (2 \pi  \epsilon ) \Gamma (\frac{1}{2}-\epsilon )^2 \Gamma (-3 \epsilon ) \Gamma (\epsilon +1) \, \mathcal{N}^{m_{1}^3 m_{2}}_{2,1}}{(\epsilon -1) (6 \epsilon -7) (6 \epsilon -5) \Gamma (3-6 \epsilon )}
  \\&\qquad\qquad\quad
  + \frac{ 2^{6\epsilon} \pi^{\frac{3}{2}} \Gamma (1-\epsilon ) \Gamma (2-\epsilon) \Gamma (\epsilon +\frac{1}{2})(2 \gamma ^2-1)^3 (-2 + 3 \epsilon) }{\Gamma(4-3 \epsilon )\sin^2(\pi\epsilon) } 
  \Bigg] |\ell|^{-4 \epsilon} \ell_{2}^{\mu}
  \\&
  -\frac{ 2^{6 \epsilon -2} \pi^{2 \epsilon +3} \sin(4\pi\epsilon) \mathcal{N}^{m_{1}^3 m_{2}}_{2,2}}{(\gamma ^2-1)^{3/2} (\epsilon -1) \sin (\pi\epsilon) \Gamma (\frac{3}{2}-3 \epsilon) \Gamma(\epsilon +\frac{1}{2})} |\ell|^{1-4\epsilon} w_{1}^{\mu}\,,
\end{aligned}\label{m1^3m2_2bracket_evaluated}
\end{equation}
where
\begin{equation}
\begin{aligned}
  \mathcal{N}^{m_{1}^3 m_{2}}_{2,1}
  =
  (3\epsilon-1)\mathcal{C}^{m_{1}^3 m_{2}}_{2,1}\,,
  \qquad
  \mathcal{N}_{2,2}^{m_{1}^3 m_{2}}
  =
  \frac{1}{\epsilon}(2 \gamma^2-1)\mathcal{C}^{m_{1}^3 m_{2}}_{2,2} \,.
\end{aligned}\label{}
\end{equation}
Expanding \eqref{m1^3m2_2bracket_evaluated} in $\epsilon$ and keeping only the terms relevant for the classical contribution,
\begin{equation}
  \frac{2i \pi ( 51-1588\gamma^2+3338\gamma^4-1696\gamma^6 ) }{105(\gamma ^2-1)^2 } \log|\ell|\ell^{\mu}
  +\frac{\pi^2 (2\gamma^{2}{-}1)(31\gamma^{2}{-}3)}{2\sqrt{\gamma ^2-1}} |\ell| w_{1}^{\mu} \,.
\label{m1^3m2_2bracket_result}\end{equation}

Let us add all the contributions in \eqref{m1^3m2_1bracket_result} and \eqref{m1^3m2_2bracket_result}
\begin{equation}
\begin{aligned}
  &\widetilde{\Delta P}{}^{\rho}_{1} (\ell)\big|^{(B)|m_{1}^{3} m_{2}}_{3} 
  \\
  &=
  m_{1}^{3} m_{2}\Bigg[\frac{2i\pi (1-16 \gamma^2 (\gamma^2-1)^2) }{(\gamma^2-1)^2 } \log|\ell|\ell^{\mu}
  +\frac{3 \pi^2 (5\gamma^2-1)(2\gamma^2-1)}{\sqrt{\gamma^2-1}}|\ell|^{2} w_{1}^{\mu} \Bigg]\,.
\end{aligned}\label{}
\end{equation}
Evaluating the Fourier integral, we obtain 
\begin{equation}
\begin{aligned}
  \Delta P^{\mu}_{1}\big|_{3}^{m_{1}^{3} m_{2}} 
  &=
  -\frac{2 \big(16 \gamma ^2 (\gamma ^2-1)^2-1\big) }{(\gamma ^2-1)^{5/2} |b|^4} b^{\mu}
  -\frac{3 \pi (5\gamma^2-1)(2\gamma^2-1)}{2 (\gamma^2-1) |b|^{3}} w_{1}^{\mu} \,.
\end{aligned}\label{}
\end{equation}
%

\subsection{The last pieces: $m_{1}^2 m_{2}^2$ terms}

We finally consider the $m_{1}^{2} m_{2}^{2}$ sector $\Delta P^{\rho}_{1} \big|_{3}^{m_{1}^{2} m_{2}^{2}}$. While only family-I integrals appeared in the previous two sectors, the $m_{1}^{2}m_{2}^{2}$ sector involves family-II and what we can call mushroom integrals. As discussed in Section~\ref{Sec:5}, Family-II integrals are divided into the potential and radiation regions, and we evaluate the corresponding integrals by the method of regions. Since there are no iterated integrals in the $m_{1}^{2}m_{2}^{2}$ sector, it is not necessary to decompose it into three parts according to the bracket structure as in the previous two cases. Instead, we treat the full $\widetilde{\Delta P}{}^{\rho}_{1}(\ell)\big|^{(B)|m_{1}^{2} m_{2}^{2}}_{3}$ at once. 

Let us consider the Family-II part. Following Section~\ref{Sec:5}, however, we do separate into the even and odd sectors. After performing the IBP reduction, the even sector of $\widetilde{\Delta P}{}^{\rho}_{1}(\ell)\big|^{(B)|m_{1}^{2} m_{2}^{2}}_{3}$ is
\small{%
\begin{equation}
\begin{aligned}
  &
  \Bigg[
    \frac{4(2\gamma^2-1)}{(\gamma^2-1)\epsilon^2}
  + \frac{7-176\gamma^2}{3(\gamma^2-1)\epsilon}
  + \frac{2(382+751\gamma^2-24\gamma^4+12\gamma^6)}{9(\gamma^2-1)}
  \Bigg] \ell^{\mu} f^{\rm II,e}_{1}
  \\&
  + \frac14(7+15\gamma^2) \ell^{\mu} |\ell|^2 f^{\rm II,e}_{2}
  + 4(5+\gamma^2) \ell^{\mu} f^{\rm II,e}_{3} 
  \\&
  -\Bigg[
    \frac{2(1+\gamma^2)}{\epsilon}
   -\frac{11{-}328\gamma^2{+}440\gamma^4{-}128\gamma^6}{6(\gamma^2-1)}
  \Bigg] \ell^{\mu} |\ell|^2 f^{\rm II,e}_{5}
  \\&
  -
  \Bigg[
   \frac{2}{\epsilon^2}
   +\frac{11{-}44\gamma^2{+}40\gamma^4{-}8\gamma^6}
          {6(\gamma^2-1)^2\epsilon}
   -\frac{2(58+133\gamma^2-431\gamma^4+246\gamma^6)}{9(\gamma^2-1)^2}
  \Bigg]
  \ell^{\mu} |\ell|^2 f^{\rm II,e}_{4}  
  \\&
  + \Bigg[
   \frac{2}{\epsilon}
   +\frac{2(6{-}11\gamma^2{+}6\gamma^4)}{3}
  \Bigg]
  \ell^{\mu} |\ell|^4 f^{\rm II,e}_{6}
  -
  \Bigg[
   \frac{2(\gamma^2{-}1)}{\epsilon}
   -\frac{(\gamma^2{-}1)(9{+}22\gamma^2)}{3}
  \Bigg] 
  \ell^{\mu} |\ell|^6 f^{\rm II,e}_{7}
  \\&
  +\frac{\gamma(1-2\gamma^2)^2(3-2\gamma^2)}{3(\gamma^2-1)}
  \ell^{\mu} |\ell|^2 \Big(f^{\rm II,e}_{8,+}+\frac{1}{2}f^{\rm II,e}_{8,-}\Big)\,,
\end{aligned}\label{even_integrand}
\end{equation}
}
and the odd sector is
{\small
\begin{equation}
\begin{aligned}
  & \Bigg[\frac{8\gamma^6 +24\gamma^4 -30\gamma^2+9}{6} f^{\rm II,o}_{1}
    + \frac{3}{2}\epsilon\,\mathcal (2\gamma^2-1)(5\gamma^2-1)
  \left(f^{\rm II,o}_{\bar 4}-f^{\rm II,o}_{4}\right)
  \Bigg] |\ell|^2 w_1^\mu
  \\&
  + \Bigg[\frac{(\gamma-1)(15\gamma^2-15\gamma+4)}{4\epsilon}\left(f^{\rm II,o}_{1}-f^{\rm II,o}_{2}\right)
  \\&\quad
  +\frac{40\gamma^8+40\gamma^7+80\gamma^6+80\gamma^5 - 321\gamma^4 + 345\gamma^3-523\gamma^2+439\gamma-100}{24(\gamma+1)} f^{\rm II,o}_{1}
  \\&\quad
  +\frac{35\gamma^7+45\gamma^6 -183\gamma^5 +88\gamma^4 -319\gamma^3 +1085\gamma^2-1133\gamma+398}{16(\gamma^2-1)} f^{\rm II,o}_{2}
  \\&\quad
  + \frac{\gamma-1}{128\epsilon}
  \Big[ \epsilon(\gamma-1)^2(55\gamma^2-14\gamma+7)
  \\&\hspace{3.3cm}
  -2(35\gamma^4-60\gamma^3+90\gamma^2-76\gamma+27)
  \Big] |\ell|^4 f^{\rm II,o}_{3}
  \\&\quad
  +\left[3\epsilon(2\gamma^2-1)(5\gamma^2-1)
  -\frac{\mathcal K(\gamma,\epsilon)}{64(\gamma^2-1)}
  \right] f^{\rm II,o}_{4}
  -\frac{\mathcal K(\gamma,\epsilon)}{64(\gamma^2-1)} f^{\rm II,o}_{\bar 4}
  \Bigg] |\ell|^2 w_2^\mu\,,
\end{aligned}\label{odd_integrand}
\end{equation}
}
where
\begin{equation}
\begin{aligned}
\mathcal K(\gamma,\epsilon)
&=
\gamma\Big[
-2(2\gamma^2-3)(35\gamma^4-30\gamma^2+11)
+\epsilon(390\gamma^6-749\gamma^4+376\gamma^2+47)
\Big]\,.
\end{aligned}
\end{equation}
where 
\begin{equation}
\begin{aligned}
  f^{\rm II,P,o}_{\bar{4}}
  &=
  - f^{\rm II,P,o}_{4}\,, 
  &\qquad
  f^{\rm II,R,o}_{\bar{4}}
  &=
  f^{\rm II,R,o}_{4}\,,
  \\
  f^{\rm II,P,e}_{8,-}
  &=
  -2f^{\rm II,P,e}_{8,+}\,.
  &\qquad
  f^{\rm II,R,e}_{8,-}
  &=
  f^{\rm II,R,e}_{8,+} \,,
\end{aligned}\label{}
\end{equation}

According to the method of regions, the master integrals can be divided into the potential and radiation regions
\begin{equation}
  f^{\rm II, e/o}_{i} = f^{\rm II,P, e/o}_{i} + f^{\rm II,R, e/o}_{i}\,,
\label{}\end{equation}
where $f^{\rm II,P, e/o}_{i}$ and $f^{\rm II,R, e/o}_{i}$ are the master integrals of the potential and radiation regions respectively. Substituting the integral values for each region and expanding the result in $\epsilon$, we derive $\widetilde{\Delta P}{}^{\mu}_{1}(\ell)\big|^{(B)|m_{1}^{2} m_{2}^{2}}_{3}$.

The potential region contribution of the even and odd sectors, \eqref{even_integrand} and \eqref{odd_integrand}, is
\begin{equation}\!\!
\begin{aligned}
  &\frac{\log|\ell|\ell^{\mu}}{48\pi^2 (\gamma^2{-}1)^{2}} \Big[ \gamma(53{-}120\gamma^2{+}90\gamma^4{-}20\gamma^6)
  -6(1-\gamma^2)^2(3+12\gamma^2-4\gamma^4)\tfrac{\cosh^{-1}\gamma}{\sqrt{\gamma^{2}-1}}
  \Big] 
  \\&
  -\frac{3i}{64\pi \sqrt{\gamma^2-1}}(1-7\gamma^2+10\gamma^4) |\ell| (w_{1}^{\mu}-w_{2}^{\mu}) \,.
\end{aligned}\label{MK_m12m22_con}
\end{equation}
and the radiation region contribution is
\begin{equation}
\begin{aligned}
&
\frac{(1{-}2\gamma^2)^2\log|\ell| \ell^{\mu}}{48 \pi^2 (\gamma^2-1)^{\frac{3}{2}}} \Big[ (5\gamma^2{-}8)+3\gamma(3{-}2\gamma^2) \tfrac{\cosh^{-1}\gamma}{\sqrt{\gamma^{2}-1}} \Big] 
  -\frac{i|\ell| w_{2}^{\mu}}{192\pi} \big(9{-}30\gamma^2{+}24\gamma^4{+}8\gamma^6\big)
  \\&
  +\frac{i |\ell| w_{1}^{\mu}}{1536\pi (\gamma^2-1)}
  \bigg[
    {-}1151{+}3388\gamma{-}3148\gamma^2{+}656\gamma^3{-}339\gamma^4{+}916\gamma^5{-}210\gamma^6
  \\&\qquad\qquad\qquad\qquad
  {-}80\gamma^7{-}80\gamma^9+3\gamma(3-2\gamma^2)(11-30\gamma^2+35\gamma^4) \tfrac{\cosh^{-1}\gamma}{\sqrt{\gamma^{2}-1}}
  \\&\qquad\qquad\qquad\qquad
  +6(1-\gamma^2)(5-76\gamma+150\gamma^2-60\gamma^3-35\gamma^4) \log \left(\tfrac{1+\gamma}{2}\right)
  \bigg] \,.
\end{aligned}\label{MK_m12m22_diss}
\end{equation}

We next consider the radiation-integral part. After IBP reduction, the integrand reduces to
\begin{equation}
\begin{aligned}
  &\frac{1-2\gamma^{2}}{(3-2\epsilon)|\ell|^{2}}\ell^{\mu}
  \Big[
      M_{0,1,1,-2,2}-M_{1,1,0,-2,2} -2\gamma^{2}|\ell|^{2} M_{1,1,1,-2,2}
  \\&\qquad\qquad\qquad
    + 2(1-3\gamma^2)\big(M_{0,1,1,-1,1}-M_{1,1,0,-1,1}\big)
    -2(1-\gamma^{2})M_{1,1,1,-1,1}
  \Big]
  \\&
  + (1-2\gamma^{2})^{2}W^{\mu} \Big[
    2\big(M_{0,1,1,0,1}+M_{1,1,0,0,1} -|\ell|^{2} M_{1,1,1,0,1} \big)
  \\&\qquad\qquad\qquad\qquad
    + \tfrac{1}{2(3-2\epsilon)} \big(M_{0,1,1,-1,2}+M_{1,1,0,-1,2}-|\ell|^{2}M_{1,1,1,-1,2} \big)
  \Big]
  \\&
  + \frac{4}{3-2\epsilon} \Big[
   \big(3-2\epsilon
       -2( 6-5\epsilon)\gamma^{2}
       +4(1-\epsilon)\gamma^{4}\big)W^{\mu}
       +2\gamma w_2^{\mu} \Big] M_{1,1,1,-2,1}
  \\&\quad
  +\frac{2}{3-2\epsilon}\Big[
   (2-\epsilon)(1-4\gamma^{4}) w_1^{\mu}
   +\gamma
     \big(1-\epsilon+4\gamma^{2}  -4(3-\epsilon)\gamma^{4}\big) w_2^{\mu}
  \Big] M_{1,1,1,-3,2} \,.
\end{aligned}\label{Integrand_mush}
\end{equation}
where $W^{\mu}=w_1^{\mu}+\gamma w_2^{\mu}$.
Substituting the values of the radiation integrals derived in Section \ref{sec:mushroom} and expanding in $\epsilon$, we have 
\begin{equation}
\begin{aligned}
  \frac{i}{384\pi} 
    \Big[2(9-30\gamma^2+24\gamma^4+8\gamma^6) w_{2}^{\mu}
      + \gamma(13-51\gamma^2+40\gamma^4+20\gamma^6) w_{1}^{\mu} 
    \Big] |\ell|\,.
\end{aligned}\label{MK_m12m22_mush}
\end{equation}

We now collect all the previous results. First, what we can call the conservative sector of $\Delta P^{\rho}_{1} \big|_{3}^{m_{1}^{2}m_{2}^{2}}$ is derived by the 2-dimensional Fourier transform of \eqref{MK_m12m22_con}
\begin{equation}
\begin{aligned}
  &\Delta P^{\mu}_{1} \big|_{3, \rm conservative}^{m_{1}^{2}m_{2}^{2}}
  \\&=
  -\frac{8 m_{1}^{2}m_{2}^{2} \cosh^{-1}\!\gamma(3 + 12 \gamma ^2 -4\gamma^4) b^{\mu}}{(\gamma ^2-1) |b|^4}
  +\frac{4 m_{1}^{2}m_{2}^{2} \gamma  (53-120 \gamma ^2+90 \gamma ^4-20 \gamma ^6) b^{\mu}}{3 (\gamma ^2-1)^{5/2} |b|^4}
  \\&\quad
  - \frac{3 \pi m_{1}^{2}m_{2}^{2} (1-7 \gamma^2+10 \gamma^4)   (w_{1}^{\mu}-w_{2}^{\mu}) }{2 (\gamma^2 -1) |b|^{3}} \,.
\end{aligned}\label{}
\end{equation}
Similarly, the dissipative sector of $\Delta P^{\mu}_{1} \big|_{3}^{m_{1}^{2}m_{2}^{2}}$ is given by the 2-dimensional Fourier transform of the sum of \eqref{MK_m12m22_diss} and \eqref{MK_m12m22_mush}
\begin{equation}
\begin{aligned}
  &\Delta P^{\mu}_{1} \big|_{3, \rm dissipative}^{m_{1}^{2}m_{2}^{2}}
  \\
  &= m_{1}^{2} m_{2}^{2}
\frac{4(2\gamma^{2}-1)^{2}}{3(\gamma^{2}-1)^{\frac{5}{2}}}
  \Big[
  \sqrt{\gamma^{2}-1}(5\gamma^{2}-8) +3\gamma(3-2\gamma^{2})\cosh^{-1}\!\gamma
  \Big] \frac{b^{\mu}}{|b|^{4}}
  \\&\quad
  +\frac{m_{1}^{2} m_{2}^{2}\pi}{48(\gamma^{2}-1)^{\frac{3}{2}}}
\Big[
-210\gamma^{6}+552\gamma^{5}-339\gamma^{4}
+912\gamma^{3}-3148\gamma^{2}+3336\gamma-1151
\\
&\qquad\qquad\qquad\qquad
+6(35\gamma^{6}+60\gamma^{5}-185\gamma^{4}
+16\gamma^{3}+145\gamma^{2}-76\gamma+5)
\log \big(\tfrac{1+\gamma}{2}\big)
\\
&\qquad\qquad\qquad\qquad
-3\gamma(70\gamma^{6}-165\gamma^{4}
+112\gamma^{2}-33)\tfrac{\cosh^{-1}\gamma}{\sqrt{\gamma^{2}-1}}
\Big]
\frac{w_{1}^{\mu}}{|b|^{3}}  \,.
\end{aligned}\label{}
\end{equation}

\subsubsection{Adding up all pieces}
If we, rather arbitrarily, define the total conservative sector by the sum of  $\Delta P^{\mu}_{1} \big|_{3}^{m_{1}^{1}m_{2}^{3}}$, $\Delta P^{\mu}_{1} \big|_{3}^{m_{1}^{3}m_{2}^{1}}$ and $\Delta P^{\mu}_{1} \big|_{3, \rm conservative}^{m_{1}^{2}m_{2}^{2}}$ pieces, we get
\begin{equation}
\begin{aligned}
  &\Delta P^{\mu}_{1} \big|_{3}^{\rm conservative}
  \\
  &=
  -\frac{2 m_{1} m_{2} (m_{1}^{2}+m_{2}^{2}) (16 \gamma ^2 (\gamma ^2-1)^2-1)}{(\gamma^2-1)^{5/2} |b|^4} b^{\mu}
  \\&\quad
  -\frac{8 m_{1}^{2}m_{2}^{2} \cosh^{-1}\!\gamma (3 + 12\gamma^2 -4\gamma^4) b^{\mu}}{(\gamma ^2-1) |b|^4}
  +\frac{4 m_{1}^{2}m_{2}^{2} \gamma (53-120 \gamma^2+90 \gamma ^4-20 \gamma ^6) b^{\mu}}{3 (\gamma ^2-1)^{5/2} |b|^4}
  \\&\quad
  -\frac{3 \pi m_{1} m_{2} (5\gamma^{2}-1)(2\gamma^{2}-1)}{2 (\gamma^2-1) |b|^{3}} 
  \Big[
    \big( m_{1}^{2} w_{1}^{\mu} - m_{2}^{2} w_{2}^{\mu} \big)
    + m_{1}m_{2} (w_{1}^{\mu}-w_{2}^{\mu}) 
  \Big] \,.
\end{aligned}\label{3PMcons}
\end{equation}
Note that this so-defined conservative part of the 3PM momentum kick is invariant under the exchange $m_{1} \leftrightarrow m_{2}$ and $w_{2}^{\mu} \leftrightarrow - w_{1}^{\mu}$.

The only remaining piece is then what we using the terminology introduced above could call the dissipative part to this order, and which is responsible for the subtle radiation back-reaction
\begin{equation}
\begin{aligned}
  &\Delta P^{\mu}_{1} \big|_{3}^{\rm dissipative}
  \\
  &= m_{1}^{2} m_{2}^{2}
\frac{4(2\gamma^{2}-1)^{2}}{3(\gamma^{2}-1)^{\frac{5}{2}}}
  \Big[
  \sqrt{\gamma^{2}-1}(5\gamma^{2}-8) +3\gamma(3-2\gamma^{2})\cosh^{-1}\!\gamma
  \Big] \frac{b^{\mu}}{|b|^{4}}
  \\&\quad
  +\frac{m_{1}^{2} m_{2}^{2}\pi}{48(\gamma^{2}-1)^{\frac{3}{2}}}
  \Big[
    -210\gamma^{6}+552\gamma^{5} -339\gamma^{4} +912\gamma^{3} -3148\gamma^{2} +3336\gamma -1151
  \\&\qquad\qquad\qquad\qquad
  +6(35\gamma^{6}+60\gamma^{5}-185\gamma^{4}
  +16\gamma^{3}+145\gamma^{2}-76\gamma+5)
  \log \big(\tfrac{1+\gamma}{2}\big)
  \\&\qquad\qquad\qquad\qquad
  -3\gamma(70\gamma^{6}-165\gamma^{4}
  +112\gamma^{2}-33)\tfrac{\cosh^{-1}\gamma}{\sqrt{\gamma^{2}-1}}
  \Big]
  \frac{w_{1}^{\mu}}{|b|^{3}} \,.
\end{aligned}\label{3PMdiss}
\end{equation}  
Note that the dissipative part of the 3PM momentum kick does not have terms that are proportional to $w_{2}^{\mu}$.  The sum of eqs. \eqref{3PMcons} and \eqref{3PMdiss} agrees
with the known result in the literature, see $.e.g.$ ref. \cite{Dlapa:2023hsl} for a comprehensive review.

\section{Conclusion}

We have provided a set of algebraic relations that solve the Einstein equations of motion and non-spinning matter equations of motion order by order in Newton's constant $G$ and expanded around the flat Minkowski metric. This strategy, the post-Minkowskian expansion computed entirely from the Einstein equations of motion coupled to matter sources, is of course the original proposal from general relativity. New developments from scattering amplitude and worldline approaches could have seemed to leave this original program behind. 
It appears that the
reason it did not move beyond second post-Minkowskian order for black hole scattering was due to sheer computational complexity involved in solving the equations. Much has been learned since, and our aim here
has been to show that this original formulation of the post-Minkowskian expansion is ready for a revival. There are many reasons for this. If we consider the milestone of
Westpfahl's computation to second post-Minkowskian order from 1985, it is clear that symbolic manipulation programs would be needed to go beyond his remarkable calculation at that time. Other developments were also slow in being introduced into general relativity, including the use of dimensional regularization. The rewriting of the Einstein-Hilbert
action in forms more suitable for an expansion in metric perturbations is another crucial ingredient. But clearly the most important new ingredient is the significant progress in
evaluation of momentum integrals required for this problem.

Our suggestion is that we may be able to profit from the number of recent theoretical developments by combining them in the framework of the original classical equations of
motion of general relativity. The advantages are the following. First, the order-by-order solution in an expansion in $G$ is based on recursion relations that build new orders from combinations of lower orders. In the metric perturbation sector this is particularly striking because the needed momentum integrals literally become iterative. For the geodesic
equation the solution is still recursive, but the iterations based on combinations of lower-order pieces become more complex because the momentum integrals are not 
straightforwardly iterative. Nevertheless, our organization of momentum integrals differs from that currently used by amplitude and worldline methods, thus providing a new 
approach. Second, the choice of variables has been found to be of crucial importance. At the core of our approach is the doubling of metric degrees of freedom at intermediate steps
to bypass the need of expanding the inverse metric. The choice of harmonic gauge, already well known to simplify the post-Minkowskian expansion in general relativity is another
example.

Finally, the virtue of using the equations of motion as the only input, is that it does away with effective field theory notions that can overshadow the simplicity involved in solving
classical differential equations. As we have shown, the equations of motion can be written down in arbitrary-order form, and there are only algebraic equations to be solved: no
diagrams are needed. A symbolic manipulation program can straightforwardly provide the fixed-order equations, and they becomes automatically expressed in recursive form. The same can be achieved in other approaches (also they can be automatized, albeit with some effort) but here it is built in from the outset.

The formulation provided in this paper is very likely not the simplest possible. Our set-up is complete: the precise solution is provided to arbitrarily high order in the expansion.  A next step is to attack the current bottleneck, and simplify the integrations as much as possible. A clear aim is to search for more factorized forms of the integrals; some examples have been given in this paper but it should be explored further. We have here pursued our program only up to (and including) third post-Minkowskian order for two
reasons: (1) from fourth post-Minkowskian order the needed integration techniques become significantly more involved, just as in worldline and amplitude-based approaches
\cite{Dlapa:2022lmu,Damgaard:2023ttc,Jakobsen:2023hig}, and (2) the third post-Minkowskian order is sufficient to demonstrate that the coupling of the
matter fields to the gravitational field is correctly taken into account by solving the differential equations with causal Green functions. Radiation back-reaction occurs when
Green functions can become singular, and thus in need of a proper $i\epsilon$-prescription, here that of the retarded Green functions. In this way, at the most fundamental level of the equations of motion of general relativity there is no separation into conservative and dissipative dynamics, just as in the amplitude and wordline approaches.

In the Introduction we stressed that in post-Minkowskian perturbation theory the scattering to any finite order in $G$ does not correspond to the scattering of black holes. but only to
finite-order approximations of those. For scatterings far from the horizon scales the picture of two separate black holes interacting gravitationally should become a better and
better approximation as the order of perturbation theory increases. For closer encounters, and when the energy contained in the gravitational field increases and becomes even comparable to the masses of the original scatterers, it seems less
clear how perturbation theory correctly will account for the space-time interpretation in terms of scattering of two fixed-mass objects that radiate to infinity. 
There are thus still conceptual issues to clarify, and it may be that this is best achieved through comparisons with the methods of numerical general relativity. These are
interesting problems, but such work is left for the future.

\acknowledgments
The work of KL is supported by the National Research Foundation of Korea(NRF) grant funded by the Korea government(MSIT) RS-2025-24533223 and KIAS Individual Grant QP106001. 

\newpage
\appendix


\section{Useful Integrals}
\subsection{Fourier integral}

To express the momentum kick $\Delta P$, obtained from the loop integrals of the currents, in terms of the impact parameter $b^{\mu}$, we need to compute two-dimensional Fourier integrals such as \eqref{Fourier_integrand}. The scalar and vector integrals are required for this as
\begin{equation}
\begin{aligned}
  \int \frac{d^{D} \ell}{(2\pi)^{D}} e^{i \ell \cdot b} \hat{\delta}( \ell \cdot v_{1} ) \hat{\delta}( \ell \cdot v_{2} ) |\ell|^{n}
  &=
  \frac{2^{n}}{\pi^{\frac{D-2}{2}}\sqrt{\gamma^2-1} }\frac{\Gamma(\frac{D-2+n}{2})}{\Gamma(-\frac{n}{2})} \frac{1}{ |b|^{D-2+n}}\,,
  \\
  \int \frac{d^{D} \ell}{(2\pi)^{D}} e^{i \ell \cdot b} \hat{\delta}( \ell \cdot v_{1} ) \hat{\delta}( \ell \cdot v_{2} ) |\ell|^{n} \ell^{\mu}
  &=
  \frac{i 2^{n+1}}{\pi^{\frac{D-2}{2}}\sqrt{\gamma^2-1} }\frac{\Gamma(\frac{D+n}{2})}{\Gamma(-\frac{n}{2})} \frac{b^{\mu}}{ |b|^{D+n}}\,.
\end{aligned}\label{}
\end{equation}
For 2PM order, the necessary integrals are
\begin{equation}
\begin{aligned}
  \int \frac{d^{D} \ell}{(2\pi)^{D}} e^{i \ell \cdot b} \hat{\delta}( \ell \cdot v_{1} ) \hat{\delta}( \ell \cdot v_{2} ) |\ell|^{\delta} 
  &=
  \frac{2^{\epsilon } \Gamma \big(\frac{\delta+2}{2}\big)}{\pi  \sqrt{\gamma ^2-1}\, \Gamma \big(-\frac{\delta}{2}\big)} |b|^{-\delta -2}\,,
  \\
  \int \frac{d^{D} \ell}{(2\pi)^{D}} e^{i \ell \cdot b} \hat{\delta}( \ell \cdot v_{1} ) \hat{\delta}( \ell \cdot v_{2} ) |\ell|^{-1} \ell^{\mu}
  &=
  \frac{ib^{\mu}}{2 \pi \sqrt{\gamma^2-1} |b|^3}\,,
\end{aligned}\label{}
\end{equation}
Taking $D=4$ and $\delta\to 0$ limitm the first integral reduces 
\begin{equation}
  \int \frac{d^{4} \ell}{(2\pi)^{4}} e^{i \ell \cdot b} \hat{\delta}( \ell \cdot v_{1} ) \hat{\delta}( \ell \cdot v_{2} ) \log|\ell|
  =
  -\frac{1 }{2 \pi \sqrt{\gamma ^2-1} |b|^{2}} 
\label{}\end{equation}
For 3PM order, the necessary integrals are
\begin{equation}
\begin{aligned}
  \int \frac{d^{4} \ell}{(2\pi)^{4}} e^{i \ell \cdot b} \hat{\delta}( \ell \cdot v_{1} ) \hat{\delta}( \ell \cdot v_{2} ) |\ell| 
  &=
  - \frac{1}{2 \pi \sqrt{\gamma^2-1} |b|^{3} }\,,
  \\
  \int \frac{d^{4} \ell}{(2\pi)^{4}} e^{i \ell\cdot b} \hat{\delta}( \ell \cdot v_{1} ) \hat{\delta}( \ell \cdot v_{2} ) |\ell|^{\delta} \ell^{\mu}
  &=
  \frac{i2^{\delta}(2+\delta)}{\pi \sqrt{\gamma^2-1}} \frac{\Gamma(\frac{2+\delta}{2})}{\Gamma(-\frac{\delta}{2})}\frac{b^{\mu}}{|b|^{4+\delta} }\,.
\end{aligned}\label{}
\end{equation}
Taking $D=4$ and $\delta \to 0$ limit, the second equation becomes
\begin{equation}
  \int \frac{d^{4} \ell}{(2\pi)^{4}} e^{i \ell\cdot b} \hat{\delta}( \ell \cdot v_{1} ) \hat{\delta}( \ell \cdot v_{2} ) \log|\ell| \ell^{\mu}
  =
  -\frac{i }{\pi \sqrt{\gamma^2-1} |b|^{4} } b^{\mu}\,.
\label{}\end{equation}
%

\subsection{1-loop integrals}
We now consider the $ D$-dimensional one-loop triangle integral
\begin{equation}
  I^{D}_{n_{1},n_{2},n_{3}}
  = 
  \int\frac{\mathrm{d}^{D} k}{(2\pi)^{D}} 
  \frac{\hat{\delta}(k\cdot v_{2})}{|k|^{2n_{1}}|\ell-k|^{2n_{2}} \big( v_{1} \cdot k - i0_{+}\big)^{n_{3}}} \,.
\label{}\end{equation}
If we require the orthogonality condition $\ell\cdot v_{1} = 0$, it is straightforward to evaluate the integral using the Schwinger parametrization as
\begin{equation}
\begin{aligned}
  I^{D}_{n_{1},n_{2},n_{3}}
  &=
  \frac{i^{n_{3}}
  \Gamma \big(\frac{D-1-n_{3}}{2}- n_{1}\big) \Gamma \big(\frac{D-1-n_{3}}{2}- n_{2}\big)
  \Gamma \big(n_{1}+n_{2}+\tfrac{n_{3}}{2}-\tfrac{D-1}{2}\big)
  }{
  2^{D-1} \pi^{\frac{D-2}{2}}
  \Gamma (n_{1}) \Gamma (n_{2}) \Gamma \big(\frac{n_{3}+1}{2}\big) \Gamma(D-1-n_{1}-n_{2}-n_{3})} 
  \\&\quad \times
  (\gamma ^2-1)^{-\frac{n_{3}}{2}} |\ell|^{D-1-2n_{1}-2n_{2}-n_{3}}
\end{aligned}\label{triangle_integral}
\end{equation}
Taking the $n_{3}\to 0$ limit, we obtain the bubble integral formula.
\begin{equation}
  I^{D}_{n_{1},n_{2},0} =
  \frac{
  \Gamma \big(\frac{D -1}{2}-n_{1}\big)\Gamma \big(\frac{D -1}{2}-n_{2}\big) \Gamma \big(n_{1}+n_{2}-\frac{D -1}{2}\big)
  }{
  2^{D -1} \pi ^{\frac{D -1}{2}} \Gamma (n_{1}) \Gamma (n_{2}) \Gamma (D-1 -n_{1}-n_{2})
  }
  |\ell|^{D-1- 2n_{1} -2n_{2}}\,.
\label{bubble_integral}\end{equation}
If we discard the orthogonality condition, $\ell \cdot v_{1} \neq 0$, there is no known closed formula. However, we can derive a closed formula when $n_1 = 0$,
\begin{equation}
  I^{D}_{0,n_{2},n_{3}}
  =
  (\gamma^{2}-1)^{n_{2}-\frac{d}{2}}\frac{i^{d-2n_{2}} \Gamma\big(\frac{d}{2}-n_{2}\big)
  \Gamma(2n_{2}+n_{3}-d)}
  {2^{2n_{2}+D-1}\pi^{\frac{D-1}{2}}\Gamma(n_{2})\Gamma(n_{3})}
  (\ell \cdot v_{1}-i0_{+})^{d-2 n_{2}-n_{3}} \,,
\label{}\end{equation}
where $D=d+1$.
We next consider the rank-$n$ tensor integral generalization
\begin{equation}
\begin{aligned}
  I^{D|\mu_{1}\mu_{2}\cdots\mu_{n}}_{n_{1},n_{2},n_{3}}
  =
  \int \frac{\text{d}^{D}k}{(2\pi)^{D}} \frac{\hat{\delta}(k\cdot v_{1})k^{\mu_{1}}k^{\mu_{2}}\cdots k^{\mu_{n}}}{
  	|k|^{2n_{1}}
  	[\ell-k|^{2n_{2}} 
  	\big[ k \cdot v_{2} - i0_{+}\big]^{n_{3}}}
\end{aligned}\label{}
\end{equation}
In this work, we need up to rank-3 tensor integrals. We derived the integrals using the so-called dimensional shift method \cite{Davydychev:1991va}. Defining the variables
\begin{equation}
  z_{1}^{\mu}=\frac{\gamma v_{1}^{\mu} - v_{2}^{\mu}}{2} \,, 
  \qquad
  z_{2}^{\mu}=\frac{\gamma v_{2}^{\mu} - v_{1}^{\mu}}{2} \,, 
  \qquad
  P^{\mu\nu}= \eta^{\mu\nu}+v_{1}^{\mu}v_{1}^{\nu}\;.
\end{equation}
Here we do not assume any orthogonality between $\ell_{\alpha}$ and $ v_{\beta} $, i.e. $\ell_{\alpha} \cdot v_{\beta} \neq 0$ for $ \alpha, \beta = 1, 2$.

The vector integral is
\begin{equation}
\begin{aligned}
  I^{D|\mu}_{n_{1},n_{2},n_{3}}
  =
    4\pi n_{2}\ell^{\mu} I^{D+2}_{n_{1},n_{2}+1,n_{3}}
  + 4\pi n_{3} z^{\mu}_{1} I^{D+2}_{n_{1},n_{2},n_{3}+1} \,.
\end{aligned}\label{}
\end{equation}
The rank-2 tensor integral is
\begin{equation}
\begin{aligned}
  I^{D|\mu\nu}_{n_{1},n_{2},n_{3}}
  &= 2\pi P^{\mu\nu} I^{D+2}_{n_{1},n_{2},n_{3}}
  +32\pi^2 n_{2} n_{3} \ell^{(\mu} z^{\nu)}_{1} I^{D+4}_{n_{1},n_{2}+1,n_{3}+1}
  \\&\quad
  + 16\pi^2 \Big[ n_{2} (n_{2}+1) \ell^{\mu} \ell^{\nu} I^{D+4}_{n_{1},n_{2}+2,n_{3}}+n_{3} (n_{3}+1) z^{\mu}_{1} z^{\nu}_{1} I^{D+4}_{n_{1},n_{2},n_{3}+2} \Big]\,.
\end{aligned}\label{}
\end{equation}
Finally the rank-3 tensor integral is
\begin{equation}
\begin{aligned}
  &I^{D|\mu\nu\rho}_{n_{1},n_{2},n_{3}}
  \\
  &= 24\pi^2 \Big[
    n_{2} P^{(\mu\nu}\ell^{\rho)} I^{D+4}_{n_{1},n_{2}+1,n_{3}}
  + n_{3} P^{(\mu\nu} z_{1}^{\rho)} I^{D+4}_{n_{1},n_{2},n_{3}+1}
  \Big]
  \\&\quad
  + 64\pi^3 \Big[ n_{2} (n_{2}{+}1) (n_{2}{+}2)\ell^{\mu} \ell^{\nu}\ell^{\rho} I^{(D+6)}_{n_{1},n_{2}+3,n_{3}} +n_{3} (n_{3}{+}1) (n_{3}{+}2) z^{\mu}_{1} z^{\nu}_{1} z^{\rho}_{1} I^{D+6}_{n_{1},n_{2},n_{3}+3} \Big]
  \\&\quad
  +192\pi^3 \Big[
    n_{2} (n_{2}{+}1) n_{3} \ell^{(\mu} \ell^{\nu} z^{\rho)}_{1} I^{D+6}_{n_{1},n_{2}+2,n_{3}+1}
  + n_{2} n_{3} (n_{3}{+}1) \ell^{(\mu} z^{\nu}_{1} z^{\rho)}_{1} I^{D+6}_{n_{1},n_{2}+1,n_{3}+2}
  \Big]\,.
\end{aligned}\label{}
\end{equation}
%

\section{Evaluation of Master Integrals}
\label{App:B}
In this appendix, we present the explicit form of the series expansion of the master integrals in the potential region defined in Section \ref{Sec:Potential Region}.


\subsection{Evaluation of the Master integrals of the Family $\rm \overline{II}$}\label{Sec:B}

The master integrals of the family $\rm \overline{II}$ are given by
\begin{equation}
\begin{aligned}
  J^{\rm \overline{II}}_{1,1,1,0,0,0,0}\,,\qquad 
  J^{\rm \overline{II}}_{0,1,1,0,1,2,0}\,,\qquad 
  J^{\rm \overline{II}}_{1,1,0,1,1,0,0}\,,\qquad 
  J^{\rm \overline{II}}_{0,0,1,1,1,1,1}\,,\qquad 
  J^{\rm \overline{II}}_{1,1,0,1,1,1,1}\,,
\end{aligned}\label{}
\end{equation}
for the even sector and
\begin{equation}
\begin{aligned}
  J^{\rm \overline{II}}_{0,0,1,1,1,0,1}\,,\qquad
  J^{\rm \overline{II}}_{0,1,1,0,1,1,0}\,,\qquad
  J^{\rm \overline{II}}_{1,1,0,1,1,0,1}\,.
\end{aligned}\label{}
\end{equation}
for the odd sector. Among them, the relevant master integrals for evaluating the master integrals for the family II are
\begin{equation}
\begin{aligned}
  \text{even}&:\qquad
  f^{\rm \overline{II},e}_{1}
  =J^{\rm \overline{II}}_{1,1,1,0,0,0,0}\,,\qquad 
  f^{\rm \overline{II},e}_{2}
  =J^{\rm \overline{II}}_{1,1,0,1,1,0,0}\,,\qquad 
  f^{\rm \overline{II},e}_{3}
  =J^{\rm \overline{II}}_{0,0,1,1,1,1,1}\,,\qquad 
  \\
  \text{odd}&:\qquad
  f^{\rm \overline{II},o}_{1}
  =J^{\rm \overline{II}}_{0,0,1,1,1,0,1}\,,\qquad
  f^{\rm \overline{II},o}_{2}
  =J^{\rm \overline{II}}_{0,1,1,0,1,1,0}\,,\qquad
  f^{\rm \overline{II},o}_{3}
  =J^{\rm \overline{II}}_{1,1,0,1,1,0,1}\,.
\end{aligned}\label{}
\end{equation}
We can evaluate these master integrals by iterating one-loop integrals.

The explicit form of $J^{\rm \overline{II}}_{1,1,1,0,0,0,0}$ is
\begin{equation}
  f^{\rm \overline{II},e}_{1} = J^{\rm \overline{II}}_{1,1,1,0,0,0,0} = \int_{\vec{k}_{1}} \frac{1}{\big|\vec{k}_{1}\big|^{2}} \int_{\vec{k_{2}}} \frac{1}{\big|\vec{k}_{2}\big|^{2} \big|\vec{\ell}-\vec{k}_{1}-\vec{k}_{2}\big|^{2}}\,.
\label{}\end{equation}
This is a double-bubble integral, and one can easily evaluate this using the bubble integral formula \eqref{bubble_integral}
\begin{equation}
  f^{\rm \overline{II},e}_{1}
  =
  \frac{(4 \pi )^{2 \epsilon -3} \Gamma \big(\frac{1}{2}-\epsilon \big)^3 \Gamma (2 \epsilon )}{\Gamma \big(\frac{3}{2}-3 \epsilon \big)} \big|\vec{\ell}\, \big|^{-4 \epsilon }
\label{}\end{equation}
It is expanded in $\epsilon$ as
\begin{equation}
  f^{\rm \overline{II},e}_{1}
  = (4 \pi )^{2 \epsilon -2} e^{-2 \gamma \epsilon}
  \bigg(\frac{1}{4\epsilon} 
  +\frac{3}{2} + \left(9-\frac{7 \pi ^2}{24}\right) \epsilon
  + \cdots
  \bigg) 
  \big|\vec{\ell}\, \big|^{-4 \epsilon }
\label{}\end{equation}

The second master integral, $f^{\rm \overline{II},e}_{2}$ is represented by
\begin{equation}
  f^{\rm \overline{II},e}_{2}
  =
  \int_{k_{1}} \frac{1}{\big|\vec{k}_{1}\big|^{2} \big|\vec{\ell}-\vec{k}_{1}\big|^{2}}
  \int_{k_{2}} \frac{1}{\big|\vec{k}_{2}\big|^{2} \big|\vec{\ell}-\vec{k}_{2}\big|^{2}}\,.
\label{}\end{equation}
Since this is the square of the bubble integral \eqref{bubble_integral}, we have
\begin{equation}
  f^{\rm \overline{II},e}_{2}
  =
  \frac{(4\pi) ^{2 \epsilon -3} \Gamma \big(\frac{1}{2}-\epsilon \big)^4 \Gamma\big(\epsilon +\frac{1}{2}\big)^2}{\Gamma (1-2 \epsilon )^2}
  |\vec{\ell}\, |^{-4 \epsilon -2}
\label{}\end{equation}
The series expansion of $f^{\rm \overline{II},e}_{2}$ in $\epsilon$ is given by
\begin{equation}
  f^{\rm \overline{II},e}_{2}
  =
  (4 \pi )^{2 \epsilon -2} e^{-2 \gamma \epsilon}
  \bigg( 
  	\frac{\pi ^2}{4} + \epsilon\, \pi^2 \log 2 + \cdots
  \bigg) |\vec{\ell}\, |^{-4 \epsilon -2} + \mathcal{O}(\epsilon^{2})
\label{}\end{equation}

We consider the third master integral in the even sector, $f^{\rm \overline{II},e}_{3}$, which is represented by
\begin{equation}
  f^{\rm \overline{II},e}_{3}
  =
  \int_{k_{1},k_{2}}
  \frac{1}{
  	\big|\vec{\ell}-\vec{k}_{1}\big|^{2} 
  	\big|\vec{\ell}-\vec{k}_{2}\big|^{2} 
  	\big|\vec{\ell}-\vec{k}_{1}-\vec{k}_{2}\big|^{2}
  	\big(\vec{k}_{1}\cdot \vec{q} - i\epsilon\big)
  	\big(-\vec{k}_{2}\cdot \vec{q} - i\epsilon\big)
  }\,.
\label{}\end{equation}
Shifting the loop momenta, $\vec{k}_{1} \to \vec{\ell} - \vec{k}_{1}$ and $\vec{k}_{2} \to \vec{\ell} - \vec{k}_{2}$, we have
\begin{equation}
  f^{\rm \overline{II},e}_{3}
  =
 \int_{k_{1},k_{2}}
  \frac{1}{
  	\big|\vec{k}_{1}\big|^{2} 
  	\big|\vec{k}_{2}\big|^{2} 
  	\big|\vec{\ell}-\vec{k}_{1}-\vec{k}_{2}\big|^{2}
  	\big(-\vec{k}_{1}\cdot \vec{q} - i\epsilon\big)
  	\big(\vec{k}_{2}\cdot \vec{q} - i\epsilon\big)
  }\,.  
\label{}\end{equation}
In fact, this integral is identical with \eqref{f_I_3} without $\frac{1}{\gamma^{2}-1}$ factor because $v_{2}$ is replaced to $q$. 
\begin{equation}
  f^{\rm \overline{II},e}_{3}
  =
  \frac{ \Gamma(-\epsilon)^{3}\Gamma(1+2\epsilon)}{6\times (4\pi)^{2-2\epsilon}\Gamma(-3\epsilon)} |\vec{\ell}\,|^{-2-4\epsilon}\,.
\label{}\end{equation}
The series expansion of $f^{\rm \overline{II},e}_{2}$ in $\epsilon$ is given by
\begin{equation}
  f^{\rm \overline{II},e}_{3}
  =
  (4 \pi )^{2 \epsilon -2} e^{-2 \gamma \epsilon}
  \bigg(
    \frac{1}{2\epsilon ^2} - \frac{\pi ^2}{12} + \cdots
  \bigg) |\vec{\ell}\,|^{-2 -4 \epsilon} + \mathcal{O}(\epsilon^{2})
\label{}\end{equation}

Similarly, the master integrals in the odd sector can be evaluated, and we obtain 
\begin{equation}
\begin{aligned}
  f^{\rm \overline{II},o}_{1}
  &=
  \frac{i 4^{2 \epsilon -3} \pi ^{2 \epsilon -\frac{5}{2}} \Gamma \left(\frac{1}{2}-2 \epsilon \right) \Gamma \left(\frac{1}{2}-\epsilon \right)^2 \Gamma (-\epsilon ) \Gamma \left(2 \epsilon +\frac{1}{2}\right)}{\Gamma \left(\frac{1}{2}-3 \epsilon \right) \Gamma (1-2 \epsilon )} |\vec\ell\,|^{-1-4 \epsilon}\,,
  \\
  f^{\rm \overline{II},o}_{2}
  &=
  \frac{i e^{-2 i \pi  \epsilon } (4 \pi )^{2 \epsilon -3} \Gamma \left(\frac{1}{2}-2 \epsilon \right)^2 \Gamma \left(\frac{1}{2}-\epsilon \right) \Gamma (\epsilon ) \Gamma \left(2 \epsilon +\frac{1}{2}\right)}{\Gamma (1-4 \epsilon )} |\vec\ell\,|^{-1-4 \epsilon}\,,
  \\
  f^{\rm \overline{II},o}_{3}
  &=
  \frac{i 4^{2 \epsilon -3} \pi ^{2 \epsilon -\frac{5}{2}}\Gamma \left(\frac{1}{2}-\epsilon \right)^2 \Gamma (-\epsilon )^2 \Gamma \left(\epsilon +\frac{1}{2}\right) \Gamma (\epsilon +1)}{\Gamma (1-2 \epsilon ) \Gamma (-2 \epsilon )} |\vec\ell\,|^{-3-4 \epsilon}\,.
\end{aligned}\label{}
\end{equation}
%

\subsection{Family II master integrals}
First, let us consider $f^{\rm II,P,e}_{1} = J^{\rm II,P, \pm}_{0,1,1,0,1,0,0}$ and $f^{\rm II,P,e}_{2} = J^{\rm II,P, \pm}_{1,1,0,1,1,0,0}$, which is a trivial part. The small-$x$ expansion 
of the first master integral $f^{\rm II,e}_{1}$ is given by
\begin{equation}
   f^{\rm II,P,e}_{1}
   =
   \sum_{m=0}^{\infty} 2^{m} (-x)^{m}
   J^{\rm \overline{II},P}_{0,1,1+m,0,1,-m,-m}\,.
\label{}\end{equation}
We have checked up to a truncation of this expansion to 10th order and integration by parts reductions
\begin{equation}
   f^{\rm II,P,e}_{1} = 0\,.
\label{}\end{equation}
On the other hand, $f^{\rm II,P,e}_{2}$ is the square of the one-loop 
bubble integral, and it can thus be evaluated by using eq. 
\eqref{bubble_integral},
\begin{equation}
\begin{aligned}
f^{\rm II,P,e}_{2}
&= (I^{D}_{1,1,0})^{2}\,,
\\
&=
\frac{(4 \pi )^{2 \epsilon -3} \Gamma \left(\frac{1}{2}-\epsilon 
\right)^4 \Gamma \left(\epsilon +\frac{1}{2}\right)^2}{\Gamma (1-2 
\epsilon )^2}  |\vec{\ell}\,|^{-4 \epsilon -2}\,.
\end{aligned}\label{}
\end{equation}

Next, truncating the above series expansion up to order $x^{10}$ and performing integration by parts reductions, we find
\begin{equation}
\begin{aligned}
f^{\rm II,P,e}_{3}
&=
\Big(C^{\ord{0}}_{3,1} + \mathcal{O}(\epsilon^{1})\Big) f^{\rm \overline{II},P,e}_{1}|\vec{\ell}\,|^{-2}\,,
\\
&=
(4\pi)^{2\epsilon-2}e^{-2\gamma_{\rm E}\epsilon}
\bigg[
\frac{1}{4\epsilon\sqrt{\gamma^{2}-1}}\cosh^{-1}\gamma + \mathcal{O}(\epsilon^{0})
\bigg] |\vec{\ell}\, |^{-2-4\epsilon}\,,
\\
f^{\rm II,P,e}_{4}
&=
\Big( C^{\ord{1}}_{4,1}\epsilon + C^{\ord{2}}_{4,1}\epsilon^{2} + 
\mathcal{O}(\epsilon^{3})\Big) f^{\rm \overline{II},P,e}_{1}|\vec{\ell}\,|^{-2}\,,
\\
&= -(4\pi)^{2\epsilon-2}e^{-2\gamma_{\rm E}\epsilon}
\bigg[
- \frac{\gamma}{2}
+ \epsilon\left(\gamma -\sqrt{\gamma ^2-1} \cosh^{-1}\gamma \right)
+ \mathcal{O}(\epsilon^{2})
\bigg] |\vec{\ell}\, |^{-2-4\epsilon}\,,
\\
f^{\rm II,P,e}_{5}
&=
\Big(C^{\ord{1}}_{5,1} \epsilon + \mathcal{O}(\epsilon^{2})\Big) 
f^{\rm \overline{II},P,e}_{1}|\vec{\ell}\,|^{-2}
\\
&=
-(4\pi)^{2\epsilon-2}e^{-2\gamma_{\rm E}\epsilon} \bigg[\frac{1}{2\sqrt{\gamma^{2}-1}} \cosh^{-1}\gamma+ 
\mathcal{O}(\epsilon^{1})\bigg] |\vec{\ell}\,|^{-2-4\epsilon}
\end{aligned}\label{}
\end{equation}
where the coefficients of the master integrals are
\begin{equation}
\begin{aligned}
C^{\ord{0}}_{3,1}
&=
1{-}\tfrac{x}{3}
{+}\tfrac{2 x^2}{15}
{-}\tfrac{2 x^3}{35}
{+}\tfrac{8 x^4}{315}
{-}\tfrac{8 x^5}{693}
{+}\tfrac{16 x^6}{3003}
{-}\tfrac{16 x^7}{6435}
{+}\tfrac{128 x^8}{109395}
{-}\tfrac{128 x^9}{230945}
{+}\tfrac{256 x^{10}}{969969}
{+}\cdots \,,
\\
C^{\ord{1}}_{4,1}
&=
- 2(1+x) \,,
\\
C^{\ord{2}}_{4,1}
&=
x{+}\tfrac{x^2}{6}
{-}\tfrac{x^3}{30}
{+}\tfrac{x^4}{105}
{-}\tfrac{x^5}{315}
{+}\tfrac{4 x^6}{3465}
{-}\tfrac{4 x^7}{9009}
{+}\tfrac{8 x^8}{45045}
{-}\tfrac{8 x^9}{109395}
{+}\tfrac{64 x^{10}}{2078505}+\cdots
\\
C^{\ord{1}}_{5,1}
&=
{-}2{+}\tfrac{2 x}{3}
{-}\tfrac{4 x^2}{15}
{+}\tfrac{4 x^3}{35}
{-}\tfrac{16 x^4}{315}
{+}\tfrac{16 x^5}{693}
{-}\tfrac{32 x^6}{3003}
{+}\tfrac{32 x^7}{6435}
{-}\tfrac{256 x^8}{109395}
{+}\tfrac{256 x^9}{230945}
{-}\tfrac{512 x^{10}}{969969} {+}\cdots \,.
\end{aligned}\label{}
\end{equation}

Next we consider $f^{\rm II,e}_{6} = J^{\rm II, \pm}_{1,1,1,1,1,0,0}$ 
and $f^{\rm II,e}_{7} = J^{\rm II, \pm}_{1,1,2,1,1,0,0}$. These are also 
expanded as
\begin{equation}
\begin{aligned}
   f^{\rm II,P,e}_{6}
   &=
   \sum_{m=0}^{\infty} 2^{m} x^{m}
   J^{\rm \overline{II},P}_{1,1,1+m,1,1,-m,-m}\,,
   \\
   f^{\rm II,P,e}_{7}
   &=
   \sum_{m=0}^{\infty} 2^{m} (m+1) x^{m}
   J^{\rm \overline{II},P}_{1,1,2+m,1,1,-m,-m}\,,
\end{aligned}\label{}
\end{equation}
By use of integration by parts reduction, we find
\begin{equation}
\begin{aligned}
f^{\rm II,P,e}_{6}
&=
\Big(C^{\ord{0}}_{6,1} + C^{\ord{1}}_{6,1} \epsilon + \mathcal{O}(\epsilon^{2})\Big) |\vec{\ell}\, |^{-4} f^{\rm \overline{II},P,e}_{1} + \Big( C^{\ord{1}}_{6,2}\epsilon + \mathcal{O}(\epsilon^{2})\Big) |\vec{\ell}\,|^{-2} f^{\rm \overline{II},P,e}_{2} \,,
\\
&=
\frac{(4 \pi )^{2 \epsilon -2} e^{-2 \gamma_{\rm E} \epsilon}}{|\vec{\ell}\,|^{4+4\epsilon}} \left[ \frac{\cosh^{-1}\!\gamma}{2 \epsilon \sqrt{\gamma^2-1}}
+ \frac{(\cosh^{-1}\!\gamma)^2 +\text{Li}_2\big({-}2 \gamma^2+2 \sqrt{\gamma^2-1} \gamma +2\big)}{2 \sqrt{\gamma ^2-1}}\right]\,,
   \\
f^{\rm II,P,e}_{7}
&=
\Big(C^{\ord{0}}_{7,1} + C^{\ord{1}}_{7,1} \epsilon + 
\mathcal{O}(\epsilon^{2})\Big) |\vec{\ell}\,|^{-6} f^{\rm 
\overline{II},P,e}_{1}
   + \Big( C^{\ord{1}}_{7,2}\epsilon + \mathcal{O}(\epsilon^{2})\Big) 
|\vec{\ell}\,|^{-4} f^{\rm \overline{II},P,e}_{2} \,,
   \\
   &=
   \frac{(4 \pi )^{2 \epsilon -2} e^{-2 \gamma_{\rm E} \epsilon}}{|\vec{\ell}\,|^{4 \epsilon +6}} \Bigg[ \frac{\gamma (5 \epsilon +1)}{2\epsilon \left(\gamma^2-1\right)} - \frac{\left(2 \gamma ^2 \epsilon +\epsilon +1\right) \cosh^{-1}\gamma}{2 \left(\gamma ^2-1\right)^{3/2} \epsilon }
\\&\qquad\qquad\qquad\qquad
- \frac{ \big(\cosh^{-1}\gamma \big)^2 + \text{Li}_2\big(2 \gamma 
(\sqrt{\gamma ^2-1}-\gamma)+2\big)}{2 \left(\gamma ^2-1\right)^{3/2} }\bigg]
\end{aligned}\label{}
\end{equation}
where the coefficients are
\begin{equation}
\begin{aligned}
C^{\ord{0}}_{6,1}
&=
{-}2{+}\tfrac{2x}{3}
{-}\tfrac{4 x^2}{15}
{+}\tfrac{4 x^3}{35}
{-}\tfrac{16 x^4}{315}
{+}\tfrac{16 x^5}{693}
{-}\tfrac{32 x^6}{3003}
{+}\tfrac{32 x^7}{6435}
{-}\tfrac{256 x^8}{109395}
{+}\tfrac{256 x^9}{230945}
{-}\tfrac{512 x^{10}}{969969} + \cdots
\\
C^{\ord{1}}_{6,1}
&=
4\Big(2 {-}\tfrac{8 x}{9}
{+}\tfrac{89 x^2}{225}
{-}\tfrac{221 x^3}{1225}
{+}\tfrac{8306 x^4}{99225}
{-}\tfrac{18878 x^5}{480249}
{+}\tfrac{503764 x^6}{27054027}
{-}\tfrac{367676 x^7}{41409225}
{+}\tfrac{50930576 x^8}{11967266025}
\\&\qquad\quad
{-}\tfrac{109248816 x^9}{53335593025}
{+}\tfrac{4653484384 x^{10}}{4704199304805} + \cdots\Big)
\\
C^{\ord{1}}_{6,2}
&=
{-}4{+}\tfrac{4 x}{3}
{-}\tfrac{8 x^2}{15}
{+}\tfrac{8 x^3}{35}
{-}\tfrac{32 x^4}{315}
{+}\tfrac{32 x^5}{693}
{-}\tfrac{64 x^6}{3003}
{+}\tfrac{64 x^7}{6435}
{-}\tfrac{512 x^8}{109395}
{+}\tfrac{512 x^9}{230945}
{-}\tfrac{1024 x^{10}}{969969}
+\cdots
\end{aligned}\label{}
\end{equation}
and
\begin{equation}
\begin{aligned}
C^{\ord{0}}_{7,1}
&=
\tfrac{4}{3}
{-}\tfrac{4 x}{5}
{+}\tfrac{16 x^2}{35}
{-}\tfrac{16 x^3}{63}
{+}\tfrac{32 x^4}{231}
{-}\tfrac{32 x^5}{429}
{+}\tfrac{256 x^6}{6435}
{-}\tfrac{256 x^7}{12155}
{+}\tfrac{512 x^8}{46189}
{-}\tfrac{512 x^9}{88179}
{+}\tfrac{2048 x^{10}}{676039}
{+}\cdots
\\
C^{\ord{1}}_{7,1}
&=
{-}\tfrac{52}{9}
{+}\tfrac{196 x}{75}
{-}\tfrac{5008 x^2}{3675}
{+}\tfrac{14576 x^3}{19845}
{-}\tfrac{318992 x^4}{800415}
{+}\tfrac{834448 x^5}{3864861}
{-}\tfrac{33762304 x^6}{289864575}
{+}\tfrac{83144704 x^7}{1329696225}
\\&\quad
{-}\tfrac{3207107456 x^8}{96004067445}
{+}\tfrac{7597911424 x^9}{427654482255}
{-}\tfrac{709602499072 x^{10}}{75409740370965}
{+}\cdots
\\
C^{\ord{1}}_{7,2}
&=
\tfrac{8}{3}
{-}\tfrac{8x}{5}
{+}\tfrac{32 x^2}{35}
{-}\tfrac{32 x^3}{63}
{+}\tfrac{64 x^4}{231}
{-}\tfrac{64 x^5}{429}
{+}\tfrac{512 x^6}{6435}
{-}\tfrac{512 x^7}{12155}
{+}\tfrac{1024 x^8}{46189}
{-}\tfrac{1024 x^9}{88179}
{+}\tfrac{4096 x^{10}}{676039}
{+}\cdots
\end{aligned}\label{}
\end{equation}

Finally, we consider the last master integral in the even sector, 
$f^{\rm II,e}_{8} = J^{\rm II, \pm}_{1,1,1,0,0,1,1}$, which has an expansion
\begin{equation}
   f^{\rm II,P,e}_{8}
   =
   \frac{1}{\gamma^{2}-1} \sum_{m=0}^{\infty} 2^{m} (-x)^{m}
   J^{\rm \overline{II},P}_{1,1,1+m,0,0,1-m,1-m}\,.
\label{}\end{equation}
Truncating up to order $x^{10}$, we find by integration by parts 
reductions the following relation for $f^{\rm II,P,e}_{8}$:
\begin{equation}
\begin{aligned}
f^{\rm II,P,e}_{8}
&=
\frac{1}{\gamma^{2}-1} \bigg[
f^{\rm \overline{II},e}_{3}
+ \Big( C^{\ord{1}}_{8,1} \epsilon + \mathcal{O}(\epsilon^{2}) \Big) 
|\vec{\ell}\,|^{-2} f^{\rm \overline{II},e}_{1}\bigg]\,,
\\
&=
\frac{(4 \pi )^{2 \epsilon -2} e^{-2 \gamma_{\rm E} 
\epsilon}}{|\vec{\ell}|^{2+4 \epsilon}} \bigg[ - \frac{1}{2\epsilon^{2}(\gamma^{2}-1)} + \mathcal{O}(\epsilon^{0})\bigg]\,,
\end{aligned}\label{}
\end{equation}
where the coefficient is given by
\begin{equation}
C^{\ord{1}}_{8,1}
=
4x
{-}\tfrac{2 x^2}{3}
{+}\tfrac{8 x^3}{45}
{-}\tfrac{2 x^4}{35}
{+}\tfrac{32 x^5}{1575}
{-}\tfrac{16 x^6}{2079}
{+}\tfrac{64 x^7}{21021}
{-}\tfrac{8 x^8}{6435}
{+}\tfrac{512 x^9}{984555}
{-}\tfrac{256 x^{10}}{1154725}
{+}\cdots\,.
\label{}\end{equation}

We now proceed to evaluate the master integrals in the odd sector. As 
mentioned earlier, among the six master integrals identified initially, 
only the four integrals, $\{f^{\rm II,o}_{1},\cdots, f^{\rm II,o}_{4}\}$ 
defined in \eqref{master_Integrals_II_odd2}, appear in the subsequent 
computations. We will therefore restrict our attention to these four 
integrals.

Let us first consider the three integrals $f^{\rm II,P,o}_{1} = 2J^{\rm II,P, \pm}_{0,1,2,0,1,0,-1,1,1}$, $f^{\rm II,P,o}_{2} = 2J^{\rm II,P, \pm}_{1,1,2,0,0,0,-1,1,1}$ and $f^{\rm II,P,o}_{3} = 2J^{\rm II,P, \pm}_{1,1,2,1,1,0,-1,1,1}$. These expand as follows:
\begin{equation}
\begin{aligned}
   f^{\rm II,P,o}_{1}
   &=
   \sqrt{\gamma^{2}-1}
   \sum_{m=0}^{\infty}
   \frac{2^{m+1} \Gamma(m+2)}{\Gamma(m+1)} (-x)^{m}
   J^{\rm \overline{II},P}_{0,1,2+m,0,1,-m,-1-m}\,,
   \\
   f^{\rm II,P,o}_{2}
   &=
   \sqrt{\gamma^{2}-1}
   \sum_{m=0}^{\infty}
   \frac{2^{m+1} \Gamma(m+2)}{\Gamma(m+1)} (-x)^{m}
   J^{\rm \overline{II},P}_{1,1,2+m,0,0,-m,-1-m}\,,
   \\
   f^{\rm II,P,o}_{3}
   &=
   \sqrt{\gamma^{2}-1}
   \sum_{m=0}^{\infty}
   \frac{2^{m+1} \Gamma(m+2)}{\Gamma(m+1)} (-x)^{m}
   J^{\rm \overline{II},P}_{1,1,2+m,1,1,-m,-1-m}\,,
\end{aligned}\label{}
\end{equation}
and by integration by parts reductions we find that they are all 
trivial,
\begin{equation}
   f^{\rm II,P,o}_{1} = 0\,,
   \qquad
   f^{\rm II,P,o}_{2} = 0\,,
   \qquad
   f^{\rm II,P,o}_{3} = 0\,.
\label{}\end{equation}

Next, we evaluate $f^{\rm II,P,o}_{4} = J^{\rm II,P, 
\pm}_{0,0,1,1,1,0,1,1,1}$, which expands as follows:
\begin{equation}
\begin{aligned}
   f^{\rm II,P,o}_{4}
   &=
   \frac{1}{\sqrt{\gamma^{2}-1}} \sum_{m=0}^{\infty} 2^{m} (-x)^{m}
   J^{\rm \overline{II},P}_{0,0,1+m,1,1,0-m,1-m}\,,
\end{aligned}\label{}
\end{equation}
and its integration by parts reduction is given by
\begin{equation}
\begin{aligned}
   f^{\rm II,P,o}_{4}
   &=
   \frac{1}{\sqrt{\gamma^{2}-1}}f^{\rm \overline{II},P,o}_{1}\,,
   \\
   &=
   \frac{i}{\sqrt{\gamma^{2}-1}} \frac{4^{2 \epsilon -3} \pi ^{2 
\epsilon -\frac{5}{2}}
   \Gamma \left(\frac{1}{2}-2 \epsilon \right) \Gamma 
\left(\frac{1}{2}-\epsilon \right)^2 \Gamma (-\epsilon ) \Gamma \left(2 
\epsilon +\frac{1}{2}\right)}{\Gamma \left(\frac{1}{2}-3 \epsilon 
\right) \Gamma (1-2 \epsilon )~|\vec{\ell}\,|^{1+4 \epsilon}} \,.
\end{aligned}\label{}
\end{equation}
We now proceed to evaluate the master integrals in the odd sector. As 
mentioned earlier, among the six master integrals identified initially, 
only the four integrals, $\{f^{\rm II,o}_{1},\cdots, f^{\rm II,o}_{4}\}$ 
defined in \eqref{master_Integrals_II_odd2}, appear in the subsequent 
computations. We will therefore restrict our attention to these four 
integrals.

\section{Currents}
In this appendix we present the explicit forms of the 2PM and 3PM currents omitted in the main text.

\subsection{2PM}\label{Appendix:2PM}

We first consider the subcurrents consisting the 2PM graviton currents. We solve their recursions and evaluate 1-loop integrals for the zero-mode sector. First, $W$ currents are
\begin{equation}
\begin{aligned}
  W^{\mu\nu}\big|^{2}_{\ell_{1},0}
  &=0\,,
  \\
  W^{\mu\nu}\big|^{2}_{0,\ell_{2}}
  &=0\,,
  \\
  W^{\mu\nu}\big|^{2}_{\ell_{1},\ell_{2}}
  &=
  \frac{8\hat{\mu}_{1,2} }{|\ell_{1}|^{2} |\ell_{2}|^{2} } \Big(2v_{1}^{(\mu} v_{2}^{\nu)}(\ell_{1}\cdot v_{2})(\ell_{2}\cdot v_{1}) 
    -v_{1}^{\mu} v_{1}^{\nu} (\ell_{1}\cdot v_{2})^{2} 
    -v_{2}^{\mu} v_{2}^{\nu} (\ell_{2}\cdot v_{1})^{2}
    \Big)\,,
\end{aligned}\label{}
\end{equation}
and $Z$-currents are given by
\begin{equation}
\begin{aligned}
  Z^{\mu\nu\rho\sigma}\big|^{2}_{\ell_{1},0} 
  &=
  -\frac{m_{1}^2 4^{2 \epsilon +1} \pi ^{\epsilon +2} \sec(\pi  \epsilon)
  \hat{\delta} (\ell_{1}\cdot v_{1})}{\Gamma (2-\epsilon) |\ell_{1}|^{1+2\epsilon}}
  v_{1}^{\nu}v_{1}^{\sigma}
  \Big(|\ell_{1}|^{2} (\tilde{\eta}^{\mu \rho} +v_{1}^{\mu} v_{1}^{\rho})+(1-2 \epsilon ) \ell_{1}^{\mu}\ell_{1}^{\rho}\Big)\,,
  \\
  Z^{\mu\nu\rho\sigma}\big|^{2}_{0,\ell_{2}} 
  &=
  -\frac{m_{2}^2 4^{2 \epsilon +1} \pi^{\epsilon +2} \sec(\pi  \epsilon) \hat{\delta} (\ell_{2}\cdot v_{2}) }{\Gamma (2-\epsilon ) |\ell_{2}|^{1+2\epsilon}}
  v_{2}^{\nu}v_{2}^{\sigma}
  	\Big(
  		  |\ell_{2}|^{2} (\tilde{\eta}^{\mu \rho} +v_{2}^{\mu} v_{2}^{\rho})
  		+ (1-2 \epsilon ) \ell_{2}^{\mu}\ell_{2}^{\rho}
  	\Big)\,,
  \\
  Z^{\mu\nu\rho\sigma}\big|^{2}_{\ell_{1},\ell_{2}} 
  &=
  -\frac{8 \gamma \hat{\mu}_{1,2}}{|\ell_{1}|^2 |\ell_{2}|^2} \Big( \ell_{2}^{\mu} v_{2}^{\nu} \ell_{1 \rho} v_{1 \sigma} + \ell_{1}^{\mu}v_{1}^{\nu} \ell_{2\rho} v_{2\sigma} \Big)\,,
\end{aligned}\label{}
\end{equation}
and
\begin{equation}
\begin{aligned}
  &Z^{\mu\nu}\big|^{2}_{\ell_{1},0}\!
  =\!
  \frac{2^{4 \epsilon{+}1}m_{1}^2 \pi^{\epsilon +2} 
  \hat{\delta} (\ell_{1}\cdot v_{1})}{\Gamma(2{-}\epsilon)\cos(\pi\epsilon) |\ell_{1}|^{1+2\epsilon}}
  \Big[
  	|\ell_{1}|^{2} \big(\tilde{\eta}^{\mu\nu} {+}2 (\epsilon {-}1) \eta^{\mu\nu} {+}(8 \epsilon {-}7) v_{1}^{\mu}v_{1}^{\nu}\big)
  	{+}(1{-}2 \epsilon ) \ell_{1}^{\mu}\ell_{1}^{\nu}
  \Big]\,,
  \\
  &Z^{\mu\nu}\big|^{2}_{0,\ell_{2}}\!
  =\!
  \frac{2^{4 \epsilon{+}1}m_{2}^2 \pi^{\epsilon +2} 
  \hat{\delta} (\ell_{2}\cdot v_{2})}{\Gamma (2{-}\epsilon)\cos(\pi\epsilon) |\ell_{2}|^{1+2\epsilon}}
  \Big[
  	|\ell_{2}|^{2} \big(\tilde{\eta}^{\mu\nu} {+}2 (\epsilon {-}1) \eta^{\mu\nu} {+}(8 \epsilon {-}7) v_{2}^{\mu}v_{2}^{\nu}\big)
  	{+}(1{-}2 \epsilon ) \ell_{2}^{\mu}\ell_{2}^{\nu}
  \Big]\,,
  \\
  &Z^{\mu\nu}\big|^{2}_{\ell_{1},\ell_{2}}\!
  \\
  &=
  \frac{\hat{\mu}_{1,2}}{|\ell_{1}|^{2}|\ell_{2}|^{2}}
  \Big[
    8\gamma \big(
   	 	  2\ell_{1}^{\mu} v_{2}^{\nu} (\ell_{2}\cdot v_{1})
    	+ 2\ell_{2}^{\mu} v_{1}^{\nu} (\ell_{1}\cdot v_{2})
   	 	- 2v_{1}^{(\mu} v_{2}^{\nu)} (\ell_{1}\cdot \ell_{2})
    	- \eta^{\mu\nu} (\ell_{1}\cdot v_{2})(\ell_{2}\cdot v_{1})
  \big)
  \\&\qquad\qquad\qquad
    + 4\gamma^{2}\big( 2\ell_{1}^{(\mu} \ell_{2}^{\nu)}- \eta^{\mu\nu} (\ell_{1}\cdot \ell_{2})\big)
  \Big]\,.
\end{aligned}\label{}
\end{equation}
where $\tilde{\eta}$ is an auxiliary metric which has the same properties except that the trace is $4-2\epsilon$, $\tilde{\eta}^{\mu\nu} \tilde{\eta}_{\mu\nu} = 4-2\epsilon$. We also defined $\hat{\mu}_{\alpha,\beta}$ as
\begin{equation}
\begin{aligned}
  \hat{\mu}_{\alpha,\beta}
  &=
  32\pi^{2} m_{\alpha}m_{\beta}\hat{\delta}(\ell_{1}\cdot v_{\alpha}) \hat{\delta}(\ell_{2}\cdot v_{\beta}) \,.
\end{aligned}\label{J_current}
\end{equation}
Finally, $d$-currents are
\begin{equation}
\begin{aligned}
  d^{\mu\nu}\big|^{2}_{\ell_{1},0}
  &=
  -\frac{m_{1}^2 \pi^{\epsilon +2} \sec(\pi \epsilon) \hat{\delta} (\ell_{1}\cdot v_{1})}{2^{1-4\epsilon} \Gamma (2-\epsilon ) |\ell_{1}|^{2\epsilon+1}}
  \Big[ |\ell_{1}|^{2} \big(\tilde{\eta}^{\mu\nu} +2 (\epsilon -1) \eta^{\mu\nu} +v_{1}^{\mu}v_{1}^{\nu}\big)
  	+ (1-2 \epsilon ) \ell_{1}^{\mu} \ell_{1}^{\nu}
  \Big]\,,
  \\
  d^{\mu\nu}\big|^{2}_{0,\ell_{2}}
  &=
  -\frac{m_{2}^2 \pi^{\epsilon +2} \sec(\pi \epsilon) \hat{\delta} (\ell_{2}\cdot v_{2})}{2^{1-4\epsilon} \Gamma (2-\epsilon ) |\ell_{2}|^{2\epsilon+1}}
  \Big[|\ell_{2}|^{2} \big(\tilde{\eta}^{\mu\nu} +2 (\epsilon -1) \eta^{\mu\nu} +v_{2}^{\mu}v_{2}^{\nu}\big)
  	+ (1-2 \epsilon ) \ell_{2}^{\mu} \ell_{2}^{\nu}
  \Big]\,,
  \\
  d^{\mu\nu}\big|^{2}_{\ell_{1},\ell_{2}}
  &=
  -\frac{\hat{\mu}_{1,2} }{|\ell_{1}|^{2}|\ell_{2}|^{2}}
  \Big( 
  2\ell_{1}^{(\mu}\ell_{2}^{\nu)}
  -\eta^{\mu\nu} (\ell_{1}\cdot \ell_{2})
  \Big)\,.
\end{aligned}\label{}
\end{equation}

Now we consider the worldline currents. Here we only focus on $X^{\mu}_{1}$. $X^{\rho}_{2}$ can be derived by swapping the label $1\leftrightarrow 2$. The zero-mode sector is
\begin{equation}
\begin{aligned}
  X^{\rho}_{1}\big|^{2}_{0,\ell_{2}}
  &=
  \bigg[
  \tfrac{i(8(1-2 \gamma^2)^2 \epsilon ^2 
    -4(3 \gamma^2-1) (4 \gamma^2-3) \epsilon 
    +3 (5 \gamma^2-1)(\gamma^2-1))\ell_{2}^{\rho}}{\left(1-\gamma^2\right) \Gamma (2-\epsilon )  \cos(\pi  \epsilon ) |\ell_{2}|}
  + \tfrac{4 \sqrt{\pi} (1-2 \gamma ^2)^2 w_{2}^{\rho}}{(\gamma^2-1)^{\frac{3}{2}} \Gamma (\frac{1}{2}-\epsilon ) \sin(\pi\epsilon )}
  \\&\qquad
  + \tfrac{4 \sqrt{\pi } \gamma  (1-2 \gamma ^2)^2 (\ell_{2}\cdot v_{1})\ell_{2}^{\rho}}{\left(\gamma ^2-1\right)^{3/2} \Gamma \left(\frac{1}{2}-\epsilon \right) \sin (\pi  \epsilon ) \ell_{2}^{2}}
  +\tfrac{4 i \gamma^2 ((2 \gamma ^2-1)^2 \epsilon-5 \gamma^4+6 \gamma ^2-2) (\ell_{2}\cdot v_{1})v_{1}^{\rho}}{\cos(\pi \epsilon)\left(\gamma ^2-1\right)^2 \Gamma (2-\epsilon ) |\ell_{2}|}
  \\&\qquad
  -\tfrac{4 i (1-2 \gamma ^2)^2 (\ell_{2}\cdot v_{1}) w_{1}^{\rho} }{\left(\gamma ^2-1\right)^2 \Gamma (1-\epsilon )\cos (\pi  \epsilon )  |\ell_{2}|}
  \bigg] \tfrac{m_{2}^2\pi^{2+\epsilon}\hat{\delta} (\ell_{2}\cdot v_{2})}{2^{1-4 \epsilon}|\ell_{2}|^{2\epsilon}(\ell_{2}\cdot v_{1})^{2}}\,,
\end{aligned}\label{}
\end{equation}
and the mixed-mode case is 
\begin{equation}
\begin{aligned}
  X^{\rho}\big|^{2}_{\ell_{1},\ell_{2}}\!
  &=\!
  \bigg(
  \Big[4 - \tfrac{2D_{12} D_5 \gamma}{D_{4}^{3}}
    +\tfrac{(1-2\gamma^{2})\left(D_{12}(D_{12}-2D_{2})-D_{1}(3D_{12}+4D_{4}^{2})\right)}{4D_{4}^{4}} 
  \Big] \ell_{1}^{\rho}
  \\&
  + \Big[
      \tfrac{2D_{5}(D_{1}\gamma-D_{4}D_{5})}{D_{4}^{3}}
    + \tfrac{(D_{1}^{2}+2D_{2}D_{4}^{2})(1-2\gamma^{2})}{2D_{4}^{4}}
    + \tfrac{D_{2}(D_{12}-2D_{2}) (1-2\gamma^{2})^{2} }{4D_{4}^{2}D_{5}^{2}}
  \Big] \ell_{12}^{\rho}
  \\&
  + \Big[
     \tfrac{2\gamma\left(4D_{5}^{2}-D_{2}(1-2\gamma^{2})\right)}{D_{4}D_{5}}
    -\tfrac{D_{1}D_{12}(1-2\gamma^{2})}{4D_{4}^{4}}
    +\tfrac{\left(D_{1} + 2(D_{4}^{2}-D_{12})\right)(1-2\gamma^{2})}{D_{4}^{2}}
  \Big]\ell_{2}^{\rho}
  \\&
  +\Big[\tfrac{ \left(2D_{1}^{2} -3D_{1}D_{12} +D_{12}^{2} -2D_{12}D_{2} +4D_{2}D_{4}^{2}\right)(1-2\gamma^{2})
  }{2D_{4}^{3}}
  + \tfrac{ 4 D_{5}\left((D_{1}-D_{12})\gamma-D_{4}D_{5}\right)}{D_{4}^{2}}
  \Big] v_{1}^{\rho}
  \\&
  +8D_{5} v_{2}^{\rho} 
  {+} \Big[
      \tfrac{2(2D_{4}^{2}-D_{12})}{D_{4}} 
    - \tfrac{4\gamma D_2}{D_5} 
    + \tfrac{D_{2}(D_{12}-2D_{2})(1-2\gamma^{2})}{2D_{4}D_{5}^{2}}
    \Big] \big(v_{1}^{\rho} {-}2\gamma v_{2}\big)
  \bigg) \tfrac{i \hat{\mu}_{12}}{D_{1}D_{12}D_{2}}\,.
\end{aligned}
\label{}\end{equation}
where
\begin{equation}
\begin{aligned}
  D_{1} = |\ell_{1}|^{2} \,,
  \quad
  D_{2} = |\ell_{2}|^{2} \,,
  \quad
  D_{12} = |\ell_{12}|^{2} \,,
  \quad
  D_{4} = \ell_{1}\cdot v_{2}\,,
  \quad 
  D_{5} = \ell_{2}\cdot v_{1}\,.
\end{aligned}\label{}
\end{equation}
%

\subsection{3PM graviton currents}\label{3PM_currents}

We now consider 3PM graviton currents and their associated subcurrents. We present the zero-mode sector only. In this sector the recursions for the $W, Z$ and $d$ currents generate one-loop integrals, and it is straightforward to evaluate them. First, $W$ currents are
\begin{equation}
\begin{aligned}
  W^{\mu\nu}\big|^{3}_{\ell,0}
  &=
  \frac{ m_{1}^3 2^{6 \epsilon +1} \pi ^{2 \epsilon +\frac{7}{2}} \csc (\pi  \epsilon ) \sec ^3(\pi  \epsilon )\hat{\delta} (\ell \cdot v_{1}) }{3 \Gamma (\frac{3}{2}-3 \epsilon ) \Gamma (1-\epsilon ) \Gamma (\epsilon -\frac{1}{2}) \Gamma (\epsilon +\frac{3}{2})}  |\ell|^{2-4 \epsilon } v_{1}^{\mu} v_{1}^{\nu}\,,
  \\
  W^{\mu\nu}\big|^{3}_{0,\ell}
  &=
  \frac{ m_{2}^3 2^{6 \epsilon +1} \pi ^{2 \epsilon +\frac{7}{2}} \csc (\pi  \epsilon ) \sec ^3(\pi  \epsilon )\hat{\delta} (\ell \cdot v_{2}) }{3 \Gamma (\frac{3}{2}-3 \epsilon ) \Gamma (1-\epsilon ) \Gamma (\epsilon -\frac{1}{2}) \Gamma (\epsilon +\frac{3}{2})}  |\ell|^{2-4 \epsilon } v_{2}^{\mu} v_{2}^{\nu}\,,
\end{aligned}\label{}
\end{equation}
and the $Z$ currents are
\begin{equation}
\begin{aligned}
  Z^{\mu\nu\rho\sigma}\big|^{3}_{0,\ell}
  &=
  \frac{m_{2}^3 \hat{\delta} (\ell \cdot v_{2})\Gamma (\frac{3}{2}-\epsilon )}{\Gamma(\frac{7}{2}-3 \epsilon)} \bigg[
  \frac{ 4^{3 \epsilon +1} \pi^{2 \epsilon +\frac{5}{2}} \csc (2 \pi  \epsilon ) \sec (\pi  \epsilon ) }{ \Gamma (1-\epsilon ) \Gamma (\epsilon +\frac{1}{2})}  (\tilde{\eta}^{\mu\rho} + v_{2}^{\mu}v_{2}^{\rho} )
  \\&\qquad\qquad\qquad\qquad\qquad\quad
  - \frac{8^{2 \epsilon +1} \pi ^{2 \epsilon +\frac{3}{2}} \sec^2(\pi \epsilon)\Gamma(\epsilon)}{\Gamma(\epsilon -\frac{1}{2})} 
	\frac{\ell^{\mu} \ell^{\rho}}{|\ell|^{2}} 
  \bigg]  |\ell|^{2-4 \epsilon}v_{2}^{\nu} v_{2}^{\sigma}\,,
\end{aligned}\label{}
\end{equation}
and
\begin{equation}
\begin{aligned}
  Z^{\mu\nu}\big|^{3}_{0,\ell}
  &=
  \frac{ 64^{\epsilon} m_{2}^3  \pi^{2 \epsilon +\frac{3}{2}}\hat{\delta} (\ell\cdot v_{2})}{\Gamma (\frac{7}{2}-3 \epsilon ) |\ell|^{4\epsilon-2}} 
  \Bigg[
    \bigg(\frac{3\pi^{2}(4 \epsilon -3)}{\Gamma (\epsilon -\frac{1}{2})} v_{2}^{\mu} v_{2}^{\nu}
  	- \frac{\pi \Gamma (\frac{3}{2}-\epsilon )}{\sec(\pi  \epsilon )} \tilde{\eta}^{\mu\nu}
  	-\frac{\pi ^{2} (\frac{5}{2}-3 \epsilon )}{\Gamma (\epsilon -\frac{1}{2})} \eta^{\mu\nu}
    \bigg) 
    \\&\qquad\qquad\qquad\qquad\qquad\times
    \frac{\csc(\pi\epsilon)\sec^3(\pi\epsilon )}{\Gamma(1-\epsilon) \Gamma(\epsilon +\frac{1}{2})} 
  +\frac{4  \sec ^2(\pi  \epsilon ) \Gamma (\frac{3}{2}-\epsilon ) \Gamma (\epsilon ) }{\Gamma (\epsilon -\frac{1}{2})|\ell|^{2}} \ell^{\mu} \ell^{\nu}
  \Bigg]\,.
\end{aligned}\label{}
\end{equation}
Finally, the $d$ currents are 
\begin{equation}
\begin{aligned}
  d^{\mu}\big|^{2}_{0,\ell}
  &=
  \frac{i m_{2}^2 16^{\epsilon} \pi^{\epsilon +2}  \sec (\pi \epsilon) \hat{\delta} (\ell \cdot v_{2})}{\Gamma (1-\epsilon )|\ell|^{2\epsilon +1}}
   \ell^{\mu}\,,
\end{aligned}\label{}
\end{equation}
and
\begin{equation}
\begin{aligned}
  d^{\mu\nu}\big|^{3}_{0,\ell}
  &=
  \frac{64^{\epsilon } \pi ^{2 \epsilon +\frac{5}{2}} m_{2}^3 \csc (2 \pi  \epsilon ) \sec (\pi  \epsilon ) \Gamma (\frac{3}{2}-\epsilon )  \hat{\delta} (\ell\cdot v_{2})}{\Gamma (\frac{7}{2}-3 \epsilon ) \Gamma (1-\epsilon ) \Gamma (\epsilon +\frac{1}{2})} |\ell|^{2-4 \epsilon} \tilde{\eta}^{\mu\nu}
  \\&\quad
  +\frac{ 2^{6 \epsilon -1} \pi ^{2 \epsilon +\frac{7}{2}} m_{2}^3 \csc (\pi  \epsilon ) \sec ^3(\pi  \epsilon )  \hat{\delta} (\ell \cdot v_{2})}{\Gamma (\frac{5}{2}-3 \epsilon ) \Gamma (1-\epsilon ) \Gamma(\epsilon -\frac{1}{2}) \Gamma (\epsilon +\frac{1}{2})} |\ell|^{2-4 \epsilon} \eta^{\mu\nu}
  \\&\quad
  +\frac{2^{6 \epsilon +1} \pi ^{2 \epsilon +\frac{7}{2}} m_{2}^3  \csc (\pi  \epsilon ) \sec ^3(\pi  \epsilon )  \hat{\delta} (\ell\cdot v_{2})}{\Gamma (\frac{7}{2}-3 \epsilon ) \Gamma (1-\epsilon ) \Gamma (\epsilon -\frac{1}{2})^2} |\ell|^{-4 \epsilon} \ell^{\mu} \ell^{\nu}
  \\&\quad
  +\frac{ 64^{\epsilon } \pi ^{2 \epsilon +\frac{5}{2}} m_{2}^3 (3 \epsilon -2)  \csc (\pi  \epsilon ) \sec ^2(\pi  \epsilon ) \Gamma (\frac{3}{2}-\epsilon ) \hat{\delta} (\ell\cdot v_{2})}{\Gamma (\frac{7}{2}-3 \epsilon ) \Gamma (1-\epsilon ) \Gamma (\epsilon +\frac{1}{2})} |\ell|^{2-4 \epsilon} v_{2}^{\mu} v_{2}^{\nu}
\end{aligned}\label{}
\end{equation}
Collecting the results, the graviton currents are given by
\begin{equation}
\begin{aligned}
  \mathfrak{J}^{\mu\nu}\big|^{3}_{0,\ell}
  &=
  \frac{2^{6 \epsilon +1} \pi ^{2 \epsilon +\frac{5}{2}} m_{2}^3 (4 \epsilon +3)   \csc (\pi  \epsilon ) \sec ^2(\pi  \epsilon ) \Gamma (-\epsilon -\frac{1}{2})  \hat{\delta} (\ell\cdot v_{2})}{\Gamma (\frac{5}{2}-3 \epsilon ) \Gamma (1-\epsilon ) \Gamma (\epsilon -\frac{1}{2})} |\ell|^{-4 \epsilon } v_{2}^{\mu} v_{2}^{\nu}\,,
\end{aligned}\label{}
\end{equation}
and the corresponding $\tilde{\mathfrak{J}}_{\mu\nu}\big|^{3}_{0,\ell}$, $H^{\ord{3}}$ and $\tilde{J}_{\mu\nu}\big|^{3}_{0,\ell}$ are
\begin{equation}
\begin{aligned}
  &\tilde{\mathfrak{J}}_{\mu\nu}\big|^{3}_{0,\ell}
  =-\frac{ 4^{3 \epsilon +1} \pi ^{2 \epsilon +\frac{7}{2}} m_{2}^3 (5 \epsilon +3)  \csc (\pi  \epsilon ) \sec ^3(\pi  \epsilon )  \hat{\delta} (\ell_{2}\cdot v_{2})}{\Gamma (\frac{5}{2}-3 \epsilon ) \Gamma (1-\epsilon ) \Gamma (\epsilon -\frac{1}{2}) \Gamma (\epsilon +\frac{3}{2})} |\ell|^{-4 \epsilon } v_{2}^{\mu} v_{2}^{\nu}\,,
  \\
  &H\big|^{3}_{0,\ell}
  =-\frac{2^{6 \epsilon +1} \pi ^{2 \epsilon +\frac{7}{2}} m_{2}^3 \csc (\pi  \epsilon ) \sec ^3(\pi  \epsilon )  \hat{\delta} (\ell_{2}\cdot v_{2})}{\Gamma (\frac{5}{2}-3 \epsilon ) \Gamma (\epsilon -\frac{1}{2}) \Gamma (-\epsilon ) \Gamma (\epsilon +\frac{3}{2})} |\ell|^{-4 \epsilon }\,.
  \\
  &\tilde{J}_{\mu\nu}\big|^{3}_{0,\ell}
  =
  \frac{m_{2}^3 2^{6 \epsilon -1} \pi^{2 \epsilon +\frac{3}{2}}
  (2 \epsilon +1) \csc (2 \pi  \epsilon ) \Gamma(-\epsilon -\frac{1}{2})^2 
  \hat{\delta}(\ell_{2}\cdot v_{2}) }{3 \Gamma \left(\frac{3}{2}-3 \epsilon \right) \Gamma (2-\epsilon )|\ell|^{2+4\epsilon}}
  \\&\quad \times
  \Big[4 \epsilon  (2 \epsilon -1) \ell^{\mu} \ell^{\nu} -|\ell^{2}|
  \big(2 (\epsilon {+}1) \tilde{\eta}^{\mu\nu} +(9{+}13 \epsilon{-}22 \epsilon^2) \eta^{\mu\nu} +2 (1 {+}5 \epsilon {-}4 \epsilon^2) v_{2}^{\mu}v_{2}^{\nu}\big)\Big]\,.
\end{aligned}\label{3PM_tJ_zero_mode}
\end{equation}


\begin{thebibliography}{99}  
 

\bibitem{Damgaard:2024fqj}
P.~H.~Damgaard and K.~Lee,
``Schwarzschild Black Hole from Perturbation Theory to All Orders,''
Phys. Rev. Lett. \textbf{132} (2024) no.25, 251603
[arXiv:2403.13216 [hep-th]].
%
\bibitem{Damgaard:2026kqg}
P.~H.~Damgaard, H.~Lee, K.~Lee and T.~Rahnuma,
``Gravitational Metric of a Star,''
[arXiv:2603.16493 [hep-th]].
%
\bibitem{Damgaard:2026ocb}
P.~H.~Damgaard, H.~Lee, K.~Lee and T.~Rahnuma,
``Iterative Solution of the Kerr Black Hole Metric,''
[arXiv:2605.19948 [hep-th]].
%
\bibitem{Bel:1981be}
L.~Bel, T.~Damour, N.~Deruelle, J.~Ibanez and J.~Martin,
``Poincar\'e-invariant gravitational field and equations of motion of two pointlike objects: The postlinear approximation of general relativity,''
Gen. Rel. Grav. \textbf{13} (1981), 963-1004
%
\bibitem{Westpfahl:1985tsl}
K.~Westpfahl,
``High-Speed Scattering of Charged and Uncharged Particles in General Relativity,''
Fortsch. Phys. \textbf{33} (1985) no.8, 417-493
%
\bibitem{Damour:2016gwp}  
T.~Damour,  
``Gravitational scattering, Post-Minkowskian approximation and Effective One-Body theory,''  
Phys. Rev. D \textbf{94} (2016) no.10, 104015;  
[arXiv: 1609.00354 [gr-qc]].  
%
\bibitem{Damour:2017zjx}  
T.~Damour,  
``High-energy gravitational scattering and the general relativistic two-body problem,''  
Phys. Rev. D \textbf{97} (2018) no.4, 044038;  
[arXiv:1710.10599 [gr-qc]].  
%
\bibitem{Bjerrum-Bohr:2018xdl}  
N.~E.~J.~Bjerrum-Bohr, P.~H.~Damgaard, G.~Festuccia, L.~Plant\'e and P.~Vanhove,  
``General Relativity from Scattering Amplitudes,''  
Phys. Rev. Lett. \textbf{121} (2018) no.17, 171601;  
[arXiv:1806.04920 [hep-th]].  
%
\bibitem{Cheung:2018wkq}  
C.~Cheung, I.~Z.~Rothstein and M.~P.~Solon,  
``From Scattering Amplitudes to Classical Potentials in the Post-Minkowskian Expansion,''  
Phys. Rev. Lett. \textbf{121} (2018) no.25, 251101;  
[arXiv:1808.02489 [hep-th]].  
%
\bibitem{Bern:2019nnu}  
Z.~Bern, C.~Cheung, R.~Roiban, C.~H.~Shen, M.~P.~Solon and M.~Zeng,  
``Scattering Amplitudes and the Conservative Hamiltonian for Binary Systems at Third Post-Minkowskian Order,''  
Phys. Rev. Lett. \textbf{122} (2019) no.20, 201603;  
[arXiv:1901.04424 [hep-th]]. 
 
\bibitem{Bern:2019crd} 
Z.~Bern, C.~Cheung, R.~Roiban, C.~H.~Shen, M.~P.~Solon and M.~Zeng, 
``Black Hole Binary Dynamics from the Double Copy and Effective Theory,'' 
JHEP \textbf{10} (2019), 206 
[arXiv:1908.01493 [hep-th]]. 
 
\bibitem{Damour:2019lcq} 
T.~Damour, 
``Classical and Quantum Scattering in Post-Minkowskian Gravity,'' 
Phys. Rev. D \textbf{102} (2020) no.2, 024060 
[arXiv:1912.02139 [gr-qc]]. 
 
\bibitem{DiVecchia:2020ymx}  
P.~Di Vecchia, C.~Heissenberg, R.~Russo and G.~Veneziano,  
``Universality of Ultra-Relativistic Gravitational Scattering,''  
Phys. Lett. B \textbf{811} (2020), 135924  
[arXiv:2008.12743 [hep-th]].  

\bibitem{Damour:2020tta}
T.~Damour,
``Radiative contribution to classical gravitational scattering at the third order in $G$,''
Phys. Rev. D \textbf{102} (2020) no.12, 124008
doi:10.1103/PhysRevD.102.124008
[arXiv:2010.01641 [gr-qc]].
 
 
\bibitem{DiVecchia:2021ndb}
P.~Di Vecchia, C.~Heissenberg, R.~Russo and G.~Veneziano,
``Radiation Reaction from Soft Theorems,''
Phys. Lett. B \textbf{818} (2021), 136379
[arXiv:2101.05772 [hep-th]].

\bibitem{DiVecchia:2021bdo}
P.~Di Vecchia, C.~Heissenberg, R.~Russo and G.~Veneziano,
``The eikonal approach to gravitational scattering and radiation at $ \mathcal{O} $(G$^{3}$),''
JHEP \textbf{07} (2021), 169
[arXiv:2104.03256 [hep-th]]. 
 
\bibitem{Bjerrum-Bohr:2021vuf} 
N.~E.~J.~Bjerrum-Bohr, P.~H.~Damgaard, L.~Plant\'e and P.~Vanhove, 
``Classical gravity from loop amplitudes,'' 
Phys. Rev. D \textbf{104} (2021) no.2, 026009 
[arXiv:2104.04510 [hep-th]]. 
 
\bibitem{Bjerrum-Bohr:2021din} 
N.~E.~J.~Bjerrum-Bohr, P.~H.~Damgaard, L.~Plant\'e and P.~Vanhove, 
``The amplitude for classical gravitational scattering at third Post-Minkowskian order,'' 
JHEP \textbf{08} (2021), 172 
[arXiv:2105.05218 [hep-th]]. 
 
 
\bibitem{DiVecchia:2022owy} 
P.~Di Vecchia, C.~Heissenberg and R.~Russo, 
``Angular momentum of zero-frequency gravitons,'' 
JHEP \textbf{08} (2022), 172 
[arXiv:2203.11915 [hep-th]]. 
 
\bibitem{DiVecchia:2022piu} 
P.~Di Vecchia, C.~Heissenberg, R.~Russo and G.~Veneziano,
``Classical gravitational observables from the Eikonal operator,''
Phys. Lett. B \textbf{843} (2023), 138049
[arXiv:2210.12118 [hep-th]].
%
\bibitem{Damgaard:2021rnk}
P.~H.~Damgaard and P.~Vanhove,
Phys. Rev. D \textbf{104} (2021) no.10, 104029
doi:10.1103/PhysRevD.104.104029
[arXiv:2108.11248 [hep-th]]. 
 %
\bibitem{Cristofoli:2019neg} 
A.~Cristofoli, N.~E.~J.~Bjerrum-Bohr, P.~H.~Damgaard and P.~Vanhove, 
``Post-Minkowskian Hamiltonians in general relativity,'' 
Phys. Rev. D \textbf{100} (2019) no.8, 084040 
[arXiv:1906.01579 [hep-th]]. 
 %
 \bibitem{Cristofoli:2020uzm}  
A.~Cristofoli, P.~H.~Damgaard, P.~Di Vecchia and C.~Heissenberg,  
``Second-order Post-Minkowskian scattering in arbitrary dimensions,''  
JHEP \textbf{07} (2020), 122;  
[arXiv:2003.10274 [hep-th]].  
%
\bibitem{Kalin:2019rwq} 
G.~K\"alin and R.~A.~Porto, 
``From Boundary Data to Bound States,'' 
JHEP \textbf{01} (2020), 072 
[arXiv:1910.03008 [hep-th]]. 
 
 
 
\bibitem{Bjerrum-Bohr:2019kec} 
N.~E.~J.~Bjerrum-Bohr, A.~Cristofoli and P.~H.~Damgaard, 
``Post-Minkowskian Scattering Angle in Einstein Gravity,'' 
JHEP \textbf{08} (2020), 038 
[arXiv:1910.09366 [hep-th]]. 
 %
 \bibitem{Damgaard:2022jem}
P.~H.~Damgaard, J.~Hoogeveen, A.~Luna and J.~Vines,
``Scattering angles in Kerr metrics,''
Phys. Rev. D \textbf{106} (2022) no.12, 124030
doi:10.1103/PhysRevD.106.124030
[arXiv:2208.11028 [hep-th]].
  
 
\bibitem{Kosower:2018adc} 
D.~A.~Kosower, B.~Maybee and D.~O'Connell, 
``Amplitudes, Observables, and Classical Scattering,'' 
JHEP \textbf{02} (2019), 137 
[arXiv:1811.10950 [hep-th]]. 
 
\bibitem{Maybee:2019jus} 
B.~Maybee, D.~O'Connell and J.~Vines, 
``Observables and amplitudes for spinning particles and black holes,'' 
JHEP \textbf{12} (2019), 156 
[arXiv:1906.09260 [hep-th]]. 
%
\bibitem{Cristofoli:2021vyo} 
A.~Cristofoli, R.~Gonzo, D.~A.~Kosower and D.~O'Connell, 
``Waveforms from amplitudes,'' 
Phys. Rev. D \textbf{106} (2022) no.5, 056007 
[arXiv:2107.10193 [hep-th]]. 
%
\bibitem{Damgaard:2023vnx}
P.~H.~Damgaard, E.~R.~Hansen, L.~Plant\'e and P.~Vanhove,
``The relation between KMOC and worldline formalisms for classical gravity,''
JHEP \textbf{09} (2023), 059
doi:10.1007/JHEP09(2023)059
[arXiv:2306.11454 [hep-th]].
%
\bibitem{Damgaard:2023ttc}
P.~H.~Damgaard, E.~R.~Hansen, L.~Plant\'e and P.~Vanhove,
``Classical observables from the exponential representation of the gravitational S-matrix,''
JHEP \textbf{09} (2023), 183
doi:10.1007/JHEP09(2023)183
[arXiv:2307.04746 [hep-th]].
 
 
\bibitem{KoemansCollado:2019ggb}  
A.~Koemans Collado, P.~Di Vecchia and R.~Russo,  
``Revisiting the Second Post-Minkowskian Eikonal and the Dynamics of Binary Black Holes,''  
Phys. Rev. D \textbf{100} (2019) no.6, 066028  
[arXiv:1904.02667 [hep-th]].  
    
\bibitem{Parra-Martinez:2020dzs}  
J.~Parra-Mart{\'\i ne}z, M.~S.~Ruf and M.~Zeng,  
``Extremal Black Hole Scattering at $\mathcal{O}(G^3)$: Graviton Dominance, Eikonal Exponentiation, and Differential Equations,''  
JHEP \textbf{11} (2020), 023  
[arXiv:2005.04236 [hep-th]]. 
 
\bibitem{Bellazzini:2022wzv} 
B.~Bellazzini, G.~Isabella and M.~M.~Riva, 
``Classical vs quantum eikonal scattering and its causal structure,'' 
JHEP \textbf{04} (2023), 023 
[arXiv:2211.00085 [hep-th]]. 
%
\bibitem{Damgaard:2021ipf} 
P.~H.~Damgaard, L.~Plante and P.~Vanhove, 
``On an exponential representation of the gravitational S-matrix,'' 
JHEP \textbf{11} (2021), 213 
[arXiv:2107.12891 [hep-th]]. 
 
\bibitem{Damgaard:2019lfh} 
P.~H.~Damgaard, K.~Haddad and A.~Helset, 
``Heavy Black Hole Effective Theory,'' 
JHEP \textbf{11} (2019), 070 
[arXiv:1908.10308 [hep-ph]]. 
 
\bibitem{Aoude:2020onz} 
R.~Aoude, K.~Haddad and A.~Helset, 
``On-shell heavy particle effective theories,'' 
JHEP \textbf{05} (2020), 051 
[arXiv:2001.09164 [hep-th]]. 
 
\bibitem{Brandhuber:2021eyq} 
A.~Brandhuber, G.~Chen, G.~Travaglini and C.~Wen, 
``Classical gravitational scattering from a gauge-invariant double copy,'' 
JHEP \textbf{10} (2021), 118 
[arXiv:2108.04216 [hep-th]]. 
 
\bibitem{Cho:2021nim}
K.~Cho, K.~Kim and K.~Lee,
``The off-shell recursion for gravity and the classical double copy for currents,''
JHEP \textbf{01}, 186 (2022)
[arXiv:2109.06392 [hep-th]].
%
\bibitem{Lee:2022aiu}
K.~Lee,
``Quantum off-shell recursion relation,''
JHEP \textbf{05}, 051 (2022)
[arXiv:2202.08133 [hep-th]].
%
\bibitem{Cho:2022faq}
K.~Cho, K.~Kim and K.~Lee,
``Perturbations of general relativity to all orders and the general n$^{th}$ order terms,''
JHEP \textbf{03}, 112 (2023)
[arXiv:2209.11424 [hep-th]].
%
\bibitem{Adamo:2023cfp}
T.~Adamo, A.~Cristofoli, A.~Ilderton and S.~Klisch,
``Scattering amplitudes for self-force,''
[arXiv:2307.00431 [hep-th]].
%
\bibitem{Lee:2023zuu}
H.~Lee, K.~Lee and S.~Lee,
``Poincar\'e generators at second post-Minkowskian order,''
JHEP \textbf{10}, 044 (2023)
[arXiv:2307.05626 [hep-th]].
%
\bibitem{Cho:2023kux}
K.~Cho, K.~Kim and K.~Lee,
``Binary Black Holes and Quantum Off-Shell Recursion,''
[arXiv:2311.01284 [hep-th]].
%
\bibitem{Tao:2023yxy}
Y.~X.~Tao,
``Berends-Giele currents for extended gravity,''
Phys. Rev. D \textbf{108}, no.12, 125020 (2023)
[arXiv:2309.15657 [hep-th]].
%
\bibitem{Cheung:2023lnj}
C.~Cheung, J.~Parra-Martinez, I.~Z.~Rothstein, N.~Shah and J.~Wilson-Gerow,
``Effective Field Theory for Extreme Mass Ratio Binaries,''
Phys. Rev. Lett. \textbf{132} (2024) no.9, 9
doi:10.1103/PhysRevLett.132.091402
[arXiv:2308.14832 [hep-th]].
%
\bibitem{Cheung:2024byb}
C.~Cheung, J.~Parra-Martinez, I.~Z.~Rothstein, N.~Shah and J.~Wilson-Gerow,
``Gravitational scattering and beyond from extreme mass ratio effective field theory,''
JHEP \textbf{10} (2024), 005
doi:10.1007/JHEP10(2024)005
[arXiv:2406.14770 [hep-th]].
%
 
\bibitem{Kalin:2020mvi} 
G.~K\"alin and R.~A.~Porto, 
``Post-Minkowskian Effective Field Theory for Conservative Binary Dynamics,'' 
JHEP \textbf{11} (2020), 106 
[arXiv:2006.01184 [hep-th]]. 
 
\bibitem{Kalin:2020fhe} 
G.~K\"alin, Z.~Liu and R.~A.~Porto, 
``Conservative Dynamics of Binary Systems to Third Post-Minkowskian Order from the Effective Field Theory Approach,'' 
Phys. Rev. Lett. \textbf{125} (2020) no.26, 261103 
[arXiv:2007.04977 [hep-th]]. 
 
\bibitem{Kalin:2020lmz} 
G.~K\"alin, Z.~Liu and R.~A.~Porto, 
``Conservative Tidal Effects in Compact Binary Systems to Next-to-Leading Post-Minkowskian Order,'' 
Phys. Rev. D \textbf{102} (2020), 124025 
[arXiv:2008.06047 [hep-th]]. 
 
\bibitem{Mogull:2020sak} 
G.~Mogull, J.~Plefka and J.~Steinhoff, 
``Classical black hole scattering from a worldline quantum field theory,'' 
JHEP \textbf{02} (2021), 048 
[arXiv:2010.02865 [hep-th]]. 
 
\bibitem{Jakobsen:2021smu} 
G.~U.~Jakobsen, G.~Mogull, J.~Plefka and J.~Steinhoff, 
``Classical Gravitational Bremsstrahlung from a Worldline Quantum Field Theory,'' 
Phys. Rev. Lett. \textbf{126} (2021) no.20, 201103 
[arXiv:2101.12688 [gr-qc]]. 
 
\bibitem{Mougiakakos:2021ckm} 
S.~Mougiakakos, M.~M.~Riva and F.~Vernizzi, 
``Gravitational Bremsstrahlung in the post-Minkowskian effective field theory,'' 
Phys. Rev. D \textbf{104} (2021) no.2, 024041 
[arXiv:2102.08339 [gr-qc]]. 
 
\bibitem{Riva:2021vnj} 
M.~M.~Riva and F.~Vernizzi, 
``Radiated momentum in the post-Minkowskian worldline approach via reverse unitarity,'' 
JHEP \textbf{11} (2021), 228 
[arXiv:2110.10140 [hep-th]]. 
 
\bibitem{Riva:2022fru} 
M.~M.~Riva, F.~Vernizzi and L.~K.~Wong, 
``Gravitational bremsstrahlung from spinning binaries in the post-Minkowskian expansion,'' 
Phys. Rev. D \textbf{106} (2022) no.4, 044013 
[arXiv:2205.15295 [hep-th]]. 
 
\bibitem{Liu:2021zxr} 
Z.~Liu, R.~A.~Porto and Z.~Yang, 
``Spin Effects in the Effective Field Theory Approach to Post-Minkowskian Conservative Dynamics,'' 
JHEP \textbf{06} (2021), 012 
[arXiv:2102.10059 [hep-th]]. 
 
\bibitem{Dlapa:2021npj} 
C.~Dlapa, G.~K\"alin, Z.~Liu and R.~A.~Porto, 
``Dynamics of binary systems to fourth Post-Minkowskian order from the effective field theory approach,'' 
Phys. Lett. B \textbf{831} (2022), 137203 
[arXiv:2106.08276 [hep-th]]. 
 
\bibitem{Jakobsen:2021lvp} 
G.~U.~Jakobsen, G.~Mogull, J.~Plefka and J.~Steinhoff, 
``Gravitational Bremsstrahlung and Hidden Supersymmetry of Spinning Bodies,'' 
Phys. Rev. Lett. \textbf{128} (2022) no.1, 011101 
[arXiv:2106.10256 [hep-th]]. 
 
\bibitem{Jakobsen:2021zvh} 
G.~U.~Jakobsen, G.~Mogull, J.~Plefka and J.~Steinhoff, 
``SUSY in the sky with gravitons,'' 
JHEP \textbf{01} (2022), 027 
[arXiv:2109.04465 [hep-th]]. 
 
\bibitem{Dlapa:2021vgp} 
C.~Dlapa, G.~K\"alin, Z.~Liu and R.~A.~Porto, 
``Conservative Dynamics of Binary Systems at Fourth Post-Minkowskian Order in the Large-Eccentricity Expansion,'' 
Phys. Rev. Lett. \textbf{128} (2022) no.16, 161104 
[arXiv:2112.11296 [hep-th]]. 
 
\bibitem{Jakobsen:2022fcj} 
G.~U.~Jakobsen and G.~Mogull, 
``Conservative and Radiative Dynamics of Spinning Bodies at Third Post-Minkowskian Order Using Worldline Quantum Field Theory,'' 
Phys. Rev. Lett. \textbf{128} (2022) no.14, 141102 
[arXiv:2201.07778 [hep-th]]. 
 
\bibitem{Jakobsen:2022psy} 
G.~U.~Jakobsen, G.~Mogull, J.~Plefka and B.~Sauer, 
``All things retarded: radiation-reaction in worldline quantum field theory,'' 
JHEP \textbf{10} (2022), 128 
[arXiv:2207.00569 [hep-th]]. 
 
\bibitem{Kalin:2022hph} 
G.~K\"alin, J.~Neef and R.~A.~Porto, 
``Radiation-reaction in the Effective Field Theory approach to Post-Minkowskian dynamics,'' 
JHEP \textbf{01} (2023), 140 
[arXiv:2207.00580 [hep-th]]. 
 
\bibitem{Dlapa:2022lmu} 
C.~Dlapa, G.~K\"alin, Z.~Liu, J.~Neef and R.~A.~Porto, 
``Radiation Reaction and Gravitational Waves at Fourth Post-Minkowskian Order,'' 
Phys. Rev. Lett. \textbf{130} (2023) no.10, 101401 
[arXiv:2210.05541 [hep-th]]. 
 
\bibitem{Jakobsen:2022zsx} 
G.~U.~Jakobsen and G.~Mogull, 
``Linear response, Hamiltonian, and radiative spinning two-body dynamics,'' 
Phys. Rev. D \textbf{107} (2023) no.4, 044033 
[arXiv:2210.06451 [hep-th]]. 
 
 \bibitem{Dlapa:2023hsl}
C.~Dlapa, G.~K{\"a}lin, Z.~Liu and R.~A.~Porto,
``Bootstrapping the relativistic two-body problem,''
JHEP \textbf{08} (2023), 109
doi:10.1007/JHEP08(2023)109
[arXiv:2304.01275 [hep-th]].
 
%
\bibitem{Jakobsen:2023hig}
G.~U.~Jakobsen, G.~Mogull, J.~Plefka and B.~Sauer,
``Dissipative Scattering of Spinning Black Holes at Fourth Post-Minkowskian Order,''
Phys. Rev. Lett. \textbf{131} (2023) no.24, 241402
doi:10.1103/PhysRevLett.131.241402
[arXiv:2308.11514 [hep-th]].
%

%
\bibitem{Ajith:2024fna}
S.~Ajith, Y.~Du, R.~Rajagopal and D.~Vaman,
``Worldline formalism, eikonal expansion and the classical limit of scattering amplitudes,''
Nucl. Phys. B \textbf{1025} (2026), 117367
doi:10.1016/j.nuclphysb.2026.117367
[arXiv:2409.17866 [hep-th]].

%
\bibitem{Bini:2024hme}
D.~Bini, T.~Damour and A.~Geralico,
``Explicit solution of the gravitational two-body problem at the second post-Minkowskian order,''
Phys. Rev. D \textbf{110} (2024) no.10, 104051
doi:10.1103/PhysRevD.110.104051
[arXiv:2408.17193 [gr-qc]].

\bibitem{Lee:2013mka}
R.~N.~Lee,
``LiteRed 1.4: a powerful tool for reduction of multiloop integrals,''
J. Phys. Conf. Ser. \textbf{523} (2014), 012059
doi:10.1088/1742-6596/523/1/012059
[arXiv:1310.1145 [hep-ph]].

%
\bibitem{Davydychev:1991va}
A.~I.~Davydychev,
``A Simple formula for reducing Feynman diagrams to scalar integrals,''
Phys. Lett. B \textbf{263} (1991), 107-111
doi:10.1016/0370-2693(91)91715-8
 
  
\end{thebibliography}
\end{document}